\documentclass[a4paper,twoside]{book}
\usepackage[dvips]{graphicx}	%Zum Einfuegen von Graphiken
\usepackage[]{amssymb}
\usepackage[]{latexsym}
\usepackage[]{amsfonts}
\begin{document}
%===============

\newcommand{\be}{\begin{equation}}
\newcommand{\ee}{\end{equation}}
\newcommand{\ba}{\begin{eqnarray}}
\newcommand{\ea}{\end{eqnarray}}
\newcommand{\eref}[1]{(\ref{#1})}
\def\C{\mathbb{C}}
\def\R{\mathbb{R}}
\def\Z{\mathbb{Z}}
\def\N{\mathbb{N}}
\renewcommand{\v}[1]{{\bf #1}}
\def\tr{\mathop{\rm Tr}}
\def\mod{\mathop{\rm mod}}
\def\arsinh{\mathop{\rm arsinh}}
\def\arccot{\mathop{\rm arccot}}
\def\up{\uparrow}
\def\down{\downarrow}
\newcommand{\ud}{\mathrm{d}}
\def\Min{\mathop{\rm Min}}
\def\Max{\mathop{\rm Max}}
\def\P{\mathop{\rm P}}
\def\sign{\mathrm{sign}}

\def\Map#1{\smash{\mathop{\hbox to 35 pt{$\,$\rightarrowfill$\,\,$}}
                             \limits^{\scriptstyle#1}}}
\def\MapRight#1{\smash{\mathop{\hbox to 35pt{\rightarrowfill}}\limits^{#1}}}

% Abbildungen, wo man die Laenge waehlen kann und  drueber und drunter beschriften: 
\def\MapRightLong#1#2#3{\smash{\mathop{\hbox to #1mm{\rightarrowfill}}
   \limits^{\scriptstyle#2}_{\scriptstyle#3}}}

\newcommand{\Rahmen}[1]{\vspace{0.2cm}\par%
   \setlength{\dimen0}{\textwidth}
   \addtolength{\dimen0}{-0.3cm}
   \hspace{-0.6cm}\fbox{\parbox{\dimen0}{#1}}\vspace{0.2cm}\par}
% macht einen Rahmen um mehrzeiligen Text ueber ganze Zeilenlaenge.

\newcommand{\Qed}{\quad\nobreak\hspace*{\fill}\mbox{$\square$}\par}
% quod erat demonstrandum: nicht verwenden innerhalb von $$..$$

%\title{Weak-Coupling Instabilities of Two-Dimensional Lattice Electrons}

%\title{Instabilit\'es \`a couplage faible de syst\`emes \'electroniques sur un r\'eseau bidimensionnel}

%\author{Benedikt Binz}
%\date{}

\frontmatter
%\maketitle
%#########

%\selectlanguage{french}
\begin{titlepage}
\begin{center}
\parbox{4.8cm}{
\begin{center}Universit\'e de Fribourg (Suisse)

D\'epartement de Physique\end{center}}\hfill et \hfill
\parbox{5.2cm}{
\begin{center}Universit\'e Paris 7 - Denis Diderot

\textsc{UFR} de Physique\end{center}}

\vfill

\vfill

{\LARGE\bf Weak-Coupling Instabilities of Two-Dimensional Lattice Electrons}

\vfill

\vfill

{\Large\bf {Th\`ese}}

\vfill

%pr\'esent\'e pour l'obtention du grade de {\scshape {\it DOCTOR RERUM NATURALIUM} de la Facul\'e des Sciences de l'Universit\'e de Fribourg (Suisse)}

pour l'obtention du grade de {\scshape Doctor rerum naturalium} 
%\newline  

{\scshape de la Facul\'e des Sciences de l'Universit\'e de Fribourg (Suisse)}

et

pour l'obtention du Dipl\^ome de {\scshape Docteur de l'Universit\'e Paris 7

 Sp\'ecialit\'e: physique th\'eorique}

\vspace{3mm}
present\'ee par 

\vfill

{%\bfseries 
\LARGE Benedikt \textsc{Binz}}

%\vfill

%dans le cadre d'une cotutelle de th\`ese

\vfill

\vfill

{\bf Directeurs de th\`ese:}

Prof. Dionys \textsc{Baeriswyl} (Universit\'e de Fribourg)

Dr. Benoit \textsc{Dou\c cot} (Universit\'e Paris 7)

\vfill

\textsc{ \bf Jury}
\vspace{3mm}

\parbox{5.5cm}{Prof. T. Maurice \textsc{Rice}

Prof. Vincent \textsc{Rivasseau}

Prof. Bernard \textsc{Roulet}

Prof. Peter \textsc{Schurtenberger}}\parbox{2cm}{Rapporteur

Rapporteur

Examinateur

Pr\'esident} 

\vfill

Date et lieu de la soutenance publique: 15 avril 2002 \`a Fribourg 

\vfill

 Num\'ero de la th\`ese (Universit\'e de Fribourg): 1374

 Imprimerie St. Paul, Fribourg

 2002

\end{center}

\newpage

 \vspace{5mm}

 Accept\'ee par la Facult\'e des Sciences de l'Universit\'e de Fribourg (Suisse) sur la proposition du Jury.

 \vspace{5mm}

 Fribourg, le 15 avril 2002

 \vspace{15mm}

\parbox{10cm}{Directeur de th\`ese:

\vspace{1cm}

Prof. D. Baeriswyl }\parbox{5cm}{Le Doyen:

\vspace{1cm}

Prof. A. von Zelewsky}

% Prof. D. Baeriswyl  \hfill \hfill  Le Doyen:  \hfill \hspace{0.01cm}

\end{titlepage}

\chapter{Remerciements}
%======================

En tout premier lieu, je remercie mon ``Doktorvater'' Dionys Baeriswyl qui m'a guid\'e et soutenu avec beaucoup d'engagement et une grande comp\'etence du d\'ebut jusqu'\'a la fin. Mais ma th\`ese n'aurait pas sa valeur sans mon directeur de th\`ese parisien Beno\^it Dou\c cot. J'ai \'enorm\'ement profit\'e de son savoir, de son intuition et de sa capacit\'e \`a toujours me poser les bonnes questions. 

\hspace{1mm}

Je remercie aussi 
\hspace{1mm}

\begin{itemize}
\item les membres du jury T. Maurice Rice, Vincent Rivasseau et Bernard Roulet  pour leurs remarques et questions utiles.   

\item les membres du LPTHE pour leur hospitalit\'e pendant mes s\'ejours \`a Paris. Je voudrais tout particuli\`erement mentionner Drazen Zanchi avec qui j'ai beaucoup appris.

\item le groupe de physique th\'eorique \`a Fribourg (membres anciens et pr\'esents) pour leur coll\'egialit\'e et pour le bon climat intellectuel et personnel. 

\item Xavier Bagnoud pour la correction de divers textes fran\c cais.

\item  le Fonds National Suisse et la Conf\'erence Universitaire Suisse pour le financement.

\item N. Dupuis, A. Ferraz, F. Gebhard, F. Guinea, G. I. Japaridze, A. A. Katanin, W. Metzner, M. Salmhofer, F. Vistulo de Abreu, D. Vollhardt and M. A. H. Vozmediano pour des discussions utiles et stimulantes.
\end{itemize}

\hspace{1mm}

Heureusement qu'il y a des gens qui me font constamment d\'ecouvrir la vie en dehors de la physique. Je remercie donc Karla et Marc Strebel-Pauchard avec lesquels j'ai partag\'e l'appartement pendant longtemps. Et surtout, je remercie ma partenaire de vie Lotti Egger, simplement pour tout.

%\selectlanguage{german}
\chapter{Zusammenfassung}
%=========================

Das Problem wechselwirkender Elektronen in zwei Dimensionen ist sp\"atestens seit der Entde"ckung der Hoch-Temperatur-Supraleitung von grosser Aktualit\"at und Gegenstand intensiver For\-schungs\-t\"atig\-keit. In der vorliegenden Arbeit wird ein erweitertes Hubbard-Modell auf einem Quadratgitter untersucht. Obwohl man weiss, dass die Leitungselektronen in den Kupraten relativ stark miteinander wechselwirken, beschr\"anke ich mich hier bewusst auf den Grenzfall einer schwachen Wechselwirkung zwischen den Elektronen. 

Selbst eine beliebig schwache Wechselwirkung f\"uhrt bei gen\"ugend tiefer Temperatur zur Bildung von starken Korrelationen und schliesslich zum Phasen\"ubergang in einen geordneten Zustand. Die Natur dieser Instabilit\"at h\"angt stark von den Eigenschaften der Fermi-Fl\"ache ab. Wenn keine besondere Situation vorliegt, ist es ein \"Ubergang in einen supraleitenden Zustand. Dies obwohl die Wechselwirkung der Elektronen rein repulsiv ist und keine phononischen Freiheitsgrade ber\"u"cksichtigt werden. Eine m\"ogliche Interpretation dieses Resultats ist, dass der Austausch antiferromagnetischer Spinfluktuationen auf eine effektive Anziehung zwischen den Quasiteilchen f\"uhrt.  

Im Spezialfall halber Bandf\"ullung erf\"ullt die Fermi-Fl\"ache die "perfect nesting"-Eigenschaft. Gleichzeitig hat die Dispersionsrelation der Elektronen einen Sattelpunkt am Fermi-Niveau, was zu einer Van-Hove-Singularit\"at in der Zustandsdichte f\"uhrt. Diese beiden Eigenschaften bewirken, dass es zu einem Wettbewerb zwischen sechs verschiedenen Instabilit\"aten kommt. Neben $s$- und $d$-Wellen Supraleitung kann eine Spindichtewelle oder Ladungsdichtewelle auftreten. Die zwei weiteren Instabilit\"aten sind sogenannte Flussphasen, die durch spontan zirkulierende Ladungsstr\"ome, respektive Spinstr\"ome charakterisiert sind. 

Um die konkurrierenden Instabilit\"aten vorurteilslos gegeneinander abzuw\"agen, ben\"otigt man die Technik der Renormierungsgruppe. Der n\"otige Formalismus wird hier auf einem m\"oglichst elementaren Niveau pr\"asentiert. Die Idee ist dabei, ausgehend von der naiven St\"orungsrechnung die dominanten Terme (d.h. die Klasse der Parkett-Diagramme) konsistent aufzusummieren. Das modernere Konzept der Wilson'schen effektiven Wirkung wird ebenfalls kurz diskutiert und zu dem hier verwendeten Ansatz in Bezug gebracht.

Als Resultat erhalte ich ein Phasendiagramm als Funktion der Modell-Wechselwirkung, die neben dem \"ublichen Hubbard-Term auch eine Dichte-Dichte- und Spin-Spin-Wechselwirkung zwischen benachbarten Gitterpl\"atzen enth\"alt. F\"ur das repulsive Hubbard-Modell erh\"alt man wie erwartet eine Spindichtewelle. Dieses Phasendiagram ist im Grenzfall unendlich schwacher Wechselwirkung vermutlich exakt, denn auf der Grenzlinie zwischen zwei benachbarten Phasen weist das System dann jeweils eine besondere Symmetrie auf, die die Entartung der beiden benachbarten Phasen garantiert. Das physikalische Bild dieser Instabilit\"aten wird erg\"anzt durch Gr\"ossen wie die uniforme Spin-Suszeptibilit\"at und die Ladungs-Kompressibilit\"at, deren Verhalten bei tiefen Temperaturen qualitativ vorausgesagt wird. Dabei zeigt sich eine allgemeine Tendenz zur spontanen Verformung der Fermi-F\"ache (Pomeranchuk-Instabilit\"at), die jedoch nicht notwendigerweise mit einer Zerst\"orung der "perfect nesting"-Eigenschaft einhergeht. 

Obwohl sich die konkurrierenden Supraleitungs-, Spin- und Ladungskorrelationen gegenseitig beeinflussen, kann sich die f\"uhrende Instabilit\"at letztlich als einzige ungehindert entwi"ckeln. Dieses Resultat ist im Widerspruch zu \"alteren L\"osungsvorschl\"agen, wo die Spindichtewelle jeweils zusammen mit der $d$-Wellen Supraleitung und der Ladungs-Flussphase auftrat.

Abschliessend stellt man fest, dass sich zwei wesentliche Eigenschaften der supraleitenden Kuprate im schwach wechselwirkenden zweidimensionalen Hubbard-Modell wiederfinden. N\"amlich eine antiferromagnetisch geordnete Phase bei halber Bandf\"ullung und das Auftreten von $d$-Wellen Supraleitung im dotierten Material. Leider ergibt sich aus der vorliegenden Theorie, die von schwacher Wechselwirkung ausgeht, keine Erkl\"arung der aussergew\"ohnlichen Eigenschaften in der metallischen Phase des optimal dotierten und unterdotierten Bereichs. Starke Wechselwirkung scheint daher ein unabdingbarer Bestandteil f\"ur eine ad\"aquate Theorie dieser Materialien zu sein. 

Trotzdem k\"onnte sich eine konsequente Weiterf\"uhrung des hier verwendeten st\"orungstheoretischen Ansatzes als durchaus fruchtbar erweisen, wobei unter anderem die Selbstenergie der Elektronen wichtig wird. Die dazu notwendige Technik ist jedoch noch nicht ausreichend entwickelt.

%\selectlanguage{french}

\chapter{R\'esum\'e}
%===================

Les syst\`emes \'electroniques bidimensionnels sont d'une grande actualit\'e tout particuli\`erement depuis la d\'ecouverte de la supraconductivit\'e \`a haute temp\'erature. Ici, on se restreint \`a l'\'etude d'un mod\`ele de Hubbard \'etendu, \`a la limite d'un couplage faible. En g\'en\'eral, le gaz \'electronique subit  une instabilit\'e supraconductrice m\^eme sans phonons. Cependant, dans le cas sp\'ecial d'une bande demi-remplie, la surface de Fermi est embo\^it\'ee et se trouve \`a une singularit\'e de Van Hove. Cette situation conduit \`a une comp\'etition entre six instabilit\'es diff\'erentes. Outre la supraconductivit\'e en onde $s$ et $d$, on trouve des ondes de densit\'es de spin et de charge ainsi que deux phases qui sont caract\'eris\'ees par des courants circulaires de charge et de spin respectivement. Le formalisme du groupe de renormalisation est pr\'esent\'e en reliant l'id\'ee de la ``sommation parquet`` au concept plus moderne de l'action effective de Wilson. Comme r\'esultat on obtient un diagramme de phases riche en fonction de l'interaction du mod\`ele. Ce diagramme de phase est exact dans la limite d'une interaction infiniment faible, puisque dans ce cas les lignes de transitions sont fix\'ees par des sym\'etries du mod\`ele. Les comportements \`a basse temp\'erature de la susceptibilit\'e de spin ainsi que de la compressibilit\'e de charge compl\`etent l'image physique de ces instabilit\'es. Il s'av\`ere que la surface de Fermi \`a une tendence g\'en\'erale de se d\'eformer spontan\'ement, mais l'embo\^itement n'est pas d\'etruit. En r\'esum\'e, le mod\`ele de Hubbard \`a couplage faible reproduit deux propri\'et\'es essentielles des cuprates: une phase antiferromagnetique \`a demi remplissage et la supraconductivit\'e en onde $d$ dans le cas dop\'e. Mais elle n'explique pas les propri\'et\'es inhabituelles de l'\'etat m\'etallique dans le r\'egime sous-dop\'e. Une extension syst\'ematique de l'approche perturbative pourrait aider \`a mieux comprendre ces propri\'et\'es, mais reste difficile puisque les techniques n\'ecessaires ne sont pas encore compl\`etement d\'evelopp\'ees.

%\selectlanguage{english}
\chapter{Abstract}
%=================

%\begin{center}\section*{Abstract}\end{center}

Interacting electrons in two dimensions are of particular interest in relation to high-temperature superconductivity. In this thesis, I study a two-dimensional extended Hubbard model in the weak coupling limit. Quite generally, the electron gas is unstable towards a superconducting state even in the absence of phonons. However in the special case of a half-filled band, the Fermi surface is nested and the system is at a Van Hove singularity. In this situation, there are six competing instabilities: $s$- and $d$-wave superconductivity, spin-and charge-density waves and two phases with circulating charge and spin currents, respectively. The required renormalization group formalism is presented on a most elementary level, connecting the idea of the ``parquet summation'' to the more modern concept of Wilson's effective action. As a result, a rich phase diagram is obtained as a function of the model interaction. This phase diagram is exact in the weak coupling limit, since the transition line between two neighboring phases is then fixed by symmetries. The physical picture of each instability is completed by studying the low temperature behavior of the spin susceptibility and the charge compressibility. We also observe a general trend towards a Fermi surface distortion, but the nesting is not destroyed. In summary, the weak-coupling theory of the Hubbard model reproduces two essential features of the cuprates, namely an antiferromagnetic phase at half-filling and $d$-wave superconductivity in the doped material. But it does not explain the unusual properties of the metallic state in the underdoped regime. A consequent extension of the perturbative approach to sub-leading orders which would imply self-energy corrections could reveal further insight, but the required techniques are not yet fully developed.

%234567891023456789202345678930234567894023456789502345678960234567897023456789802345678990234567100

\tableofcontents

\mainmatter
%auto-ignore

\chapter{Introduction}
%+++++++++++++++++++++

The interest in the problem of interacting electrons in two space dimensions has been increased tremendously by the discovery of quasi two-dimensional (2D) materials with novel electronic properties. 
The high-$T_c$ superconducting cuprates are certainly the most famous examples\footnote{See \cite{Waldram} for an introduction to the cuprates and superconductivity in general and \cite{Orenstein00} for a recent review}. Other unconventional superconductors with a quasi 2D electronic structure are strontium ruthenate \cite{Maeno01}, some graphite-based compounds \cite{ricardo01}, $MgB_2$ \cite{Nagamatsu01}, and doped $C_{60}$ crystals \cite{schoen00,schoen01}. Yet another class of quasi-2D materials is given by some transition metal dichalcogenides with interesting charge-density wave formations \cite{Aebi01}. 2D electronic systems have also been experimentally realized in $Si$- or $GaAs$- based semiconductor devices, where the 2D electron gas undergoes a metal-insulator transition \cite{Kravchenkho95,Hanein98}.

The metallic phase of the cuprates deviates considerably from the predictions of Landau's  Fermi liquid theory, i.e. the theory of usual metals based on the concept of quasi-particles\footnote{See for example \cite{pines}}. Impressing evidence for the breakdown of the quasi-particle concept in a cuprate material has been provided very recently by the violation of the Wiedemann-Franz law \cite{Hill01}. 

Most of the unusual properties of the cuprates are likely to be linked to the quasi-two-dimensional nature of their electronic structure close to the Fermi energy.
Therefore certain single-band 2D models of interacting electrons may be able, in principle, to account for at least part of the anomalies observed in these compounds \cite{Anderson88,Zhang88}. Angle-resolved photo-emission spectroscopy measurements support this idea, since they show a single well-defined Fermi surface at least in the over-doped regime \cite{takahashi88,olson89,campuzano90}\footnote{For recent reviews on photoemission spectroscopy of the cuprates, see \cite{johnson01,Damascelli01,golden01}}. Furthermore the undoped insulating parent compounds are well described by the 2D spin $1/2$- Heisenberg model \cite{Manousakis91}, which is also a single-band model.  

Unfortunately, even very simple models, such as the 2D Hubbard or the 2D $t-J$ model, have so far resisted a rigorous analysis. Moreover, the available numerical studies are not yet conclusive enough for making definite predictions for, e.g., the zero-temperature phase diagram of these many-electron systems. Another difficulty is that fluctuations (both thermal and 
quantum) are strong in two dimensions so that mean-field 
approximations cannot be trusted.

The coupling between electrons in the cuprates, for instance the parameter $U$ of 
the Hubbard model, is large, i.e. of the order of the bandwidth.
 Therefore it is not clear whether a ground state consisting of
 occupied Bloch orbitals with energies below $\epsilon_F$ is a good 
starting point or whether one has rather to think in terms of 
configurations of singly occupied and empty sites (doped Mott 
insulator). Actually, the successful analysis of the insulating
 phase in terms of the Heisenberg model suggests that the Mott 
insulator is the appropriate reference state \cite{Manousakis91}. 

I deliberately choose the limit of weak bare
 couplings, keeping in mind that this parameter range may 
miss completely some important characteristic aspects of 
the region of strong bare interactions. For example, the spin-density wave (SDW) instability of the half-filled Hubbard model at weak coupling, which will be studied in Chapter \ref{chapter3}, fulfills the physical picture of a Slater instability \cite{Slater51} rather than a Mott transition. The Slater instability arises through the formation of a SDW, whereas the Mott transition is characterized by the formation of local magnetic moments, which are not necessarily ordered\footnote{For a review on the Mott transition, including the distinction between Mott and Slater transitions, see \cite{Gebhard}.}. Nevertheless, it is to my opinion important to understand the weak coupling limit, which is already a non-trivial problem. It cannot be excluded that certain properties are qualitatively
 the same over the whole range of couplings, as is 
the case for the 1D Hubbard model, a Luttinger liquid for 
all positive values of $U$ and all densities except $n=1$ 
\cite{Shiba92,Voit95}. 

 Although it is not excluded that lattice vibrations (phonons) are important to understand high-temperature superconductivity, I will completely neglect them in this thesis.

The most clear picture of two-dimensional interacting 
electrons has been obtained for the 2D jellium model 
with its circular Fermi surface. A series of rigorous studies has shown that the Landau Fermi liquid theory is stable at not too low temperatures
 such that $|U\log T|<\mbox{const}$, where $U$ is the strength of the local interaction \cite{Salmhofer98,Disertori00}, but it breaks down at a critical temperature $T_c\sim \exp{-\frac{\mbox{const}}{|U|}}$. For lower temperatures, the properties of the interacting system are no longer analytically connected to those of the non-interacting system. The usual interpretation is a phase transition into a superconducting state, although there is no rigorous statement about $T<T_c$ available up to now. 

Strictly speaking, there can be no long-range order in two dimensions due to thermal fluctuations \cite{mermin66}. $T_c$ is thus interpreted as a mean-field critical temperature, which is close to the temperature of the Kosterlitz-Thouless transition \cite{kosterlitz73}. In reality, there is always some coupling of the 2D planes by an electron hopping $t_\perp$. At small temperature, the coupling between the planes and the strong correlations inside the plane will result in three-dimensional long range order\footnote{This was shown explicitly for the $XY$ model \cite{Janke90,Baeriswyl92}.}. The same remark applies to the SDW- and all other instabilities which are discussed in this thesis.

While the interpretation of $T_c$ in terms of superconductivity is very natural for attractive interactions, it is conceptually more problematic in the repulsive case. The possibility of a superconducting state in a purely repulsive system without any phonons, which was pointed out already by Kohn and Luttinger \cite{Kohn65}, is the main subject of Chapter \ref{chapter1} of this thesis. It has to be mentioned however that the interpretation of $T_c$ in terms of Kohn-Luttinger superconductivity is not accepted by everybody \cite{anderson01}.

The discussion of the superconducting instability in Chapter \ref{chapter1} proceeds with a minimum of formalism. It is based on the fact that in general there is only one kind of dominant fluctuations, the superconducting ones. The problem is more difficult in situations where density fluctuations are equally strong (actually diverging) as the superconducting ones. This can arise in 2D, due to the effects of the lattice. 

In fact, electrons hopping between the sites of a square lattice yield a spectrum that differs in two 
respects from the parabolic spectrum of the 
jellium model. First, the spectrum exhibits 
extrema and saddle points in the Brillouin 
zone. General considerations imply that there
 are at least two saddle points and two extrema
 (one maximum and one minimum). Obvious points
 are $\v P_0=(0,0)$ and $\v Q=(\pi,\pi)$ for the
 extrema and $\v P_1=(\pi,0)$ and $\v P_2=(0,\pi)$
 for the two saddle points, but more complicated 
patterns are also possible. The density of states
 has a logarithmic van Hove singularity at the 
saddle points, in strong contrast to the constant
 density of states of the parabolic spectrum of 
the jellium model. The second difference is the 
shape of the lines of constant energy. These
 are circles in the case of the jellium model, 
whereas in the case of the square lattice one 
can easily find portions with almost vanishing
 curvature. In fact, for the tight-binding model
 (with hopping restricted to nearest neighbor 
sites) the Fermi surface for the half-filled band
 is a perfect square. 

The van Hove singularities and flat Fermi surfaces lead to strong fluctuations both of the superconducting and the density type. For the  one-dimensional electron gas the same complication has  been solved successfully by the renormalization group (RG) method. 

RG concepts have been used in very different fields of modern physics and RG can have quite different meanings for different people. The notion was first introduced by St\"uckelberg and Petermann \cite{stuckelberg53} and independently by Gell-Mann and Low \cite{gell-mann54}  in the context of quantum field theories (like QED) in order to cope with infinities that appear in naive perturbation theory. 

In the early 1970s, Kadanoff and Wilson have associated the RG to the procedure of mode elimination in classical statistical mechanics (or systems of bosons) \cite{wilson71,wilson74}. In Wilson's formulation, the RG transformation consists of integrating out some degrees of freedom of the system and including them in the renormalization of some parameters (for example coupling constants). This alternative formulation of the RG idea proved tremendously successful in analyzing the critical behavior in the vicinity of second order phase transitions\footnote{see \cite{wilson75} for an early review including also a detailed discussion of the Kondo problem}. 

In general, a RG transformation is some change of the length or energy scale and the RG equation describes the response of the system as the length or energy scale is changed. In the interacting electron problem considered here, the energy scale is given by the temperature or alternatively by some cutoff $\Lambda$ in the band energy of the electrons. The single-particle states with energies far from the Fermi level, i.e. with lattice momenta far from the Fermi surface, are subsequently integrated out. The result is an effective theory for the degrees of freedom at the Fermi surface. It turns out that the effective coupling constants of the low-energy effective theory have an alternative interpretation in terms of two-particle Green's functions with an infrared cutoff $\Lambda$. In this sense, the RG equation describes the change of Green's functions as a function of the energy scale $\Lambda$. In many cases, the RG flow to low energies produces a singularity at a finite energy scale $\Lambda_c$. This is interpreted as a transition into a strongly correlated state. The latter can establish long-range order, if stabilized by a hypothetical coupling of the 2D planes in three dimensions. The energy scale $\Lambda_c$ is then comparable to the transition temperature towards the ordered state.

It turns out that the one-loop (i.e. the leading order) approximation of the RG is equivalent to the so-called parquet approximation. This method was developed by the soviet school \cite{pomeranchuk57} in order to treat different diverging fluctuations on an equal footing. It was successfully applied  to one-dimensional conductors \cite{Bychkov66,Dzyaloshinskii72,Gorkov74}, to the Kondo problem \cite{Abrikosov65} and to the $X$-ray absorption edge singularity problem in metals \cite{Roulet69,Nozieres69}. A detailed description of the method can be found in \cite{Roulet69}. In my derivation of the one-loop RG equations in Chapter \ref{chapter2}, I will follow mainly the parquet philosophy, but the relation to the Wilsonian interpretation in terms of an effective theory depending on the energy scale will also be explained.

The application of renormalization group (RG) ideas to 2D fermionic problems started in the mathematical physicists community \cite{Feldman90,Benfatto91} and has led to a considerable progress  during the last decade\footnote{Pedagogical introductions can be found in \cite{Polchinski92,Shankar94,froehlich96}. A precise relationship between Fermi liquid theory and the renormalization group has been established in \cite{chitov95,chitov98,Dupuis98,dupuis00}.}. In fact, the rigorous works \cite{Disertori00,Salmhofer98} mentioned above are based on similar ideas. 

A numerical scheme for calculating the complete flow from the bare 
action of an arbitrary microscopic model to the low-energy effective
 action as a function of a continuously decreasing energy cutoff 
$\Lambda$ has been presented by Zanchi and Schulz \cite{Zanchi,Zanchi97,Zanchi98,Zanchi00}. The application 
of this method to the Hubbard model near half filling does provide 
an appealing picture, namely a transition from an antiferromagnetically 
ordered ground state at half filling to a $d$-wave superconductor upon 
doping \cite{Zanchi00}. This result has been confirmed by Halboth and 
Metzner using a similar approach \cite{Halboth00}. Recently, numerical
 RG calculations have brought up two additional phases, one with a 
deformed Fermi surface (Pomeranchuk instability) \cite{Halboth01} 
and one with suppressed uniform spin and charge susceptibilities 
(``insulating spin liquid'') \cite{HSFR01}. 

The analytical approach presented here is complementary to these numerical RG 
calculations. I start from the same RG equations and analyze the flow in the limit of small $\Lambda$. 
The RG equation is reduced to its leading order terms in the low-energy regime $\Lambda\to0$ whereas in the regime of higher energies, the only control of the flow is by naive perturbation theory. In this sense the analytical approach is limited as compared to the numerical studies, which are formally not restricted to small values of $\Lambda$. On the other hand, the numerical methods suffer from the need to replace the continuous Fermi surface by a discrete set of points. 

 It is argued in Chapter \ref{chapter2} of this thesis that the reduction of the one-loop RG equation to the leading terms in $\Lambda\to0$ is imperative for a consistent treatment. In fact, it turns out that sub-leading contributions to the one-loop equation are comparable to higher order terms, which have been neglected in the one-loop approximation. 

 In our approach - as well as in previous RG calculations carried out to one loop order - self-energy effects are neglected. While this can be easily justified for the jellium model, the argument is more subtle in the case of lattice fermions. In fact, the second-order contribution to the self-energy is infrared divergent in the case of the half-filled nearest-neighbor tight-binding band. Nevertheless, I find that also in this case self-energy effects are of subleading order in $\Lambda$, provided that an instability (superconducting or density wave) occurs.

Four distinct cases are considered in this thesis. First the generic case with a finite density of states and no Fermi surface nesting. Second, an anisotropic model with different hopping parameters $t_x$ and $t_y$ for electron hopping between nearest neighbors in the $x$- and $y$ directions. The Fermi surface of this model is perfectly nested at half filling. Third, a situation where the Fermi surface passes through the saddle points of the dispersion at $\v P_1=(\pi,0)$ and $\v P_2=(0,\pi)$, without being nested. Finally the main emphasis is on the half-filled nearest-neighbor tight-binding model, where the Fermi surface is completely flat and contains two saddle points.

RG calculations for a model where the Fermi 
surface contains flat portions have
 been performed by various authors \cite{abreu97,abreu01,dusuel01,
Kwon97,zheleznyak97,gonzalez97}. They agree in that a 
$d$-wave superconducting instability occurs for 
repulsive interactions, due to the interplay
 of particle-particle and particle-hole correlations. 

Early scaling approaches to the problem of van Hove singularities
\cite{Schulz87,Dzyaloshinskii87,Lederer87} focussed on the 
interactions between electrons at the saddle
points, by treating these points in analogy to 
the two Fermi points of the one-dimensional 
electron gas \cite{Solyom79}. In this thesis I show that, indeed,
 the logarithmically dominant RG flow at low
 energies is controlled by the neighborhood of the
 van Hove points. However, in contrast to the one-dimensional case where the 
scattering processes can be characterized in terms of a few 
coupling {\it 
constants } 
connecting the two Fermi points, in two 
dimensions the effective couplings are {\it 
functions } 
of incoming 
and outgoing momenta, even if these are restricted to the 
Fermi surface. It turns out that this functional dependence plays 
 a crucial role in the asymptotic decoupling of competing 
 instabilities. A step in this direction has already been made in 
the parquet approach of Ref. \cite{dzyaloshinskii88,yakovenko88}.

When the Fermi level is at a van Hove 
singularity the system is not renormalizable
 in the traditional sense of field theory. 
Nevertheless, electrons near a van Hove singularity 
have been treated by applying the field theory formalism 
 to the particle-hole sector \cite{Gonzalez96,Guinea97,Gonzalez99,Gonzalez00}. No mixture with particle-particle diagrams can be treated 
within this formalism. The Wilsonian
 RG used here does not assume renormalizability and may be 
applied without constraints. 

The case where the Fermi surface contains the van Hove singularity without being nested has been addressed in many recent investigations \cite{IKK01,IK01,HS01,HSPRL01}. It turns out that (in the case of a repulsive interaction between the electrons) the leading order terms in the small parameter $\Lambda$ is not sufficient to obtain the full phase diagram. The abovementioned works are hence mainly based on subleading contributions to the RG flow\footnote{I. e. simply logarithmically diverging terms of the perturbation theory. In contrast, the leading terms diverge as $\log^2\Lambda$.}. It is not entirely clear to me whether this can be done consistently within a one-loop approach. 

In contrast, the analysis of the dominant parts of the RG equations is sufficient 
for establishing a rich phase diagram for the nearest-neighbor tight-binding band with a nested Fermi surface. This will be done in Chapter \ref{chapter3}.

%auto-ignore

\chapter{Superconducting instabilities of a general 2D Fermi surface}
%++++++++++++++++++++++++++++++++++++++++++++++++++++++++++++++++++++
\markboth{CHAPTER 2. SUPERCONDUCTING INSTABILITIES...}{}\label{chapter1}

This chapter is organized as follows. After a brief presentation of some basic notions and definitions in Section \ref{intellat}, I will illustrate the breakdown of naive perturbation theory due to infrared divergences in Section \ref{perturbation}. Section \ref{ladder} presents the so-called ladder approximation, which consists of a consistent summation of the leading diverging terms. It is shown in Section \ref{superconductivity}, how this approximation indicates superconductivity in the extended Hubbard model. Finally the Section \ref{Nesting}, where it is shown that the ladder summation is not sufficient if the Fermi surface is nested, serves as motivation for the more formal investigations of Chapter \ref{chapter2}.

\section{Interacting electrons on a lattice}\label{intellat}
%------------------------------------------
 
\subsection{The model}\label{model}
%---------------------------------

We consider a general single-band model of interacting electrons on a two-dimensional square lattice. Formally, the size of the lattice is $L\times L$ with periodic boundary conditions, but all the calculations are done in the thermodynamic limit $L\to\infty$. The lattice spacing is put equal to unity.  

The single-particle states consist of one localized Wannier state for each lattice site $\v r$ and they are labeled by a spin index $\sigma=\up,\down$. $c^\dagger_{\v r\sigma}$ and  $c^{}_{\v r\sigma}$ are the usual creation and annihilation operators. The Hamiltonian is of the form $H=H_0+H_I$, where
\be 
H_0=\sum_{\v r,\v r',\sigma}t(\v r,\v r')\,c^\dagger_{\v r\sigma}c^{}_{\v r'\sigma}
\ee
is the non-interacting part and 
\be
H_I=\frac12\sum_{\v r_1,\ldots,\v r_4}U(\v r_1,\ldots,\v r_4) \sum_{\sigma, \sigma'}c^\dagger_{\v r_1\sigma} c^\dagger_{\v r_2\sigma'}c^{}_{\v r_3\sigma'}c^{}_{\v r_4\sigma}
\ee
is the most general two-body interaction respecting the global spin rotation symmetry and particle number conservation. 

Translation-invariance is also assumed, i.e. $t(\v r,\v r')=t(\v r-\v r',0)$ and  $U(\v r_1,\ldots,\v r_4)=U(\v r_1-\v r_4,\v r_2-\v r_4,\v r_3-\v r_4,0)$. After Fourier transformation $c_{\v k,\sigma}=1/L\sum_{\v r}e^{-i\v k\v r}c_{\v r,\sigma}$, $H_0$ is diagonal 
\be
H_0=\sum_{\v k,\sigma}\epsilon_{\v k}\,c^\dagger_{\v k\sigma}c^{}_{\v k\sigma}\label{H0}
\ee
with the single-particle dispersion $\epsilon_{\v k}=\sum_{\v x}e^{i\v k\v x}t(\v x,0)$. The properties of $\epsilon_{\v k}$ are crucial for the behavior of the system at weak coupling and different specific cases will be discussed in this thesis. The interaction after Fourier transformation reads
\be 
H_I=\frac12\frac1{L^2}\sum_{\v k_1,\v k_2,\v k_3}g(\v k_1,\v k_2,\v k_3) \sum_{\sigma, \sigma'}c^\dagger_{\v k_1\sigma} c^\dagger_{\v k_2\sigma'}c^{}_{\v k_3\sigma'}c^{}_{\v k_1+\v k_2-\v k_3\,\sigma},\label{HI}
\ee
where 
\be
g(\v k_1,\v k_2,\v k_3)=\sum_{\v x_1,\v x_2, \v x_3}e^{i(\v k_1 \v x_1+\v k_2 \v x_2-\v k_3 \v x_3)}U(\v x_1,\v x_2, \v x_3,0).
\ee
The momenta $\v k$ are only defined modulo a reciprocal lattice vector (i.e. mathematically speaking $\v k\in\frac{2\pi}L\Z^2/2\pi\Z^2$) and throughout this thesis it is implicitly assumed that sums, equations and Kronecker symbols involving momenta are taken modulo $2\pi\Z^2$. For example the equation $2\v p=0$ has four solutions, $\v p=(0,0),(\pi,0),(0,\pi)$ and $(\pi,\pi)$. The summations $\sum_{\v k}$ are by convention over the first Brillouin zone $]-\pi,\pi]\times ]-\pi,\pi]$. The hopping matrix as well as the interaction is assumed to be short-ranged in real space, which leads to smooth functions $\epsilon_\v k$ and $g(\v k_1,\v k_2,\v k_3)$ in momentum space. The Hamiltonian was written in a very general form for the following reason. Even if one starts with a simple model such as the Hubbard model, the effective models which are generated by the renormalization group procedure are of the general form \eref{H0} and \eref{HI}. 

Nevertheless the effective model will always satisfy the symmetries of the original model, at least if these symmetries are not spontaneously broken. It is therefore worthwhile to list some basic symmetry relations in addition to {particle number conservation}, {spin rotation invariance} and {translation invariance} already mentioned above. 

Time reversal invariance\footnote{See appendix \ref{time}.} requires $g(\v k_1,\v k_2, \v k_3)=\overline{g(-\v k_1,-\v k_2, -\v k_3)}$, where $\overline{g}$ means the complex conjugate of $g$. Together with the spatial reflection invariance this means, that $g$ is real. In addition, we require 
\ba
g(\v k_1,\v k_2, \v k_3)&=g(\v k_2,\v k_1, \v k_1+\v k_2-\v k_3)&\mbox{permutation symmetry} \\
&=g(\v k_1+\v k_2-\v k_3,\v k_3,\v k_2)&\mbox{Hermiticity}\\
&=g(R\v k_1,R\v k_2, R\v k_3),
\ea
where $R$ is a point symmetry operation of the lattice.

There is no symmetry with respect to the operation 
$Xg(\v k_1,\v k_2,\v k_3)=g(\v k_2,\v k_1,\v k_3)=g(\v k_1,\v k_2,\v k_1+\v k_2-\v k_3)$. The symmetric part $g^{S}=\frac12(1+X)g$ and the
antisymmetric part $g^{T}=\frac12(1-X)g$ describe scattering 
of singlet and triplet pairs, as becomes clear if we write the interaction as
\ba
 H_I&=&\frac12\frac1{L^2}\sum_{\v k_1,\v k_2,\v k_3}\bigg\{ g^S(\v k_1,\v k_2,\v k_3) \phi^\dagger_S(\v k_2,\v k_1)\phi^{}_S(\v k_3,\v k_1+\v k_2-\v k_3)\nonumber\\
& &\qquad + g^T(\v k_1,\v k_2,\v k_3)
\sum_{\alpha=0,\pm1}\phi^\dagger_{\alpha}(\v k_2,\v k_1)\phi^{}_{\alpha}(\v k_3,\v k_1+\v k_2-\v k_3)\bigg\},
\ea
with 
\ba
&\phi_{S}(\v k,\v k')=\frac1{\sqrt{2}}\left(c_{\v k\up }c_{\v k'\down }-c_{\v k\down }
c_{\v k'\up}\right),&\\
&\phi_{0}(\v k,\v k')=\frac1{\sqrt{2}}\left(c_{\v k\up }c_{\v k'\down }+c_{\v k\down }
c_{\v k'\up}\right)\ ,\ 
\phi_{1}(\v k,\v k')=c_{\v k\up }c_{\v k'\up }\ ,\
\phi_{-1}(\v k,\v k')=c_{\v k\down }c_{\v k'\down}\ .&\nonumber
\ea
Sometimes one separates the term with parallel spins ($\sigma=\sigma'$) from the term with anti-parallel spins in Eq. \eref{HI}. It is clear that the coupling function for electrons with anti-parallel spins is $g$, whereas for parallel spins it is $g^T$. 

Alternatively, via the combinations $g^c=(2-X)g$ and $g^s=-Xg$ one can write $H_I$ in terms of normal ordered charge-charge and spin-spin interactions 
\be 
H_I=\frac1{2L^2}\sum_{\v k_1,\v k_2,\v k_3}\left\{g^c(\v k_1,\v k_2,\v k_3):C_{\v k_2 \v k_3}C_{\v k_1\,\v k_1+\v k_2-\v k_3}:+\,g^s(\v k_1,\v k_2,\v k_3)\,:\vec S_{\v k_2 \v k_3}\vec S_{\v k_1\,\v k_1+\v k_2-\v k_3}:\right\},
\ee
where  normal ordering is given by $:c^\dagger_2c^{}_3c^\dagger_1c^{}_4:\,=c^\dagger_1 c^\dagger_2c^{}_3c^{}_4$. $C$ and $\vec S$ are given by
\be 
C_{\v k \v k'}=\frac12\sum_\sigma c^\dagger_{\v k\sigma}c^{}_{\v k'\sigma},\qquad
\vec S_{\v k \v k'}=\frac12\sum_{\sigma\sigma'} c^\dagger_{\v k\sigma}\vec \sigma_{\sigma\sigma'} c^{}_{\v k'\sigma'},
\ee
where $\vec\sigma$ are the Pauli matrices.

For example a general interaction restricted to on-site and nearest-neighbor terms consists of 5 independent terms. The usual Hubbard term $U$, a nearest-neighbor density-density interaction $V$, a Heisenberg term $J$, a pair-hopping term $W$ and the Coulomb-assisted hopping $K$
\ba
H_I&=&U\sum_{\v r} n_{\up \v r} n_{\down \v r}+V\sum_{\langle\v 
r,\v r'\rangle}n_{\v r}n_{\v r'}+J\sum_{\langle\v r,\v r'\rangle}
{\vec S}_{\v r}{\vec S}_{\v r'}+W\sum_{\langle\v r,\v r'\rangle}\left(c^\dagger_{\v r\up}c^\dagger_{\v r\down}c^{}_{\v r'\down}c^{}_{\v r'\up}+\mbox{h.c.}\right)\nonumber\\
& &\qquad +K\sum_{\langle\v r,\v r'\rangle,\sigma}\left(c^\dagger_{\v r\sigma}(n_{\v r-\sigma}+n_{\v r'-\sigma})c^{}_{\v r'\sigma}+\mbox{h.c.}\right),\label{UVJWK}
\ea
where  $n_{\v r}=\sum_\sigma n_{\v r\sigma}=\sum_\sigma c^\dagger_{\v r\sigma}c^{}_{\v r\sigma}$ and $\vec S_{\v r}=\frac12\sum_{\sigma\sigma'} c^\dagger_{\v r\sigma}\vec \sigma_{\sigma\sigma'} c^{}_{\v r\sigma'}$ are the usual charge- and spin operators on the lattice site $\v r$, the sums $\sum_{\langle\v r,\v r'\rangle}$ are over all nearest-neighbor bonds and $U,V,J,W$ and $K$ are parameters. After Fourier transformation, this leads to Eq. \eref{HI} with the coupling function
\be
g(\v k_1,\v k_2,\v k_3)=U+(V-J/4)f_{\v k_3-\v k_2}-J/2\, f_{\v k_3-\v k_1}+W f_{\v k_1+\v k_2}+K(f_{\v k_1}+f_{\v k_2}+f_{\v k_3}+f_{\v k_1+\v k_2-\v k_3}),\label{gUVJWK}
\ee 
where $f_{\v k}=2(\cos k_x+\cos k_y)$.

\subsection{Green's functions}\label{Green}
%-----------------------------------------------------

Since we are concerned with weak coupling instabilities, the first idea is perturbation theory in $H_I$. The diagrammatic technique using the imaginary time (or Matsubara-) formalism is well established and  explained in various textbooks (for example \cite{Negele,Abrikosov,Fetter,Mahan}).

The partition function in the grand canonical ensemble is given  by
\be 
Z=\tr {e^{-\beta(H-\mu N)}},
\ee
where $\beta$ is the inverse temperature, $\mu$ the chemical potential and  
$N=\sum_{\v r\sigma}c^\dagger_{\v r\sigma}c^{}_{\v r\sigma}$ the particle number operator. The trace is to be taken over the entire Fock space. 

Thermal averages of physical observables are calculated by the formula
\be 
\langle O\rangle=\frac1Z\tr {e^{-\beta(H-\mu N)} O}.
\ee
The aim is to calculate the imaginary time Green's functions, which are not directly observable but they are analytically related to various physical response functions.

They are defined by 
\be
G^n(x_1,\ldots,x_n|x_{2n},\ldots,x_{n+1})=(-1)^n\langle T(c^{}_{x_1}\cdots c^{}_{x_n}c^\dagger_{x_{n+1}}\cdots c^\dagger_{x_{2n}})\rangle,
\ee
where $x=(\tau,\v r,\sigma)$ is a multi-index containing the imaginary time $\tau\in[0,\beta]$ in addition to the labels of the single particle states $\v r$ and $\sigma$, the imaginary time Heisenberg representation of operators is 
\be
O(\tau)=e^{\tau(H-\mu N)}\,O\,e^{-\tau(H-\mu N)},\label{imagHeis}
\ee
and $T$ is the usual time ordering operator with respect to $\tau$. Note that the Hermitian conjugate of $O(\tau)$ is $O^\dagger(-\tau)$.

Because of the translation invariance in space and imaginary time, it is convenient to work with Fourier transformed quantities. This means passing from $(\tau,\v r)$ to the $2+1$-dimensional frequency-momentum vector $k=(k_0,\v k)$. It contains the Matsubara frequency $k_0$ which runs over the odd multiples of $\pi/\beta$. 

The annihilation operators in frequency-momentum presentation are defined by
\be
c_{\sigma k}=(\beta L^2)^{-1/2}\int_0^\beta\ud\tau\, e^{ik_0\tau}\sum_{\v r}e^{-i\v k\v r}c_{(\tau,\v r,\sigma)}.
\ee
The Fourier-transformed one- and two-particle Green's functions are then
\ba
G(k)&=&\int_0^\beta\ud\tau\sum_{\v r}e^{i(k_0\tau-\v k\v r)} G(\tau\v r\sigma|0\v 0\sigma)\nonumber\\
&=&-\langle Tc^{}_{\sigma k}c^\dagger_{\sigma k}\rangle
\ea
(independent of $\sigma$) and 
\ba
G^{\sigma\sigma'}(k_1,k_2,k_3)&=&\int_0^\beta\ud\tau_1\ud\tau_2\ud\tau_3\sum_{\v r_1,\v r_2,\v r_3}e^{i(k_{01}\tau_1+k_{02}\tau_2-k_{03}\tau_3-\v k_1\v r_1-\v k_2\v r_2+\v k_3\v r_3)}G(\tau_1\v r_1\sigma,\tau_2\v r_2\sigma'|0\v 0\sigma,\tau_3\v r_3\sigma')\nonumber\\
&=&\beta L^2\,\langle Tc^{}_{\sigma k_1}c^{}_{\sigma' k_2}c^\dagger_{\sigma' k_3}c^\dagger_{\sigma k_1+k_2-k_3}\rangle
\ea

The single-particle Green's function of the non-interacting system is given by
\be
C(k)=\frac1{ik_0-\xi_{\v k}},
\ee
where $\xi_{\v k}=\epsilon_{\v k}-\mu$. It is also called free electron propagator or covariance of the fermion field. The full Green's functions are readily extracted from their one-particle irreducible (1PI) parts. For example the self-energy $\Sigma(k)$, defined as the 1PI part of $G(k)$ is related to the latter by Dyson's equation
\be
G(k)=\frac1{ik_0-\xi_{\v k}-\Sigma(k)}.
\ee
The 1PI vertex function $\Gamma^{\sigma\sigma'}(k_1,k_2,k_3)$, defined as the 1PI part of the two-particle Green's function $G^{\sigma\sigma'}(k_1,k_2,k_3)$, fulfills the relation
\be
G^{\sigma\sigma'}(k_1,k_2,k_3)=\beta L^2(\delta_{k_2,k_3}-\delta_{\sigma,\sigma'}\delta_{k_1,k_3})G(k_1)G(k_2)+\Gamma^{\sigma\sigma'}(k_1,k_2,k_3)\prod_{j=1}^4G(k_j),\label{vertexdef}
\ee
where $k_4=k_1+k_2-k_3$. 

At this point it is worthwhile to investigate the symmetries of the vertex function. A simple $SU(2)$ transformation shows that for a spin rotation invariant system,
\be
\Gamma^{\up\up}=(1-X)\Gamma^{\up\down},
\ee
where $X\Gamma(k_1,k_2,k_3)=\Gamma(k_2,k_1,k_3)$. 
Since the whole information is contained in the function $\Gamma^{\up\down}$, we will only consider the vertex of anti-parallel spins from now on and omit the spin indices (i.e. $\Gamma=\Gamma^{\up\down}$). 

The permutation symmetry gives 
\be
\Gamma(k_1,k_2,k_3)=\Gamma(k_2,k_1,k_1+k_2-k_3).
\ee
Time reversal invariance\footnote{See appendix \ref{time}.}, spin rotation invariance and parity imply
\be
\Gamma(k_1,k_2,k_3)=\Gamma(k_1+k_2-k_3,k_3,k_2). \label{tgamma}
\ee

From the behavior under complex conjugation one obtains
\be
\overline{\Sigma(k_0,\v k)}=\Sigma(-k_0,\v k)
\ee
and
\be
\overline{\Gamma(k_1,k_2,k_3)}=\Gamma(\overline{k_1+k_2-k_3},\overline{k_3},\overline{k_2}),
\ee
where $\overline{k}=(-k_0,\v k)$. It follows that both $\Sigma$ and $\Gamma$ are real if the frequencies are put to zero.

\section{Naive perturbation theory}\label{perturbation}
%---------------------------------

\begin{figure} 
\centerline{\includegraphics[width=11cm]{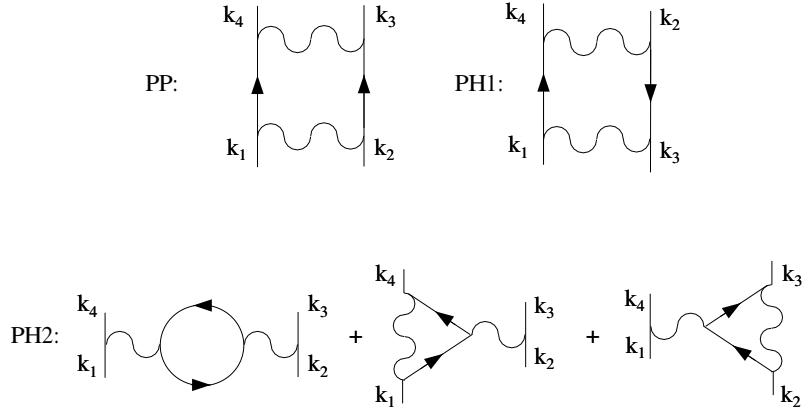}}
\caption{Second order diagrams for the 1PI vertex $\Gamma(k_1,k_2,k_3)$. $k_4$ is determined by energy-momentum conservation.}\label{diags}
\end{figure}

A calculation of the vertex function to second order in perturbation theory gives
\be 
\Gamma(k_1,k_2,k_3)=-g(\v k_1,\v k_2,\v k_3)+\mbox{PP}+\mbox{PH1}+\mbox{PH2}
\label{2ndorder}
\ee
$$ 
\mbox{PP}=\frac1{\beta L^2}\sum_p
g(\v k_1,\v k_2,\v k-\v p)C(p)C(k-p) g(\v p,\v k-\v p,\v k_3), 
$$
$$ 
\mbox{PH1}=\frac1{\beta L^2}\sum_p
g(\v k_1,\v p+\v q_1,\v k_3)C(p)C(p+q_1) g(\v p,\v k_2,\v p+\v q_1), 
$$
\begin{eqnarray*}
\mbox{PH2}&=&\frac1{\beta L^2}\sum_pC(p)C(p+q_2)
\left[-2g(\v k_1,\v p,\v p+\v q_2) g(\v p+\v q_2,\v k_2,\v k_3)\right.\\& 
 &\qquad\left.+g(\v p,\v k_1,\v p+\v q_2) g(\v p+\v q_2,\v k_2,\v k_3)+
g(\v k_1,\v p,\v p+\v q_2) g(\v p+\v q_2,\v k_2,\v p)\right],
\end{eqnarray*}
 where  $k=k_1+k_2$, $q_1=k_3-k_1$, $q_2=k_3-k_2$. The representation of the second order contributions in terms of Feynman diagrams is shown in Fig. \ref{diags}. The internal electron lines stand for free electron propagators $C$ and the wavy interaction lines stand for coupling functions $g$, that depend on the incoming and outgoing momenta. The convention is that the spin index is conserved along the fermion lines. The minus sign in the first of the three PH2 diagrams comes from the fermion loop and the factor $2$ in the same diagram from the sum over the spin index in the fermion loop. PP is referred to as the particle-particle (p-p) diagram and PH1 and PH2 as particle-hole (p-h) diagrams. 

\subsection{Divergences in the particle-particle channel}\label{divergences}
%-------------------------------------------------------

Suppose for a moment that $g$ is constant ($=U$ in the Hubbard model). The p-p diagram is then proportional to the p-p bubble
\be
B^{pp}(k)=\frac1{\beta L^2}\sum_pC(p)C(k-p),
\ee
where $k=k_1+k_2$. The p-h diagrams are proportional to the p-h bubble 
\be
B^{ph}(q)=\frac1{\beta L^2}\sum_pC(p)C(p+q),
\ee
where $q=k_3-k_1$ for PH1 and  $q=k_3-k_2$ for PH2. This section is concerned with the p-p bubble. 

Summation over the Matsubara frequencies leads to
\be
B^{pp}(k)=\frac1{L^2}\sum_{\v p}\frac{n(\xi_{\v p})+n(\xi_{\v k-\v p})-1}{ik_0-\xi_{\v p}-\xi_{\v k-\v p}},
\ee
where $n(\xi)=(1+\exp{\beta\xi})^{-1}$ is the Fermi distribution function. This quantity is diverging at $k=0$ as the temperature goes to zero. At a finite temperature $T=\beta^{-1}$ one finds
\ba
B^{pp}(0)&=&\frac12\int_{-W}^{W}\!\ud\xi\,\nu(\xi)\frac{\tanh{\frac{\beta\xi}2}}\xi\nonumber\\
&=&\nu(0)\,\log{\frac WT}+\mbox{``finite''},\label{logT}
\ea
where $W$ is the bandwidth and $\nu(\epsilon)=1/L^2\sum_{\v k}\delta(\xi_{\v k}-\epsilon)$ is the density of states, and ``finite'' is any contribution with a finite limit $T\to0$. Eq. \eref{logT} can be shown by taking the zero temperature limit of $T\frac{\ud B^{pp}(0)}{\ud T}$. The only two assumptions are reflection symmetry $\xi_{\v k}=\xi_{-\v k}$ and that the density of states $\nu(\xi)$ is an analytic function of $\xi$ at the Fermi level.

Under the same assumption it can be shown that the biggest term of order $n$ in the coupling diverges as $\log^{n-1}(W/T)$. It is clear that naive perturbation theory breaks down at large enough values of the coupling or low enough temperature, such that $|g\log(W/T)|\sim1$. Alternatively, one can organize the perturbation series in the form
\be 
\Gamma=a_1(g\log(W/T))\,g+a_2(g\log(W/T))\,g^2+\cdots.\label{dev}
\ee
We only calculate the first term in this expansion, which means that diagrams of every order in $g$ have to be calculated to the leading logarithmic precision and all the subleading contributions such as the term ``finite'' in \eref{logT} are neglected. In other words, $g$ is considered small, but $g\log(W/T)$ is not.

At any finite frequency, the zero temperature limit of the p-p bubble can be taken and the result is
 \ba
B^{pp}(k_0,0)&=&\int_{-W}^W\!\ud\xi\,\nu(\xi)\frac{\sign\xi}{2\xi-ik_0}\nonumber\\
&=&\nu(0)\,\log{\left|\frac W{ik_0}\right|}+\mbox{``finite''}.
\ea
 The finite frequency just replaces the temperature in the logarithmic expression.

For a more general coupling function $g(\v k_1,\v k_2,\v k_3)$ the p-p diagram at $k_1+k_2=0$ is of the more complicated form
\be
\mbox{PP}=\frac1{\beta L^2}\sum_pC(p)C(-p) F(\v p),
\ee
where the function $F(\v p)$ comes from the coupling functions and depends on the external momenta. 

Any higher order diagram which is obtained by replacing the two coupling functions in PP by a more complicated sub-diagram is of the same form, with a frequency dependent function $F(p)$. The frequency summation leads to
\be
\mbox{PP}=\frac1{L^2}\sum_{\v p} \frac{n(\xi_{\v p})F(p)|_{ip_0=\xi_{\v p}}-(1-n(\xi_{\v p}))F(p)|_{ip_0=-\xi_{\v p}}}{-2\xi_{\v p}}.\label{Finbubble}
\ee
We can now perform a change of variables from $\v p$ to $\xi_{\v p}$ and an additional variable $\theta_{\v p}$, which parameterizes the curves of constant band energy in the Brillouin zone. Assuming that $F$ is an analytic function of $p_0$ and $\xi$, one can Taylor expand
\be
F(\omega,\xi,\theta)=F_{00}(\theta)+F_{10}(\theta)i\omega+F_{01}(\theta)\xi+\cdots.\label{Fexpand}
\ee
It is clear, that only the first term contributes to the logarithmic divergence and the others can be neglected to leading logarithmic order.

\subsection{Cutting off the infrared divergence}\label{cutoff}
%--------------------------------------------

We have seen that naive perturbation theory breaks down at zero temperature. One can regularize the zero temperature perturbation theory by introducing artificially an infrared cutoff $\Lambda$. There are several ways to do this. One method to avoid the singularity is to replace the bare propagator by $C_\Lambda(k)=\Theta(|k_0|-\Lambda)C(k)$, where $\Theta$ is the Heavyside step function. This construction leads to
\ba
B^{pp}_\Lambda(0)&=&\frac1\pi\int_{-W}^W\!\ud\xi\,\nu(\xi)\frac{\arctan{\frac{\xi}\Lambda}}\xi\nonumber\\
&=&\nu(0)\,\log{\frac W\Lambda}+\mbox{``finite''}.
\ea
We see that $B^{pp}(T,k=0)=B^{pp}_\Lambda(T=0,k=0)|_{\Lambda=T}$ to leading logarithmic precision.  Alternatively the cutoff can be introduced in the band energy via $C_\Lambda(k)=\Theta(|\xi_{\v k}|-\Lambda)C(k)$, leading to
\ba
B^{pp}_\Lambda(0)&=&\frac12\left(\int_{-W}^{-\Lambda}\!\ud\xi+\int^{W}_{\Lambda}\!\ud\xi\right)\,\frac{\nu(\xi)}{|\xi|}\label{zeropBubble}\\
&=&\nu(0)\,\log{\frac W\Lambda}+\mbox{``finite''}.
\ea
The logarithmic term is always the same. One could also use the quantity $\sqrt{k_0^2+\xi_{\v k}^2}$ to introduce the cutoff or replace $\Theta$ by a smooth function without changing the result to leading logarithmic order. We are thus free to calculate at zero temperature, introduce the cutoff in the most convenient way for calculations and in the end of the calculation replace $\Lambda$ by $T$. Useless to say that the correspondence $\Lambda\leftrightarrow T$ is only correct to the leading logarithmic order. In a more elaborate calculation including subleading contributions, the result depends on how the cutoff is introduced. 

\markright{THE CASE OF A GENERAL FERMI SURFACE...}\section{The case of a general Fermi surface: Ladder diagrams and BCS theory}\label{ladder}\markright{THE CASE OF A GENERAL FERMI SURFACE...}
%------------------------------------------------------------

\begin{figure}[hbp]  
\centerline{\includegraphics[width=7cm]{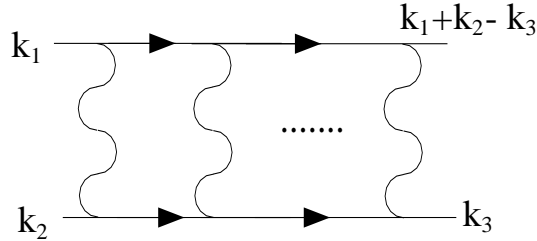}}
\caption{The class of p-p ladder diagrams. They give the leading logarithmic contribution to the vertex in general situations.}\label{ppladders}
\end{figure}

We now calculate the 1PI vertex to leading order in the expansion \eref{dev}. In a general situation without any fine tuned parameter, the only diverging terms occur in the p-p diagram. The other second order terms have a finite limit $T,\Lambda\to0$ except for special situations, which will be discussed later. The leading term in the development \eref{dev} is then given by the sum of all the p-p ladder diagrams, shown in Fig. \ref{ppladders}.

The analytical expression of this series is 
\ba
\Gamma_\Lambda(k_1,k_2,k_3)&=&-g(\v k_1,\v k_2,\v k_3)+\frac1{\beta L^2}\sum_pg(\v k_1,\v k_2,\v k-\v p)D^{pp}_{\Lambda,k}(p)\,g(\v p,\v k-\v p,\v k_3)\label{seriespp}\\
&&\!\!\!\!\!\!\!\!-\frac1{(\beta L^2)^2}\sum_{p,p'}g(\v k_1,\v k_2,\v k-\v p)D^{pp}_{\Lambda,k}(p)\,g(\v p,\v k-\v p,\v k-\v p')D^{pp}_{\Lambda,k}(p')\,g(\v p',\v k-\v p',\v k_3)+\cdots,\nonumber
\ea
where $k=k_1+k_2$ is the total frequency-momentum and $D^{pp}_{\Lambda,k}(p)=C_\Lambda(p)C_\Lambda(k-p)$ is the propagator of a pair of particles.

In principle, the internal electron propagators in Fig \ref{ppladders} could be dressed by self-energy corrections.  Such corrections have three main effects. First they change the shape and location of the Fermi surface, second they modify the properties of the single particle dispersion (the Fermi velocity) and third they lead to a reduction of the quasi particle weight. The deformation of the Fermi surface requires in general the introduction of counter-terms\footnote{See for example Section 5.7 of Nozi\`eres' book \cite{Nozieres} for a comprehensive explanation. Rigorous mathematical statements about the moving Fermi surface have been presented recently in \cite{Feldman96,Feldman98,Feldman99}.}. 
I will ignore this effect for the moment and assume, that in the weak coupling limit, the interacting Fermi surface is not so different from the non interacting one such that the difference is not crucial for the study of the weak coupling instabilities. In fact, perturbative corrections to the Fermi velocity $\nabla_{\v p}\Sigma(p)$ and the quasi particle weight $z=(1+i\partial_{p_0}\Sigma(p))^{-1}$ are finite as $T\to0$ and therefore do not contribute to the leading order.

The total frequency-momentum $k$ is the same for all vertices entering Eq. \eref{seriespp}. It is therefore useful to define the functions
\ba
\Gamma^{BCS}_{\Lambda,k}(k_1,k_4)&=&\Gamma_{\Lambda}(k_1,k-k_1,k-k_4),\nonumber\\
g^{BCS}_{\v k}(\v k_1,\v k_4)&=&g(\v k_1,\v k-\v k_1,\v k-\v k_4).
\ea
and rewrite Eq. \eref{seriespp} as
\be
\Gamma^{BCS}_{\Lambda,k}(k_1,k_4)=-g^{BCS}_{\v k}(\v k_1,\v k_4)+\frac1{\beta L^2}\sum_pg^{BCS}_{\v k}(\v k_1,\v p)D^{pp}_{\Lambda,k}(p)\,g^{BCS}_{\v k}(\v p,\v k_4)-\cdots\label{seriespp2}
\ee
In analogy with the discussion at the end of Section \ref{divergences} we emphasize that after the change of variables $\v p\to\xi,\theta$, the dependence of $g^{BCS}$ on the frequencies $k_{01},k_{04}$ and on the band energies $\xi_1, \xi_4$ is not relevant to leading logarithmic order. One can replace  $\Gamma^{BCS}_{\Lambda,k}(k_1,k_4)$ and $g^{BCS}_{\v k}(\v k_1,\v k_4)$ in Eq. \eref{seriespp} by its value at  $k_{01}=k_{04}=\xi_1=\xi_4=0$ and write
\be
\Gamma^{BCS}_{\Lambda,k}(\theta_1,\theta_4)=-g^{BCS}_{\v k}(\theta_1,\theta_4)+\int\!\ud\theta\,g^{BCS}_{\v k}(\theta_1,\theta)B^{pp}_{\Lambda,k}(\theta)g^{BCS}_{\v k}(\theta,\theta_4)-\cdots\label{seriespp3}
\ee
where we have introduced the angle-resolved p-p bubble
\be
B^{pp}_{\Lambda,k}(\theta)=\frac1\beta\sum_{p_0}\int\!\ud\xi\,J(\theta,\xi)\, D_{\Lambda,k}(p)|_{\v p=\v p(\theta,\xi)},\label{anglepp}
\ee
with the Jacobian $J(\theta,\xi)$ coming from the change of integration variables.

For $k=T=0$ with the energy-cutoff $\Lambda$ one obtains 
\be
B_\Lambda(\theta)=\frac12\left(\int_{-W}^{-\Lambda}\!\ud\xi+\int^{W}_{\Lambda}\!\ud\xi\right)\frac{J(\theta,\xi)}{\xi}
\ee
and thus
\be
\partial_\Lambda B_\Lambda(\theta)=-\frac{(J(\theta,\Lambda)+J(\theta,-\Lambda))}{2\Lambda}\ \MapRight{\Lambda\to0}\ -\frac{J(\theta,0)}{\Lambda},
\ee
from which the logarithmic behavior 
\be
B_\Lambda(\theta)=J(\theta,0)\log\frac W\Lambda+\ \mbox{``finite''}
\ee
can be deduced. The limit $\lim_{\Lambda\to0}J(\theta,\Lambda)$ can be safely taken, provided that no van Hove singularity is present.  As shown in Section \ref{perturbation}, $\Lambda$ can be replaced by $\Min\{\Lambda,T,k_0\}$, i.e. the cutoff $\Lambda$ can be interpreted as the temperature or a finite frequency to leading logarithmic precision.\footnote{The dependence on the momentum $\v k$ is less simple and depends on the form of the Fermi surface. For example if the ladder has flat portions, then the p-p bubble diverges for finite values of $\v k$ parallel to the flat Fermi surface. At the same time, the flat portions introduce divergences in the p-h channel.} 

Introducing the measure $\ud\mu(\theta)=\ud\theta J(\theta,0)$, Eq. \eref{seriespp3} can be further simplified to
\be
\Gamma^{BCS}_{\Lambda}(\theta_1,\theta_4)=-g^{BCS}(\theta_1,\theta_4)+\log\frac W\Lambda\int\!\ud\mu(\theta)\,g^{BCS}(\theta_1,\theta)g^{BCS}(\theta,\theta_4)-\cdots\label{finalseriespp}
\ee
Note that the integral over the Fermi surface $\int\!d\mu(\theta)$ equals the density of states at the Fermi level $\nu(0)$.

Eq. \eref{finalseriespp} is like a geometric series of functions $g^{BCS}(\theta,\theta')$. In order to sum the series \eref{finalseriespp}, we consider $g^{BCS}$ as the kernel of an operator acting in the space of functions $f(\theta)$ as
\be
f(\theta)\ \MapRight{g^{BCS}}\ \int d\mu(\theta')\, g^{BCS}(\theta,\theta')f(\theta'). 
\ee
This operator can be diagonalized. This means that $g^{BCS}$ has the spectral representation given by
\be
g^{BCS}(\theta,\theta')=\sum_n \lambda_n\,f_n(\theta)f_n(\theta'),\label{diagonal}
\ee
where $\lambda_n$ are the eigenvalues of the operator and $f_n(\theta)$ the corresponding eigenfunctions. The latter satisfy the orthogonality relation
\be
 \int d\mu(\theta)\,f_n(\theta) f_m(\theta)=\delta_{nm}.
\ee
It is important to note that the eigenvalues $\lambda_n$ not only depend on the interaction $g$, but also on the properties of the band structure. For example in the simple case of a constant coupling $g^{BCS}(\theta,\theta')=U$ (the Hubbard model), the only non-zero eigenvalue is given by $\lambda=\nu(0)U$.

The series \eref{finalseriespp} can now be summed easily
\ba
\Gamma^{BCS}_{\Lambda}(\theta,\theta')&=&\sum_nf_n(\theta)f_n(\theta')\left(-\lambda_n+\lambda_n^2\log\frac W\Lambda-\lambda_n^3\log^2\frac W\Lambda+\cdots\right)\nonumber\\
&=&\sum_n\frac{f_n(\theta)f_n(\theta')}{-\lambda_n^{-1}-\log\frac W\Lambda}.
\ea
Clearly, negative eigenvalues produce a singularity in the vertex as $\Lambda$ is lowered to the energy scale
\be
\Lambda_c=W e^{\frac1{\lambda_n}}.\label{Lc}
\ee

The divergence of the vertex $\Gamma^{BCS}_\Lambda$ at $\Lambda=\Lambda_c$ is usually interpreted as a signature of the onset of superconductivity. First it has to do with p-p pairs and it requires some attractive channel in the interaction. This suggests that the instability is associated with the formation of Cooper pairs. Moreover formula  \eref{Lc} for the critical energy scale corresponds exactly to the BCS formula for the critical temperature. To make the connection to BCS-theory, consider the gap equation
\be
\Delta_{\v k}=-\frac1{L^2}\sum_{\v k'}g^{BCS}(\v k,\v k')\frac{\Delta_{\v k'}}{2E_{\v k'}}\tanh(\frac{\beta E_{\v k'}}2).
\ee
Close to the critical temperature, the mean-field dispersion $E_{\v k'}=\sqrt{\Delta_{\v k}^2+\xi_{\v k}^2}$ is replaced by $|\xi_{\v k}|$. Furthermore one can neglect the dependence of $\Delta$ and $g$ on the band energies $\xi_{\v k}$ and $\xi_{\v k'}$ in the weak coupling limit. Using Eq. \eref{logT}, this leads to
\be
\Delta(\theta)=-\int\ud\theta'\,g^{BCS}(\theta,\theta')\Delta(\theta')B_{T_c}(\theta'),
\ee
where the angle-resolved p-p bubble has been defined in Eq. \eref{anglepp}. To logarithmic precision, $B_{T_c}(\theta')=J(\theta,0)\log(W/T_c)$. One finally obtains
\be
\Delta(\theta)=-\log(W/T_c)\int d\mu(\theta')\,g^{BCS}(\theta,\theta')\Delta(\theta').
\ee
This is precisely the eigenvalue equation for the operator $g^{BCS}$ introduced above. We conclude that the eigenfunctions $f(\theta)$ correspond up to a normalizing factor to the gap function ($s$-wave, $p$-wave, etc.) and find the correspondence $-\lambda^{-1}(0)=\log(W/T_c)$ for the eigenvalues.

\section{Superconductivity from repulsive interactions}\label{superconductivity}
%--------------------------------------------------------

The theory presented above predicts a superconducting instability, if one of the eigenvalues of $g^{BCS}$ is negative. However, according to Kohn and Luttinger \cite{Kohn65}, superconductivity occurs for any short-range interactions, even if it is purely repulsive. 

The ladder series \eref{finalseriespp} satisfies the following differential equation
\be
-\Lambda\partial_\Lambda\Gamma^{BCS}_\Lambda(\theta_1,\theta_3)=\int\!d\mu(\theta)\,\Gamma^{BCS}_\Lambda(\theta_1,\theta)\Gamma^{BCS}_\Lambda(\theta,\theta_3).\label{rgpp}
\ee
This equation describes how the vertex changes as we vary some energy scale, i.e. the temperature, the frequency or the artificially introduced cutoff. Quite generally, the map $\Gamma_\Lambda\to\Gamma_{\Lambda-\ud\Lambda}$ is called a renormalization group (RG) transformation and equations such as Eq. \eref{rgpp} are called RG equations.

A good strategy for the calculation of the vertex at low energy is a combination of naive perturbation theory and ladder approximation. First use simple (second order) perturbation theory to calculate $\Gamma_{\Lambda_0}$. Then integrate Eq. \eref{rgpp} from $\Lambda_0$ down to lower energies. The energy scale $\Lambda_0$ must be chosen not too small, such that perturbation theory works, i.e. $gB^{pp}_{\Lambda_0}\ll1$. On the other hand it must be small enough, such that the ladder approximation is justified, i.e.  $B^{pp}_{\Lambda_0}\ll B^{ph}_{\Lambda_0}$. 

Eq. \eref{rgpp} can be solved using the same strategy as for the summation of the series \eref{finalseriespp}, namely by writing $\Gamma^{BCS}_\Lambda$ in a spectral representation similar to Eq. \eref{diagonal}. Because of the special form of Eq. \eref{rgpp} only the eigenvalues, but not the eigenfunctions, depend on the energy scale $\Lambda$, i. e.
\be
\Gamma^{BCS}_\Lambda(\theta,\theta')=\sum_n\gamma_n(\Lambda)f_n(\theta)f_n(\theta').
\ee
This ansatz solves Eq. \eref{rgpp}, provided that the eigenvalues $\gamma_n(\Lambda)$ satisfy the RG equation
\be
-\Lambda\partial_\Lambda\gamma_n(\Lambda)=\gamma^2_n(\Lambda),\label{eigendiff}
\ee
and thus are given by
\be
\gamma_n(\Lambda)=\frac1{\gamma^{-1}_n(\Lambda_0)-\log\frac{\Lambda_0}\Lambda}.
\ee

The vertex calculated by this method does not correspond to the ladder series \eref{seriespp} but to a similar series, where $g$ is replaced by $-\Gamma^{BCS}_{\Lambda_0}$ and $D_\Lambda$ is replaced by $D_\Lambda-D_{\Lambda_0}$. Therefore the eigenvalues $\lambda_n$ in Section \ref{ladder} are replaced by $-\gamma_n(\Lambda_0)$, and $\log\frac W\Lambda$ is replaced by $\log\frac {\Lambda_0}\Lambda$.

\begin{figure}  
 \centerline{\includegraphics[width=4cm]{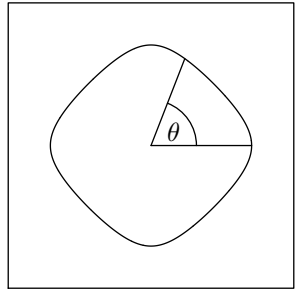}}\vspace{-11pt}
\centerline{\hspace{0.6cm}\raisebox{2.1cm}[0pt][0pt]{$\theta$}}
\caption{The Fermi surface of the nearest-neighbor tight binding band for $\mu=-0.8$ (i.e. $n\approx0.7$).}\label{genFS}
\end{figure}

I have calculated the superconducting instabilities for an extended Hubbard model away from half-filling. The electron hopping is restricted to nearest neighbors, i.e. $\xi_{\v k}=-2t(\cos k_x+\cos k_y)-\mu$. Fig. \ref{genFS} shows the Fermi surface for a typical electron density. The variable $\theta$ is defined as the radial angle. The interaction contains a nearest neighbor term  $V$ in addition to the Hubbard $U$ (See \eref{UVJWK} and  \eref{gUVJWK}). The same calculation was done before by Zanchi and Schulz \cite{Zanchi96} for the repulsive Hubbard model.

The study of superconducting instabilities involves two steps. First calculate the vertex $\Gamma^{BCS}_{\Lambda_0}(\theta_1,\theta_3)$ by second order perturbation theory. The calculations presented here have been done for $\Lambda_0=\Min\{-\mu/4,(\mu+4)/4\}$.  To make a numerical treatment feasible a  discretization of the $\theta$ variable into 24 patches was introduced.

The second step consists in writing $\Gamma^{BCS}_{\Lambda_0}$ in the form \eref{diagonal} by diagonalizing the corresponding operator. This is done by choosing an orthonormal basis $b_n(\theta)$. If the basis is suitably chosen, the infinite matrix 
\be
\Gamma_{nm}=\int\!\ud\mu(\theta)\,\int\!\ud\mu(\theta')\,\overline{b_n(\theta)}\,\Gamma^{BCS}_{\Lambda_0}(\theta,\theta')\,b_m(\theta),
\ee
can be cut off at some finite value of the indices $n,m$ and diagonalized by standard algorithms. The biggest positive eigenvalue of $\Gamma_{n,m}$ gives the leading superconducting instability.

 I have used the harmonic functions $\cos(n\theta)$ and $\sin(n\theta)$ divided by the square root of $J(\theta,0)$ as a basis. Due to the symmetry of the square lattice, the matrix has a block-diagonal form, where each block corresponds to an irreducible representation of the point group $D_4$. Table \ref{irreps}  lists the five irreducible representations and the corresponding basis functions.

\begin{table}[h]
\centerline{\begin{tabular}{c|l}
Irreducible representation&Basis functions\\
\hline
$A_1$&$\cos{4m\theta}$ \\
$A_2$&$\sin{4m\theta}$\\
$B_1$&$\cos{(4m+2)\theta}$\\
$B_2$&$\sin{(4m+2)\theta}$\\
$E$&$\left\{\begin{array}{c}\cos{(2m+1)\theta}\\ \sin{(2m+1)\theta}\end{array}\right.$
\end{tabular}}
\caption{Table of the irreducible representations of the $D_4$ point group and the according basis. In order to obtain an orthogonal system, the functions are divided by $\sqrt{J(\theta,0)}$, but this doesn't change their symmetry properties.}\label{irreps}
\end{table}

\begin{figure}   
\centerline{\includegraphics[width=11cm]{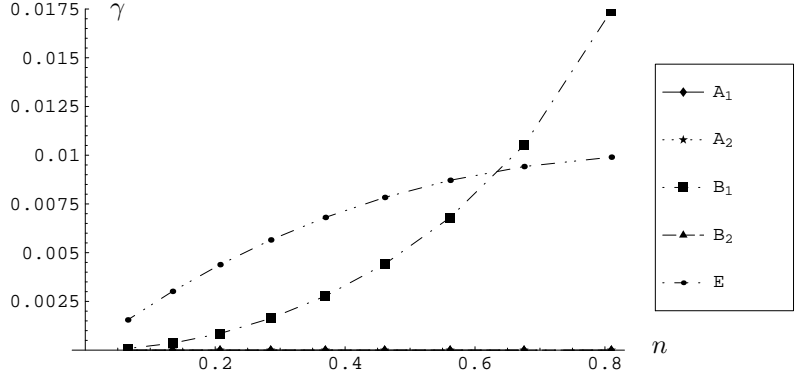}}\vspace{-11pt}
\centerline{\hspace{-0.3cm}\raisebox{5cm}[0pt][0pt]{$\gamma$}\hspace{7cm}\raisebox{0.5cm}[0pt][0pt]{$n$}}
\caption{The positive eigenvalues of $\Gamma^{BCS}$ as a function of the electron density for an attractive nearest-neighbor interaction $V=-0.05t$ and $U=0.5t$, as obtained by a first order calculation. The only positive eigenvalues are in the $E$ and $B_1$ symmetry block. The corresponding eigenfunctions are $p$-wave like (for $E$) and $d_{x^2-y^2}$-wave like (for $B_1$).}\label{Vneg}
\end{figure}

All the calculations have been done for a repulsive on-site interaction $U=0.5t$. In the case of an attractive $V<0$, the appearance of superconductivity is not surprising. In fact, the first order vertex $\Gamma=-g$ has positive eigenvalues in this case. The results can be seen in Fig. \ref{Vneg}. They show triplet $p$-wave superconductivity for $n<0.65$ and singlet  $d_{x^2-y^2}$-wave superconductivity for $0.65<n$ (close to half filling, additional instabilities arise in the p-h channel).  The second order contributions to $\Gamma^{BCS}_{\Lambda_0}(\theta_1,\theta_3)$ give only minor changes to the present result, although they introduce small positive eigenvalues in the symmetry blocks $B_2, A_1$ and $A_2$.

In contrast, in the pure Hubbard model ($V=0$, $U>0$), positive eigenvalues of $\Gamma^{BCS}_{\Lambda_0}(\theta_1,\theta_3)$ appear only via second order terms. 
Since the sum of three diagrams PH2 is zero in the Hubbard model and  PP is merely a constant (i.e. independent of the angles $\theta,\theta'$), positive eigenvalues can only be generated by the diagram PH1. Kohn and Luttinger have investigated the same diagram in their historical paper \cite{Kohn65}. The difference is that we are dealing with an anisotropic 2D system as opposed to the isotropic 3D case considered in \cite{Kohn65}.

\begin{figure}   
\centerline{\includegraphics[width=11cm]{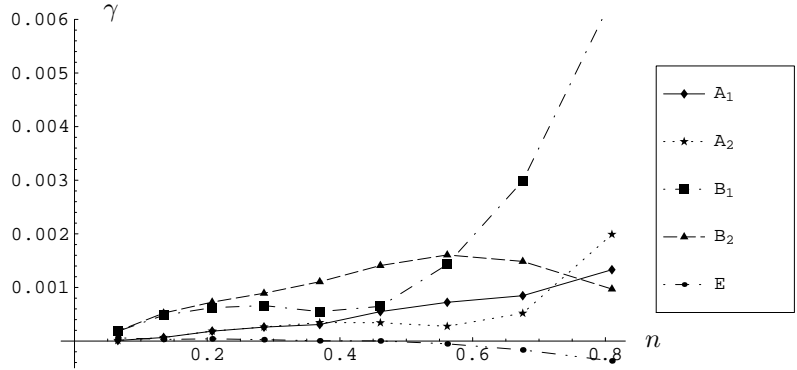}}\vspace{-11pt}
\centerline{\hspace{-0.5cm}\raisebox{5cm}[0pt][0pt]{$\gamma$}\hspace{7cm}\raisebox{0.6cm}[0pt][0pt]{$n$}}
\caption{Second order calculation for the pure Hubbard model $V=0$ and $U=0.5t$ (Kohn-Luttinger superconductivity). The biggest eigenvalue of each symmetry block is shown. The corresponding eigenfunctions are $d_{xy}$-wave like for $B_2$ and $d_{x^2-y^2}$-wave like for $B_1$.}\label{Vzero}
\end{figure}

 The results in Fig. \ref{Vzero} for the Hubbard model show $d_{xy}$-wave symmetry for $n<0.55$ and $d_{x^2-y^2}$-wave for higher densities. Note that the eigenvalues of Fig. \ref{Vzero} created by second order terms are much smaller than those in Fig \ref{Vneg} with an attractive $V$, leading to a superconductivity at much lower energy scales. The eigenvalues of the $B_1$ block ($d_{xy}$) and the $B_2$ block ($d_{x^2-y^2}$) are degenerate in the limit of low electron density, reflecting the approximate rotational symmetry at low filling. 

The appearance of superconductivity by second-order corrections to the vertex has a physical interpretation in terms of an effective attractive interaction, which is mediated by spin fluctuations. In fact, the vertex $\Gamma_\Lambda$ can be interpreted as an effective coupling function $g_\Lambda=-\Gamma_\Lambda$. This  point will be made more precise later in Section \ref{effaction}. The contribution of PH1 to $g^{BCS}_{\Lambda_0}(\theta_1,\theta_3)$ equals
\be
-U^2B^{ph}(\v k_{\theta_1}+\v k_{\theta_3})=U^2\chi_0(\v k_{\theta_1}+\v k_{\theta_3}),
\ee
where $\chi_0(\v q)$ is the spin susceptibility of the non-interacting system\footnote{See Section \ref{corr} or standard textbooks.}. The momentum dependence of the effective interaction (and thus the attraction) comes therefore from the spin susceptibility $\chi_0(\v q)$. One can thus say that the effective attraction is created by the exchange of spin fluctuations. The situation is analogous to the conventional superconductivity of metals, where an effective attraction is created through the exchange of phonons. However with the important difference that the same electrons which feel the effective interaction are also responsible for the spin fluctuations.  

The idea of spin-fluctuation-induced superconductivity has led to semi-phenomenological theories of materials with strong magnetic correlations such as the cuprates which are close to antiferromagnetism \cite{BickersPRL89,Bickers89,Moriya90,Monthoux92,Kondo99}, but more recently also for $Sr_2RuO_4$ which is nearly ferromagnetic \cite{Monthoux99}.

\begin{figure}   
\centerline{\includegraphics[width=11cm]{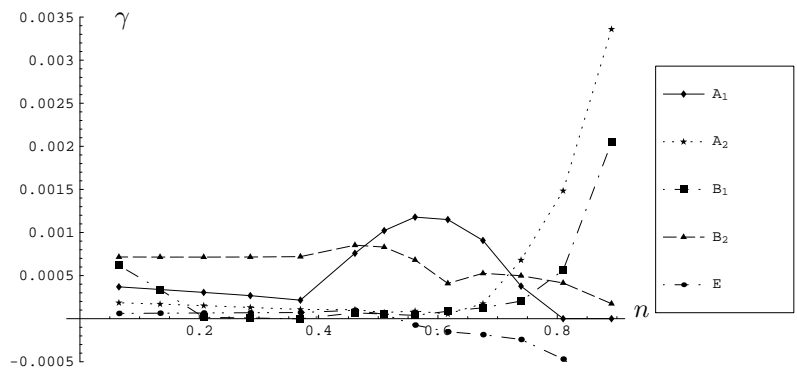}}\vspace{-11pt}
\centerline{\hspace{-0.5cm}\raisebox{4.9cm}[0pt][0pt]{$\gamma$}\hspace{6.7cm}\raisebox{1cm}[0pt][0pt]{$n$}}
\caption{Same as Fig. \ref{Vzero} for repulsive nearest-neighbor interaction  $V=0.05t$ and $U=0.5t$. The leading eigenvalues from the $A_1$ and $A_2$ blocks correspond to highly oscillating eigenfunctions with eight nodes along the Fermi surface.}\label{Vpos}
\end{figure}

If a positive nearest-neighbor term $V>0$ is added to the Hubbard model, all second order contributions PP, PH1 and PH2 to $\Gamma^{BCS}_{\Lambda_0}(\theta_1,\theta_3)$ are non zero and angle-dependent. As it is shown in Fig. \ref{Vpos}, the $V$-term has a destructive effect on superconductivity. The effect is strongest in the $d_{x^2-y^2}$-wave channel (corresponding to the $A_1$ symmetry). While the $d_{xy}$ regime for $n<0.5$ is not affected much by the  nearest-neighbor repulsion, the $d_{x^2-y^2}$-wave regime has completely disappeared. The leading instabilities arise in two highly oscillating exotic channels instead. 

These results indicate strongly, that there is indeed superconductivity in the repulsive Hubbard model, at sufficiently low temperature. Furthermore they illustrate how the RG equation Eq. \eref{rgpp} can be successfully used to investigate the weak-coupling instabilities of 2D lattice electrons. Simple perturbation theory is used to calculate the vertex at an energy scale $\Lambda_0$. They serve as initial conditions for Eq. \eref{rgpp}. Superconducting instabilities manifest themselves as poles in the vertex function, or equivalently as diverging solutions of Eq. \eref{rgpp}.

\section{Divergences in the particle-hole channel: Nesting}\label{Nesting}
%---------------------------------------------------------
 
In Sections \ref{ladder} and \ref{superconductivity}, it was assumed that the divergences in the perturbation series of the vertex are exclusively generated by the p-p bubble.

The divergence of $B^{pp}(k)$ at $\v k=0$ depends on the parity relation 
\be
\xi_{\v p}=\xi_{-\v p}.\label{parity}
\ee
Since parity is a general symmetry, this divergence is generally present.

In contrast, divergences in the p-h channel only arise in special situations. The p-h diagrams are proportional to the p-h bubble
\ba
B^{ph}(k)&=&\frac1{\beta L^2}\sum_pC_\Lambda(p)C_\Lambda(p+k)\nonumber\\
&=&\frac1{L^2}\sum_{\v p}\frac{n(\xi_{\v p})-n(\xi_{\v p+\v k})}{\xi_{\v p}-\xi_{\v p+\v k}+ik_0}\label{uniformBph}
\ea
where $k=k_3-k_1$ in PH1 and $k=k_3-k_1$ in PH2. The p-h bubble is not diverging at $k=0$, because of the numerator. In fact it is easy to see, that
\be
\lim_{\v k\to0}B^{ph}(0,\v k)=\int\!\ud\xi\,\nu(\xi)\frac{\ud n(\xi)}{\ud\xi}\ \MapRight{T\to0}\ -\nu(0)\label{qlimitbubble}
\ee
 and $B^{ph}(k_0,\v 0)=0$. 

\begin{figure}   
\centerline{\includegraphics[width=4cm]{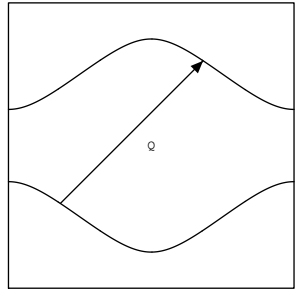}}
\caption{The Fermi surface of the nearest-neighbor $t_x$-$t_y$-band. It is perfectly nested by the vector $\v Q=(\pi,\pi)$}\label{nestedFS}
\end{figure}

Consider now the special case of the half-filled nearest-neighbor tight-binding band. In order to be more general, one can allow for an anisotropy in the hopping parameter, i.e. two different values $-t_x, -t_y$ for hopping between nearest neighbors in the $x$- and $y$-directions. The electron dispersion $\xi_{\v p}=-2t_x\cos p_x-2t_y\cos p_y$ has the special property 
\be
\xi_{\v p} =-\xi_{\v p+\v Q}\label{dispnesting}
\ee
for the nesting vector $\v Q=(\pi,\pi)$. The Fermi surface is shown in Fig. \ref{nestedFS}. The immediate consequence is the following relation between the p-p and the p-h bubbles
\be
B^{ph}(k)=-B^{pp}(Q-k),\label{bubblenesting}
\ee
where $Q=(0,\v Q)$. Therefore, everything that was said before about the p-p bubble at momentum $k=0$ applies as well for the p-h bubble at momentum $k=Q$. 

The condition \eref{dispnesting} is in general too restrictive, i.e. it is not a necessary condition for divergences in the p-h channel. We now formulate more general conditions for divergences both in the p-h and p-p channels. They depend only on the geometry of the Fermi surface and are valid for every periodic lattice in any dimension. It is then useful to remember that the Fermi surface divides the Brillouin zone in two parts, the Fermi sea with the occupied single particle states ($\xi_{\v k}<0$) and the remaining part with the empty states ($\xi_{\v k}>0$). 

\begin{itemize}
\item $B^{ph}$ is diverging at a fixed momentum $\v k$, if:
\begin{enumerate}
\item The Fermi surface $FS$ has some finite overlap with its own translation $\v k+FS$.
\item The overlap is such that the occupied side of $FS$ is covering the empty side of $\v k+FS$ and vice versa.
\end{enumerate}
This property is called p-h nesting with a nesting vector $\v k$.  For example every bipartite lattice (i.e. a lattice with electrons hopping only from one sub-lattice $A$ to another sub-lattice $B$) leads to a p-h nested Fermi surface at half filling\footnote{The reason for this is p-h symmetry as explained in Section \ref{phnesting}.}.

\item $B^{pp}$ is diverging at a fixed momentum $\v k$, if:
\begin{enumerate}
\item The Fermi surface $FS$ has some finite overlap with  $\v k-FS$
\item The overlap is such that the occupied (empty) side of $FS$ is covering the occupied (empty) side of $\v k-FS$.
\end{enumerate}
This property is called p-p nesting with a nesting vector $\v k$. For example, every Fermi surface with spatial inversion symmetry is p-p nested by $\v k=\v 0$.
\end{itemize}

These are necessary conditions if the density of states at the Fermi level is finite, i.e. in the absence of van Hove singularities.

The situation is particularly rich if the Fermi surface has flat portions. In this case there is usually a continuum of nesting vectors $\v k$ satisfying the above conditions. For example the square Fermi surface of the nearest-neighbor tight-binding model is both p-h and p-p nested by every vector $\v k\parallel(1,\pm1)$, which is parallel to one of the edges of the square.

%auto-ignore

\chapter{Renormalization group formalism}\label{chapter2}
%========================================

In this chapter I derive the RG equation for the vertex function to the lowest (one-loop) order. As opposed to some more modern formulations \cite{Shankar94,Zanchi00,Halboth00,Honerkamp},  approach chosen here is quite close to the parquet summation technique \cite{Roulet69}, where the aim is a consistent summation of the leading divergent terms  of the perturbation series (see the expansion \eref{dev}). 
We have done this in sections \ref{ladder} and \ref{superconductivity} for the case where the divergences exclusively arise in the p-p bubble. In the following it is generalized to the case where divergences appear in the p-h channels as well.

I feel that the less modern route chosen here makes the meaning of the one-loop approximation more transparent. In particular it gives a natural guideline for deciding which terms are to be included in the calculation and which ones are to be neglected. 

The chapter starts with the derivation of several exact Bethe-Salpeter equations (Section \ref{BSE}). These are used in Section \ref{floweq} to derive the one-loop RG equation for the vertex function. In Section \ref{corr}, the RG flow of the vertex is related to various generalized susceptibilities associated with pairing, density waves and flux phases. In Section \ref{Wilson}, I make the connection to Wilson's effective action and compare different one-loop RG approaches to 2D fermions known in the literature. In Section \ref{revise}, I reconsider a general (non-nested) Fermi surface and argue that in this case, the one-loop RG equation is not better than the ladder approximation of Chapter \ref{chapter1}. Finally, in Sections \ref{nestingsection} and \ref{vHsection} I consider two specific Fermi surfaces, first the case of perfect nesting in an anisotropic tight-binding model and second a non-nested case at van Hove filling. The nearest-neighbor tight-binding model, where perfect nesting and a van Hove singularity occur simultaneously, will be addressed later in Chapter \ref{chapter3}.

\section{Bethe-Salpeter equations}\label{BSE}
%--------------------------------
  
If the divergences arise in both  p-p and p-h bubbles, the ladder approximation used in section \ref{ladder} is not sufficient.  In fact, in such cases the leading term of the perturbative expansion \eref{dev} is given by the so-called parquet diagrams. These diagrams are obtained by retaining the five one-loop (or second order) diagrams shown in Fig. \ref{diags} and by replacing any bare vertex by one of the one-loop diagrams and continuing this process to any order. An example is given in Appendix \ref{parquetex}.

\begin{figure}  
\centerline{\includegraphics[width=6cm]{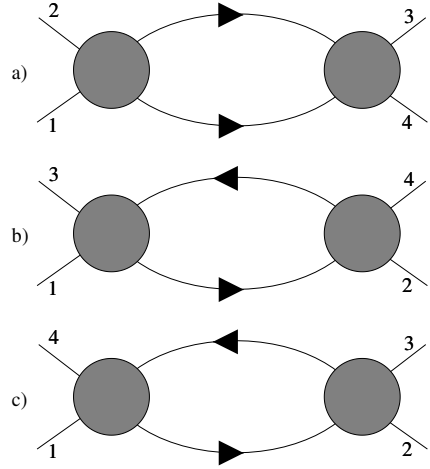}}
\caption{The structure of diagrams, which are reducible in the p-p channel (a), the p-h 1 channel (b) and the p-h 2 channel (c). The grey circles represent any sub-diagram.}\label{reducible}
\end{figure}

To keep track of all the logarithmic divergences, it is useful to define the set of two-particle reducible diagrams. The diagrams contributing to  the 1PI vertex $\Gamma_\Lambda(k_1,k_2,k_3)$ (with an infrared cutoff $\Lambda$) can be two-particle reducible in three possible ``channels''. A diagram is called reducible in the p-p channel if it has the structure shown in Fig. \ref{reducible} a), i.e. if it can be divided in two disconnected pieces by cutting two particle lines. Similarly, the diagrams of the form shown in Fig. \ref{reducible} b) and c) are called reducible in the channels p-h 1 and p-h 2, respectively.

Let us denote by

\begin{tabular}{ll}
$R^{pp}_\Lambda(k_1,k_2,k_3)$&the set of reducible diagrams in the p-p channel, with an infrared cutoff $\Lambda$,\\
$R^{ph1}_\Lambda(k_1,k_2,k_3)$&the set of reducible diagrams in the p-h 1 channel,\\
$R^{ph2}_\Lambda(k_1,k_2,k_3)$&the set of reducible diagrams in the p-h 2 channel,\\
$I_\Lambda(k_1,k_2,k_3)$&The set of two particle irreducible diagrams.
\end{tabular}\\

It is rather simple to check that a given diagram is either two-particle irreducible or reducible in only one of the three possible channels p-p, p-h 1 and p-h 2. Hence
\be
\Gamma_\Lambda=I_\Lambda+R^{pp}_\Lambda+R^{ph1}_\Lambda+R^{ph2}_\Lambda.\label{decomposition}
\ee
It is also useful to define $I^\diamond_\Lambda=\Gamma_\Lambda-R^\diamond_\Lambda$, the set of irreducible graphs in each of the three channels $\diamond=pp, ph1$ or $ph2$.

Clearly, $R^{pp}_\Lambda$ is given by an infinite series similar to the ladder summation Eq. \eref{seriespp}, where $-g$ is replaced by $I^{pp}_\Lambda$ and the single particle propagators $C_\Lambda$ are replaced by full Green's functions $G_\Lambda$. We write this series formally as
\be
R^{pp}_\Lambda=I^{pp}_\Lambda D^{pp}_\Lambda I^{pp}_\Lambda+I^{pp}_\Lambda D^{pp}_\Lambda I^{pp}_\Lambda D^{pp}_\Lambda I^{pp}_\Lambda+\ldots\ .\label{Rppseries}
\ee
 The first-order term is missing in Eq. \eref{Rppseries}, since it corresponds to an irreducible contribution. The series Eq. \eref{Rppseries} satisfies the integral equation
\be
R^{pp}_\Lambda=I^{pp}_\Lambda D^{pp}_\Lambda I^{pp}_\Lambda+I^{pp}_\Lambda D^{pp}_\Lambda R^{pp}_\Lambda
\ee
and since $\Gamma=I^{pp}+R^{pp}$
\be
R^{pp}_\Lambda=I^{pp}_\Lambda D^{pp}_\Lambda\, \Gamma_\Lambda.\label{formalpp}
\ee
This is the Bethe-Salpeter equation\footnote{See for example \cite{Abrikosov,Nozieres}. The  Bethe-Salpeter equation was originally introduced to calculate two-particle bound-states within quantum electrodynamics \cite{Salpeter51}.} With all the functional dependences included it reads
\be
R^{pp}_\Lambda(k_1,k_2,k_3)=\frac1{\beta L^2}\sum_pI^{pp}_\Lambda(k_1,k_2,k-p)D^{pp}_{\Lambda,k}(p)\,\Gamma_\Lambda(p,k-p,k_3),\label{BSpp}
\ee
where $k=k_1+k_2$ and $D^{pp}_{\Lambda,k}(p)=G_\Lambda(p)G_\Lambda(k-p)$.

\begin{figure}  
\centerline{\includegraphics[width=7cm]{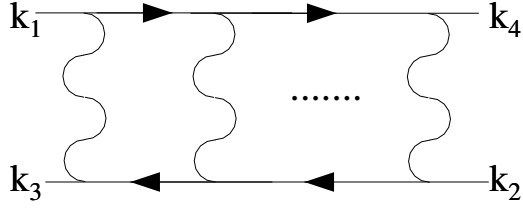}}
\caption{The p-h ladder diagrams. The wavy stand for $-I^{ph1}_\Lambda$ and the electron lines represent full propagators $G_\Lambda$.}\label{ph1ladders}
\end{figure}

There are similar equations for the two other channels. First, $R^{ph1}$ is given by a series over the p-h ladders shown in Fig. \ref{ph1ladders}, where the wavy lines stand for $-I^{ph1}_\Lambda$ and the electron lines denote the full propagators $G_\Lambda$. In complete analogy with the p-p case one finds the Bethe-Salpeter equation for the p-h 1 channel
\be
R^{ph1}_\Lambda(k_1,k_2,k_3)=\frac1{\beta L^2}\sum_pI^{ph1}_\Lambda(k_1,p+q_1,k_3)D^{ph}_{\Lambda,q_1}(p)\,\Gamma_\Lambda(p,k_2,p+q_1),\label{ph1BS}
\ee
where $q_1=k_3-k_1$ and 
\be
D^{ph}_{\Lambda,q}(p)=G_\Lambda(p)G_\Lambda(p+q).
\ee

\begin{figure}  
\centerline{\includegraphics[width=12cm]{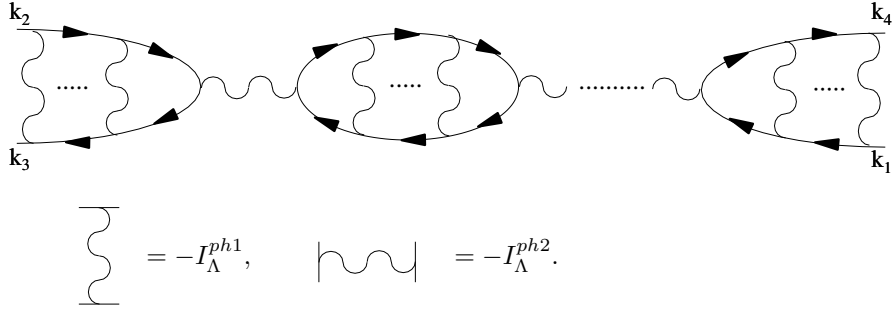}}\vspace{-11pt}
\centerline{\hspace{-2.7cm}\raisebox{1.2cm}[0pt][0pt]{$=-I_\Lambda^{ph1},$}\hspace{2.7cm}\raisebox{1.2cm}[0pt][0pt]{$=-I_\Lambda^{ph2}.$}}
\caption{The general structure of a reducible diagram in the p-h 2 channel. The wavy lines drawn horizontally stand for $-I^{ph2}_\Lambda$, but the wavy lines drawn vertically stand for $-I^{ph1}_\Lambda$. The electron lines stand for full propagators $G_\Lambda$.}\label{ph2ladders}
\end{figure}

The p-h 2 channel is more involved. In fact, a general $R^{ph2}$-diagram has the structure shown in Fig. \ref{ph2ladders}, where the wavy lines drawn horizontally stand for $-I^{ph2}_\Lambda$, but the wavy lines drawn vertically correspond to $-I^{ph1}_\Lambda$. The electron lines stand for full propagators $G_\Lambda$. Note that the diagram must have at least one horizontal wavy line (in the opposite case it is a p-h ladder and not reducible in the p-h 2 channel).

The simplest examples of p-h 2-reducible diagrams are shown in Fig. \ref{diags}. The first PH2 diagram of this figure is given by 
\be
-2\ \frac1{\beta L^2}\sum_pI^{ph2}_\Lambda(k_1,p,p+q_2)D^{ph}_{\Lambda,q_2}(p)\,I^{ph2}_\Lambda(p+q_2,k_2,k_3),\label{explicitph2}
\ee
where $q_2=k_3-k_2$. We will write this specific convolution of the functions  $I^{ph2}_\Lambda$ and  $D_\Lambda^{ph}$ as a formal product 
\be
-2\ I^{ph2}_\Lambda D_\Lambda^{ph} I^{ph2}_\Lambda. \label{formalph2}
\ee
Note that this notation differs from the formal product introduced before in the p-p case. 

With this notation one can write
\be
R^{ph2}= -2\ I^{ph2} D^{ph} I^{ph2}+(XI^{ph1}) D^{ph} I^{ph2}+I^{ph2} D^{ph} (XI^{ph1})+\mbox{``higher order terms,''}\label{2ndorderph2}
\ee
where the overall $\Lambda$-index has been omitted for the simplicity of notation and $XF(k_1,k_2,k_3)=F(k_2,k_1,k_3)$ for any function of three energy-momenta. The p-h ladder series can also be written using the same notation:
\be
XR^{ph1}=(XI^{ph1}) D^{ph} (XI^{ph1}) +\mbox{``higher order terms.''}\label{2ndorderph1}
\ee

The following definitions turn out to be very useful. 
\ba
R^c&=&2R^{ph2}-XR^{ph1},\nonumber\\
I^c&=&2I^{ph2}-XI^{ph1},\label{cdef}\\
\Gamma^c&=&I^c+R^c=(2-X)\Gamma\nonumber
\ea
and
\ba
R^s&=&-XR^{ph1},\nonumber\\
I^s&=&-XI^{ph1}\label{sdef}\\
\Gamma^s&=&I^s+R^s=-X\Gamma.\nonumber
\ea
The superscripts $c$ and $s$ refer to charge and spin, respectively. Note that $\Gamma^{c/s}=\Gamma^{\up\up}\pm\Gamma^{\up\down}$, so these vertex functions enter naturally in the calculation of charge- and spin- response functions. 

Eqs. \eref{2ndorderph2} and  \eref{2ndorderph1} can now be written in the simple form 
\be
R^{s,c}=-I^{s,c}D^{ph}I^{s,c}+\mbox{``higher order terms.''}
\ee

The exact expressions for $R^{s,c}$ are given by the infinite series
\be
R^{s,c}=-I^{s,c}D^{ph}I^{s,c}+I^{s,c}D^{ph}I^{s,c}D^{ph}I^{s,c}-\cdots\ .\label{scseries}
\ee

\subsubsection{Proof of Eq. \eref{scseries}}

We first write $R^{ph2}$ as an infinite series $R^{ph2}=\sum_{n=1}^\infty R^{ph2}_n$, where $R^{ph2}_n$ is the set of diagrams, which are exactly $n$-times reducible in the p-h 2 channel. A similar decomposition is done for $R^{ph1}$.  

\begin{figure}  
\centerline{\includegraphics[width=12cm]{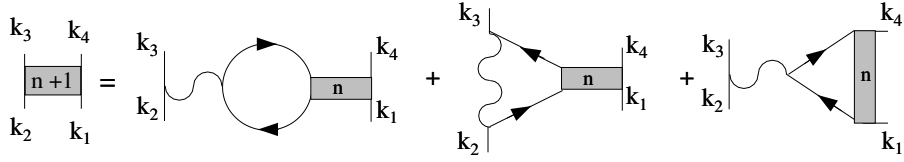}}
\caption{The recursion relation for $R^{ph2}_{n+1}$. Wavy lines drawn horizontally stand for $I^{ph2}$ and wavy lines drawn vertically stand for $I^{ph1}$. The grey rectangles with index $n$ stand for $R^{ph2}_n$ ($R^{ph1}_n$) if they are drawn horizontally (vertically).}\label{ph2recursion}
\end{figure}

\begin{figure}  
\centerline{\includegraphics[width=6cm]{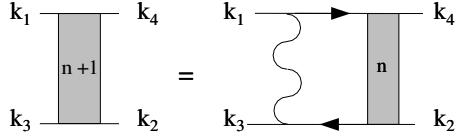}}
\caption{The recursion relation for $R^{ph1}_{n+1}$. The conventions are as in Fig. \ref{ph2recursion}}\label{ph1recursion}
\end{figure}

The recursion relations between $R^{ph1}_{n+1},R^{ph2}_{n+1}$ and  $R^{ph1}_{n},R^{ph2}_{n}$ are shown graphically in Figs. \ref{ph2recursion} and \ref{ph1recursion}. The relation for $R^{ph2}_{n}$ (Fig. \ref{ph2recursion}) can be understood as follows. A general $R^{ph2}_{n+1}$-diagram as shown in Fig. \ref{ph2ladders} can either start on the left with a vertical or a horizontal wavy line. If it starts with a vertical line, then it is given by the second diagram of Fig. \ref{ph2recursion}. If it starts with a horizontal line, there are two cases. If there are no other horizontal lines except for the starting one, the diagram is given by the third term of Fig. \ref{ph2recursion}. If there is more than one horizontal line in the diagram, it is given by the first diagram in Fig. \ref{ph2ladders}.

Analytically the equations depicted in Figs. \ref{ph2recursion} and \ref{ph1recursion} read
\ba 
R_{n+1}^{ph2}&=& -2\ I^{ph2} D^{ph} R_n^{ph2}+XI^{ph1} D^{ph} R_n^{ph2}+I^{ph2} D^{ph} XR_n^{ph1}\\
XR_{n+1}^{ph1}&=&XI^{ph1} D^{ph} R_n^{ph1},
\ea
or, with the definitions \eref{cdef} and \eref{sdef},
\be
 R_{n+1}^{s,c}=-I^{s,c} D^{ph} R_n^{s,c}.
\ee
From this, it is easy to deduce Eq. \eref{scseries}. \Qed

\subsubsection{}

In analogy to the p-p case, we find now the Bethe-Salpeter equations for $\Gamma^s$ and $\Gamma^c$ 
\be
R^{s,c}=-I^{s,c}D^{ph}\Gamma^{s,c}.\label{scBS}
\ee

\section{RG flow of the vertex}\label{floweq}
%-----------------------------

The Bethe-Salpeter equations \eref{BSpp} and \eref{scBS}, three integral equations for the three unknown functions $R^{pp}$, $R^{c}$ and $R^{s}$, are in general difficult to solve. 

We will address the more modest task of calculating the leading term in the perturbative expansion \eref{dev}. This will be accomplished via the flow equation, i. e. a differential equation for $\partial_\Lambda\Gamma$ where we keep only the leading terms in $\Lambda\to0$.

Within this approximation, the two-particle irreducible vertex $I$ is given by the bare interaction $-g$. Thus by Eq. \eref{decomposition},
\be
\dot\Gamma(k_1,k_2,k_3)=\dot R^{pp}(k_1,k_2,k_3)+\dot R^{ph1}(k_1,k_2,k_3)+\dot R^{ph2}(k_1,k_2,k_3),\label{master}
\ee
where the dot means a partial derivative with respect to $\Lambda$.

Consider the first term $\dot R^{pp}$. The derivative of the Bethe-Salpeter equation \eref{formalpp} has three contributions
\be
 \dot R^{pp}=\dot I^{pp}D^{pp}\Gamma+I^{pp}\dot D^{pp}\Gamma+I^{pp}D^{pp}\dot \Gamma.\label{Rppderiv}
\ee 
Although $\dot I^{pp}$ is by no means negligible by itself (since it contains terms which are reducible in the p-h 1 and p-h 2 channels), the first contribution, $\dot I^{pp}D^{pp}\Gamma$, can be shown to be of subleading order in $\Lambda$. I will not prove this here, but an example is discussed in detail in Appendix \ref{parquetex}. The last term is written as $I^{pp}D^{pp}\dot I^{pp}+I^{pp}D^{pp}\dot R^{pp}$ and $I^{pp}D^{pp}\dot I^{pp}$ is neglected for the same reason as $\dot I^{pp}D^{pp}\Gamma$. Therefore
\be
 \dot R^{pp}=I^{pp}\dot D^{pp}\Gamma+I^{pp}D^{pp}\dot R^{pp}.
\ee 
This equation can be iterated to give
\ba
 \dot R^{pp}&=&I^{pp}\dot D^{pp}\Gamma+I^{pp}D^{pp}I^{pp}\dot D^{pp}\Gamma+I^{pp}D^{pp}I^{pp}D^{pp}I^{pp}\dot D^{pp}\Gamma+\cdots\nonumber\\
&=&I^{pp}\dot D^{pp}\Gamma+R^{pp}\dot D^{pp}\Gamma\nonumber\\
&=&\Gamma\dot D^{pp}\Gamma\label{rg1},
\ea
or written out with the full functional dependencies
\be
\dot R^{pp}_\Lambda(k_1,k_2,k_3)=\frac1{\beta L^2}\sum_p\Gamma_\Lambda(k_1,k_2,k-p)\dot D^{pp}_{\Lambda,k}(p)\,\Gamma_\Lambda(p,k-p,k_3),\label{RGpp}
\ee
where $k=k_1+k_2$ is the total frequency-momentum.

The same kind of differential equations are obtained in an analogous way for the p-h channels\footnote{Remember that the formal products of functions mean something different in Eq. \eref{rg2} than in Eq. \eref{rg1}. The formal product in the p-p channel is defined by Eqs. \eref{formalpp} and \eref{BSpp}, whereas the formal product in the p-h channels is defined by Eqs. \eref{explicitph2} and \eref{formalph2}.}. Using Eqs. \eref{scBS} and \eref{scseries}, we obtain
\ba
 \dot R^{s,c}&=&-I^{s,c}\dot D^{ph}\Gamma^{s,c}-I^{s,c}D^{ph}\dot R^{s,c}\label{analog}\\
&=&-I^{s,c}\dot D^{ph}\Gamma^{s,c}+I^{s,c}D^{ph}I^{s,c}\dot D^{ph}\Gamma^{s,c}-\cdots\\
&=&-I^{s,c}\dot D^{ph}\Gamma^{s,c}-R^{s,c}\dot D^{ph}\Gamma^{s,c}\\
&=&-\Gamma^{s,c}\dot D^{ph}\Gamma^{s,c}.\label{rg2}
\ea
The only point to be verified is that the terms $\dot I^{s,c}D^{ph}\Gamma^{s,c}$ and $I^{s,c}D^{ph}\dot I^{s,c}$, which have been left out in Eq. \eref{analog}, are in fact of subleading order in $\Lambda$. Writing out  $\dot I^{s,c}$ in terms of $\dot I^{ph1}$ and $\dot I^{ph2}$, one can verify that all the neglected terms are of the form shown graphically in Fig. \ref{neglected}. They can be shown to be negligible in analogy with the example of Appendix \ref{parquetex}.

\begin{figure}
\centerline{\includegraphics[width=7cm]{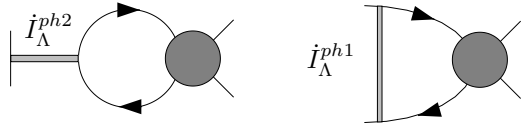}}\vspace{-11pt}
\centerline{\hspace{-2.1cm}\raisebox{1.3cm}[0pt][0pt]{$\dot I_\Lambda^{ph2}$}\hspace{3.1cm}\raisebox{0.9cm}[0pt][0pt]{$\dot I_\Lambda^{ph1}$}}
\caption{Negligible terms in the flow equation for $\partial_\Lambda\Gamma$. The grey circles stand for any sub-diagram.}\label{neglected}
\end{figure}

Eq. \eref{rg2}, expressed in terms of $R^{ph1}$ and $R^{ph2}$ leads to
\ba
\dot R^{ph1}&=&X\left((X\Gamma)\dot D^{ph}(X\Gamma)\right)\\
\dot R^{ph2}&=&-2\,\Gamma\dot D^{ph}\Gamma+(X\Gamma)\dot D^{ph}\Gamma+\Gamma\dot D^{ph}(X\Gamma),
\ea
or written out:
\ba
\dot R^{ph1}_\Lambda(k_1,k_2,k_3)&=&\frac1{\beta L^2}\sum_p\dot D^{ph}_{\Lambda,q_1}(p)\,
\Gamma_\Lambda(p,k_2,p+q_1)\Gamma_\Lambda(k_1,p+q_1,k_3)\label{RGph1}\\
\dot R^{ph2}_\Lambda(k_1,k_2,k_3)&=&\frac1{\beta L^2}\sum_p\dot D^{ph}_{\Lambda,q_2}(p)\,\left[-2\Gamma_\Lambda(k_1,p,p+q_2) \Gamma_\Lambda(p+q_2,k_2,k_3)\right.\label{RGph2}\\& 
 &\ \ \left.+\Gamma_\Lambda(p,k_1,p+q_2) \Gamma_\Lambda(p+q_2,k_2,k_3)+
\Gamma_\Lambda(k_1,p,p+q_2) \Gamma_\Lambda(k_2,p+q_2,k_3)\right],\nonumber
\ea
where $q_1=k_3-k_1$, $q_2=k_3-k_2$ are the direct and exchanged transfered momenta and $D_\Lambda^{ph}(p,q)=G_\Lambda(p)G_\Lambda(p+q)$.

Eqs. \eref{master}, \eref{RGpp}, \eref{RGph1} and \eref{RGph2} are the one-loop RG equations we were looking for. They describe the behavior of the vertex under a differential change of the energy scale $\Lambda$. Because of their central role in this thesis, I will put everything together once more:

\Rahmen{
\centerline{\bf One-loop RG equation}
$$
\dot\Gamma(k_1,k_2,k_3)=\dot R^{pp}(k_1,k_2,k_3)+\dot R^{ph1}(k_1,k_2,k_3)+\dot R^{ph2}(k_1,k_2,k_3)
$$
\ba
\dot R^{pp}_\Lambda(k_1,k_2,k_3)&=&\frac1{\beta L^2}\sum_p\dot D^{pp}_{\Lambda,k}(p)\,\Gamma_\Lambda(k_1,k_2,k-p)\Gamma_\Lambda(p,k-p,k_3)\nonumber\\
\dot R^{ph1}_\Lambda(k_1,k_2,k_3)&=&\frac1{\beta L^2}\sum_p\dot D^{ph}_{\Lambda,q_1}(p)\,
\Gamma_\Lambda(p,k_2,p+q_1)\Gamma_\Lambda(k_1,p+q_1,k_3)\label{1loop}\\
\dot R^{ph2}_\Lambda(k_1,k_2,k_3)&=&\frac1{\beta L^2}\sum_p\dot D^{ph}_{\Lambda,q_2}(p)\,\left[-2\Gamma_\Lambda(k_1,p,p+q_2) \Gamma_\Lambda(p+q_2,k_2,k_3)\right.\nonumber\\& 
 &\ \ \left.+\Gamma_\Lambda(p,k_1,p+q_2) \Gamma_\Lambda(p+q_2,k_2,k_3)+
\Gamma_\Lambda(k_1,p,p+q_2) \Gamma_\Lambda(k_2,p+q_2,k_3)\right],\nonumber
\ea
where $k=k_1+k_2$ is the total frequency-momentum and $q_1=k_3-k_1$, $q_2=k_3-k_2$ are the direct and exchanged transfered frequency-momenta, respectively. The p-p and p-h propagators are defined as $D_{\Lambda,k}^{pp}(p)=G_\Lambda(p)G_\Lambda(k-p)$ and $D_{\Lambda,q}^{ph}(p)=G_\Lambda(p)G_\Lambda(p+q)$.

 }

This equation has a very simple diagrammatic interpretation. The three terms $\dot R^{pp}$, $\dot R^{ph1}$ and $\dot R^{ph2}$ correspond to the one-loop diagrams shown in Fig. \ref{diags}, where the wavy lines now represent full vertices $\Gamma_\Lambda$ and the product of two single particle propagators has to be derived with respect to $\Lambda$. Since $\dot D^{pp/ph}_{\Lambda,q}(p)=\dot G_\Lambda(p)G_\Lambda(q\pm p)+G_\Lambda(p)\dot G_\Lambda(q\pm p)$, each diagram can be viewed as the sum of two terms, where one of the two lines represents the propagator $G_\Lambda(p)$ and the other is $\dot G_\Lambda(p)$, the propagator restricted to the energy scale $\Lambda$. 

Although we have defined $\Lambda$ as a sharp infrared cutoff in the band energy, the RG equation is also correct for a cutoff in the frequency or for smooth cutoffs. It is straightforward to use the finite temperature instead of a cutoff to regularize the theory. The RG equation for $\partial_T\Gamma_T$ is then obtained by replacing in Eq. \eref{1loop}
\be
\frac1{\beta L^2}\sum_p\dot D^\diamond(p,q)\ldots\ \to\frac1{L^2}\sum_{n\in\Z,\,\v p} \partial_T\left(\frac1\beta D^\diamond(p,q)|_{p_0=2\pi T(2n+1)}\right)\ldots,
\ee
where $\diamond=pp,ph$. 

Different regularizations of the theory give the same result to leading logarithmic order. For instance, the vertex at zero temperature and finite infrared cutoff $\Lambda$ is, within logarithmic precision, equal to the vertex at temperature $T=\Lambda$ without cutoff. The following calculations will be done at zero temperature and a finite cutoff, but at the end the parameter $\Lambda$ can be interpreted as the temperature.

Since the one-loop RG equation is only correct to leading order in $\Lambda$, only the leading terms are to be taken seriously. Taking into account subleading terms in Eq. \eref{1loop} is inconsistent and arbitrary, since these subleading terms are comparable to terms that have been neglected in Eq. \eref{1loop}.

This remark about subleading terms concerns in particular the self-energy corrections to the propagator $G_\Lambda$. As already pointed out after Eq. \eref{seriespp}, the self-energy changes the shape and location of the Fermi surface. This effect is by no means negligible in perturbation theory. However, in some special situations, that will be addressed later, the Fermi surface is fixed by exact symmetries, so the self-energy cannot affect it. Other effects, such as the renormalization of the Fermi velocity or the reduction of the quasi-particle weight are of subleading order, as will be checked later from case to case. So these effects cannot be addressed consistently within the one-loop approximation. For this reason, I will replace the dressed electron propagators  $G_\Lambda$ in Eq. \eref{1loop} by the bare ones $C_\Lambda$.

Eq. \eref{1loop} differs from the one-loop RG equation derived by Salmhofer and Honerkamp \cite{Honerkamp,HSFR01,Salmhofer01} only in such self-energy terms, which should be neglected for consistency reasons. In fact, if the cutoff is introduced by some function multiplying the bare propagator
\be
C_\Lambda(p)=\frac{\theta_\Lambda(p)}{ip_0-\xi_{\v p}},
\ee
then the full propagator is given by
\ba
G_\Lambda(p)&=&\frac{C_\Lambda(p)}{1-C_\Lambda(p)\Sigma_\Lambda(p)},\\
&=&\frac{\theta_\Lambda(p)}{ip_0-\xi_{\v p}-\theta_\Lambda(p)\Sigma_\Lambda(p)}.
\ea
The derivative with respect to $\Lambda$ gives
\be
\dot G_\Lambda(p)=\frac{\dot\theta_\Lambda(p)(ip_0-\xi_{\v p})}{\left(ip_0-\xi_{\v p}-\theta_\Lambda(p)\Sigma_\Lambda(p)\right)^2}+\frac{\theta^2_\Lambda(p)\dot\Sigma_\Lambda(p)}{(ip_0-\xi_{\v p}-\theta_\Lambda(p)\Sigma_\Lambda(p))^2}.\label{Gdot}
\ee
The first term of the right hand-side is the single-scale propagator of Ref. \cite{HSFR01}. The one-loop RG equation of Salmhofer and Honerkamp is obtained from Eq. \eref{1loop} by omitting the second term of Eq. \eref{Gdot}.

\section{Linear response}\label{corr}
%------------------------

To study the instabilities of the electron gas, we investigate the linear response of the system to a weak external perturbation. 

\subsection{Generalized susceptibilities for superconductivity}
%-------------------------------------------------------------

Let the system be coupled to a time dependent, real valued field $ \lambda(\v r,t)$, via an additional term in the Hamiltonian
\be
H_{{\rm ext},\,t}=-\sum_{\v r} \lambda(\v r,t)\left( \Delta_A^{}(\v r)+ \Delta_A^\dagger(\v r)\right),
\ee 
where
\be
 \Delta_A^{}(\v r)=\sum_{\v r'}  f_A(\v r-\v r')\,c_{\v r'\down}c_{\v r'\up},\label{DSC}
\ee
is the locally defined superconducting order parameter. The index $A$ stands for the internal wave function $  f_A(\v r-\v r')$ of the Cooper pair. This term can also be written in Fourier space as 
\be
H_{{\rm ext},\,t}=-L^2\sum_{\v k}\left(\lambda(-\v k,t)\Delta_A^{}(\v k)+\mbox{ h. c. }\right),
\ee
where 
\be
\Delta_A^{}(\v k)=\sum_{\v r}e^{i\v k\v r}\Delta_A^{}(\v r)=\sum_{\v p}f_A(\v p)\,c_{\v p\down}c_{\v k-\v p\up}
\ee
and the functions $\lambda(\v k)$ and $f_A(\v p)$ are the Fourier transforms\footnote{Throughout this thesis I use the conventions $F(\v r)=\frac1{L^2}\sum_{\v k}e^{i\v k\v r}F(\v r)$ and $F(\v k)=\sum_{\v r}e^{-i\v k\v r}F(\v k)$ for any quantity $F(\v r)$. The only exceptions are the creation, and annihilation operators. There, I use a pre-factor $\frac1L$ for both transformations. Note that $F^\dagger(\v k)$ denotes the Hermitian conjugate of $F(\v k)$ rather than the usual Fourier transform of $F^\dagger(\v r)$. The Fourier transform in (real) time is defined as $F(t)=\int\!\frac{\ud\omega}{2\pi}\,e^{-i\omega t}F(\omega)$ and $F(\omega)=\int\!\ud t\,e^{i\omega t}F(t)$.} of $\lambda(\v r)$ and $f_A(\v r)$.

One wants to test the effect of this term on the expectation value $\langle \Delta_B^{}(\v r)\rangle_t$, where the internal wave function $  f_B$, can in general be different from $  f_A$. The coupling of the system to the field $\lambda$ induces in general a non-zero value of $\langle \Delta_B^{}(\v r)\rangle$. Physically, this is a strongly idealized version of the superconducting proximity effect. The linear response is given by the retarded response function or susceptibility $ \chi^{BCS}_{{\rm ret},BA}(\v r-\v r',t-t')$, via
\be
\langle \Delta_B(\v r)\rangle_t=\int\!\ud t'\,\sum_{\v r'} \chi^{BCS}_{{\rm ret},BA}(\v r-\v r',t-t')\, \lambda(\v r',t')
\ee
or, after Fourier transformation in space and time, 
\be
\langle\Delta_B(\v k)\rangle_\omega=\chi^{BCS}_{{\rm ret},BA}(\v k,\omega)\,\lambda(\v k,\omega).
\ee
The global superconducting order parameter is given by $\Delta(\v k=0)$, but later in Section \ref{uniform}, I will also consider the more exotic case $\Delta(\v k=(\pi,\pi)$.

A more convenient quantity is the Matsubara response function given by
\be
\chi^{BCS}_{BA}(\v k,\nu)=\frac1{L^2}\int_0^\beta\!\ud\tau\,e^{i\nu\tau}\left\langle\Delta^{}_B(\v k,\tau)\Delta^\dagger_A(\v k)\right\rangle,\label{chiBCS1}
\ee
where the Bose-Matsubara frequency $\nu$ is restricted to even multiples of $2\pi/\beta$ and the imaginary time dependence of operators is given by Eq. \eref{imagHeis}. The retarded response function is obtained by analytic continuation 
\be
\chi(\nu)\ \MapRight{i\nu\to\omega+i\delta}\ \chi_{\rm ret}(\omega),
\ee
where $\delta$ is an infinitesimal quantity, which mimics dissipation.

From Eq. \eref{chiBCS1} and using  Eq. \eref{vertexdef}, we calculate 
\ba
\chi^{BCS}_{BA}(k)&=&\frac1{L^2}\int_0^\beta\!\ud\tau\,e^{i\nu\tau}\sum_{\v p,\v p'}f_B(\v p)\overline{f_A(\v p')}\,\left\langle c^{}_{\v p\down}(\tau)c^{}_{\v k-\v p\up}(\tau)c^\dagger_{\v k-\v p'\up}c^\dagger_{\v p'\down}\right\rangle\\
&=&\frac1{(\beta L^2)^2}\sum_{p,p'}f_B(\v p)\overline{f_A(\v p')}\ G^{\down\up}(p,k-p,k-p'),\\
&=&\frac1{(\beta L^2)^2}\sum_{p,p'}f_B(\v p)\left[\beta L^2\delta_{p,p'}+D^{pp}_{k}(p)\,\Gamma^{BCS}_k(p,p')\right]D^{pp}_{k}(p')\,\overline{f_A(\v p')}\label{chiBCS},
\ea
where $k=(\nu,\v k)$,  $\Gamma^{BCS}_k(p,p')=\Gamma(p,k-p,k-p')$ and $\overline{f_A(\v p')}$ is the complex conjugate of $f_A(\v p')$.

Let us perform the change of variables $\v p\,\to\,\xi_{\v p},\theta_{\v p}$, where $\xi$ is the band energy and $\theta$ is a suitable (angular) variable. In the weak coupling limit the low energy behavior is determined by the degrees of freedom close to the Fermi surface. Therefore the dependence of the form factors on $\xi$ is irrelevant (the argument is analogous to that in Eqs. \eref{Finbubble} and \eref{Fexpand}). The most elementary superconducting susceptibility is a function of two angular variables, which is obtained by putting $f_B(\v p)=\delta(\theta_{\v p}-\theta)$ and $f_A(\v p')=\delta(\theta_{\v p'}-\theta')$ in Eq. \eref{chiBCS}, 
\be
\chi^{BCS}(k,\theta,\theta')=\int\!\ud\xi\,\int\!\ud\xi'\,J(\xi,\theta)J(\xi',\theta')\,\frac1{\beta^2}\!\sum_{p_0,p'_0}G^{\down\up}(p_{(\theta,\xi)},k-p_{(\theta,\xi)},k-p_{(\theta',\xi')})\label{anglesusceptbcs}
\ee
where $J(\xi,\theta)$ is the Jacobian, such that $L^{-2}\sum_{\v p}\to\int\!\ud\theta\int\!\ud\xi J(\xi,\theta)$.

These susceptibility allows for an analysis of the superconducting instability without any prejudices on the form factor $f(\v p)$. The natural form factor is obtained by writing $\chi(\theta,\theta')$ in the diagonalized form 
\be
\chi(\theta,\theta')=\sum_n\chi_nf_n(\theta)f_n(\theta'),
\ee
where $\chi_n$ and $f_n$ are, respectively, eigenvalues and eigenfunctions of the operator
\be
f(\theta)\ \to\ \int\!\ud\theta'\,\chi(\theta,\theta') f(\theta').
\ee
If $\chi_n$  diverges at an energy scale $\Lambda_c$, this indicates a (spin-, charge- or superconducting) instability with the form factor $f_n$. 

To simplify the calculations, one often chooses by hand a certain form factor $f(\v p)$ and calculates the susceptibility for $f_A(\v p)=f_B(\v p)=f(\v p)$.

\subsection{Density waves and flux phases}
%----------------------------------------

In the same way as we coupled the system to a particle-particle pair creating operator Eq. \eref{DSC}, one can perturb the system with a term that creates particle-hole pairs. We again allow for two different form factors $f_A$ and $f_B$.
\be
\Delta_{A/B}(\v r)=\frac12\sum_{\v r',\sigma}s_\sigma f_{A/B}(\v r'-\v r)c^\dagger_{\v r'\sigma}c^{}_{\v r\sigma},
\ee
or in Fourier space
\be
\Delta_{A/B}(\v q)=\frac12\sum_{\v p,\sigma}s_\sigma f_{A/B}(\v p)c^\dagger_{\v p\sigma}c^{}_{\v p+\v q\sigma},
\ee
where either $s_\sigma=\sigma$ for the spin susceptibility $\chi^s$ or $s_\sigma=1$ for the charge susceptibility $\chi^c$.

What is the physical interpretation of the order parameter $\Delta_B(\v q)$? Writing this operator in real space gives
\be
\Delta_{B}(\v q)=\frac12\sum_{\v r,\v r',\sigma}s_\sigma e^{-i\v q\v r} f_{B}(\v r'-\v r)c^\dagger_{\v r'\sigma}c^{}_{\v r\sigma}.\label{orderparameter}
\ee
Its physical nature depends on the orbital form factor $f_B$, the spin structure $s_\sigma$ and also on the momentum $\v q$. Different cases will be discussed in this thesis.

\begin{itemize}
\item Spin- and charge-density waves

 The simple choice $f_B(\v p)=1$ yields a charge- or spin-density wave if $s_\sigma=1$ or $\sigma$, respectively. Most prominent is the checkerboard case, where $\v q=(\pi,\pi)$. In Section \ref{uniform}, we will also consider the limit $\v q\to\v 0$, where we obtain the particle number and the total spin operators, respectively. The corresponding susceptibilities are the charge compressibility and the usual uniform spin susceptibility.

\item Flux phases

If $f_B(\v p)$ is a function with $d_{x^2-y^2}$-symmetry, $\v q=(\pi,\pi)$ and $s_\sigma=1$ ($\sigma$), then the nearest-neighbor-terms  in Eq. \eref{orderparameter} yield circular charge (spin) currents flowing around
 the plaquettes of the square lattice with alternating directions (see 
Fig. \ref{flux}). These four charge and spin instabilities have been discussed a long time ago in the context of the excitonic insulator \cite{Halperin68}.

\begin{figure} 
        \centerline{\includegraphics[width=4cm]{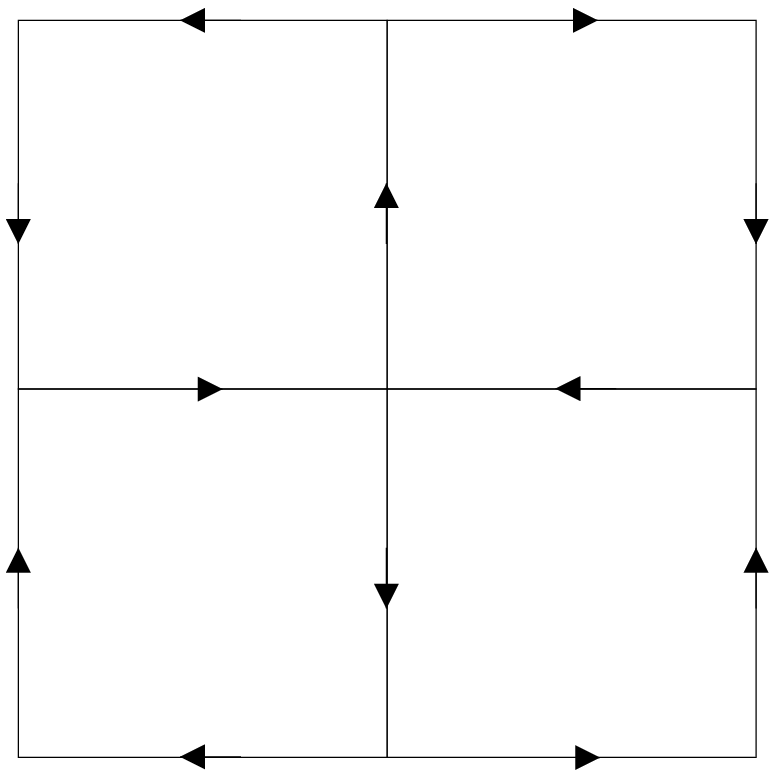}}
\caption{The pattern of charge (spin) currents along the bonds of 
the square lattice in a charge (spin) flux-phase}\label{flux}
\end{figure}

We call the phase with circulating charge currents the charge flux-phase (CF) \cite{Affleck88,Kotliar88,Zou88}, it is sometimes also called $d$-density wave, charge-current wave or orbital antiferromagnetism. The charge flux-phase (CF), closely related to the concept of the chiral spin liquid
\cite{Wen89,Laughlin89}, still plays a prominent role in the strong
 coupling $SU(2)$ theory of the $t-J$ model \cite{Wen96,Lee98}. Recently it
 was proposed to be {\it the} competing order parameter to $d$-wave
 superconductivity and responsible for the pseudo gap phase of the 
cuprates \cite{Chakravarty01}. 

The phase with circulating spin currents is called the spin flux-phase (SF). Other names encountered in the literature are ``spin current wave'' or ``spin nematic state'' (because it is a state with broken rotational symmetry and unbroken time reversal symmetry).
The low-temperature thermodynamics of both the charge and spin flux-phases have been investigated in mean field approximation in Refs. \cite{Nersesyan89,Nersesyan91}.

\item Fermi surface deformations

In Section \ref{uniform}, I will discuss the case $\v q\to\v0$ also with a $d_{x^2-y^2}$-wave form factor. In the case $s_\sigma=1$, the term $H_{\rm ext}$ modifies the dispersion relation $e_{\v k}$, i.e. it deforms the Fermi surface. It makes the electrons move preferably in the $x$-direction, than in the $y$-direction. The nearest-neighbor terms of Eq. \eref{orderparameter} then measures, the difference $T_x-T_y$, where $T_x$ ($T_y$) is the kinetic- or hopping energy of the $x$ ($y$) bonds. In the case $s_\sigma=\sigma$, the Fermi surface deformation is in the opposite direction for up- or down-spin electrons.
\end{itemize}

 The generalized charge- and spin susceptibilities are given by
\ba
\chi^{s/c}_{BA}(q)&=&\frac1{L^2}\int_0^\beta\!\ud\tau\,e^{i\nu\tau}\left\langle\Delta^{}_B(\v q,\tau)\left(\Delta^\dagger_A(\v q)+\Delta^{}_A(-\v q)\right)\right\rangle\\
&=&\frac1{4L^2}\int_0^\beta\!\ud\tau\,e^{i\nu\tau}\sum_{\v p,\v p'}f_B(\v p)\left(\overline{f_A(\v p')}+f_A(\v p'+\v q)\right)\cdot\nonumber\\
& &\qquad\qquad\cdot\,\sum_{\sigma\sigma'}s_\sigma s_{\sigma'}\left\langle c^\dagger_{\v p\sigma}(\tau)c^{}_{\v p+\v q\sigma}(\tau)c^\dagger_{\v p'+\v q\sigma'}c^{}_{\v p'\sigma'}\right\rangle,
\ea
where  $q=(\nu,\v q)$. We consider in this thesis, either $\v q\to(\pi,\pi)$ (Section \ref{squareFS})  or $\v q\to\v 0$ (Section \ref{uniform}). In both cases, $2\v q\to\v 0$, and $\Delta_{A/B}(\v q)$ can be made Hermitian if we choose $f(\v p+\v q)=\overline{f(\v p)}$. Within this choice, the spin- and charge susceptibilities can be written in perfect analogy with Eq. \eref{chiBCS}
\ba
\chi^{s/c}_{BA}(q)&=&\frac1{(\beta L^2)^2}\sum_{p,p'}f_B(\v p)\overline{f_A(\v p')}\left(G^{\up\up}(p',p+q,p)\mp G^{\up\down}(p',p+q,p)\right).\\
&=&\frac1{(\beta L^2)^2}\sum_{p,p'}f_B(\v p)\left[-\beta L^2\delta_{p,p'}+D^{ph}_q(p)\,\Gamma^{s/c}_q(p,p')\right]D^{ph}_q(p')\,\overline{f_A(\v p')},\label{chisc}
\ea
where $\Gamma^s_q(p,p')=-X\Gamma(p',p+q,p)$ and  $\Gamma^c_q(p,p')=(2-X)\Gamma(p',p+q,p)$.

\subsection{Flow equations for the susceptibilities}
%--------------------------------------------------

The RG formalism can be used to calculate the susceptibilities to leading order in the infrared cutoff $\Lambda$. For this, rewrite Eqs. \eref{chiBCS} and \eref{chisc} as
\ba
\chi^{BCS}_{B\!A,\Lambda}(k)&=&\frac1{\beta L^2}\sum_p Z^{BCS}_{B,\Lambda,k}(p)D^{pp}_{\Lambda,k}(p)f_A(\v p)\label{chiZBCS}\\
\chi^{s/c}_{B\!A,\Lambda}(q)&=&-\frac1{\beta L^2}\sum_p Z^{s/c}_{B,\Lambda,q}(p)D^{ph}_{\Lambda,q}(p)f_A(\v p),\label{chiZsc}
\ea
where the effective field vertices $Z$ are defined as 
\ba
Z^{BCS}_{B,\Lambda,k}(p)=f_B(\v p)+\frac1{\beta L^2}\sum_{p'}f_B(\v p')D^{pp}_{\Lambda,k}(p')\Gamma_{\Lambda,k}^{BCS}(p',p)\label{ZBCS}\\
Z^{s/c}_{B,\Lambda,q}(p)=f_B(\v p)-\frac1{\beta L^2}\sum_{p'}f_B(\v p')D^{ph}_{\Lambda,q}(p')\Gamma_{\Lambda,q}^{s/c}(p',p).\label{Zsc}
\ea
The vertex function $\Gamma^{BCS}$ in Eq. \eref{ZBCS} can be expressed by an infinite series in the p-p irreducible part $I^{pp}$. Formally
\ba
Z^{BCS}_B&=&f_B+f_BD^{pp}\Gamma^{BCS}\nonumber\\
&=&f_B+f_BD^{pp}I^{pp}+f_BD^{pp}I^{pp}D^{pp}I^{pp}+\cdots\nonumber\\
&=&f_B+Z^{BCS}_BD^{pp}I^{pp}\label{ZBCSequation}
\ea

To get the flow equation, we take the derivative with respect to $\Lambda$  of Eq. \eref{ZBCSequation}. It gives three terms
\be
\dot Z^{BCS}_B=\dot Z^{BCS}_BD^{pp}I^{pp}+Z^{BCS}_B\dot D^{pp}I^{pp}+Z^{BCS}_BD^{pp}\dot I^{pp}.\label{ZRG1}
\ee
The last term of Eq. \eref{ZRG1} is neglected for the same reason as in Eqs. \eref{Rppderiv}. The remaining equation is iterated to give
\ba
\dot Z^{BCS}_B&=&Z^{BCS}_B\dot D^{pp}I^{pp}+Z^{BCS}_B\dot D^{pp}I^{pp}D^{pp}I^{pp}+Z^{BCS}_B\dot D^{pp}I^{pp}D^{pp}I^{pp}D^{pp}I^{pp}+\cdots\nonumber\\
&=&Z^{BCS}_B\dot D^{pp}\,\Gamma^{BCS}\label{ZRG}
\ea
This is the RG equation for the field-vertex $Z^{BCS}_B$. The form factor $f_B$ enters as an initial condition $Z^{BCS}_{B,\Lambda_0,k}(p)\approx f_B(\v p)$.

It is now easy to obtain a RG equation for $\chi^{BCS}$ as well. From Eqs. \eref{chiZBCS} and \eref{ZRG}
\ba
\dot\chi^{BCS}&=&\dot Z^{BCS}_B D^{pp}f_A+ Z^{BCS}_B \dot D^{pp}f_A\nonumber\\
&=&Z^{BCS}_B\dot D^{pp}\,\Gamma^{BCS}D^{pp}f_A+ Z^{BCS}_B \dot D^{pp}f_A\nonumber\\
&=&Z^{BCS}_B\dot D^{pp}\left(f_A+\Gamma^{BCS}D^{pp}f_A\right)\nonumber\\
&=&Z^{BCS}_B\dot D^{pp}Z^{BCS}_A
\ea
In the last equation, we have introduced the function $Z^{BCS}_{A,\Lambda,k}(p)$, which is the same as $Z^{BCS}_{B,\Lambda,k}(p)$, except that the initial condition $f_B(\v p)$ is changed into $f_A(\v p)$.

Written out, the RG equations for the pairing susceptibility are

\Rahmen{
\ba
\dot\chi^{BCS}_{B\!A,\Lambda}(k)&=&\frac1{\beta L^2}\sum_p Z^{BCS}_{B,\Lambda,k}(p)\dot D^{pp}_{\Lambda,k}(p)\,Z^{BCS}_{A,\Lambda,k}(p)\label{chiRGfinal}\\
\dot Z^{BCS}_{\diamond,\Lambda,k}(p)&=&\frac1{\beta L^2}\sum_{p'} Z^{BCS}_{\diamond,\Lambda,k}(p')\dot D^{pp}_{\Lambda,k}(p')\,\Gamma^{BCS}_{\Lambda,k}(p',p),\label{ZRGfinal}
\ea
where $\diamond=A$ or $B$, and $\Gamma^{BCS}_{\Lambda,k}(p',p)=\Gamma_\Lambda(p',k-p',k-p)$. 

}

Once the vertex function $\Gamma^{BCS}$ is known from Eq. \eref{1loop}, one can solve the linear equation \eref{ZRGfinal} for the initial condition $Z^{BCS}_{\diamond,\Lambda_0,k}(p)=f_\diamond(\v p)$ and finally one can integrate Eq. \eref{chiRGfinal} to obtain the susceptibility. 

All the steps can be repeated for the charge- and spin susceptibilities, just by replacing  the quantities $Z^{BCS}, \Gamma^{BCS},I^{pp}$ by  $Z^{s/c}, \Gamma^{s/c},I^{s/c}$ and the p-p propagator $D^{pp}$ by $-D^{ph}$.  

\Rahmen{
\ba
\dot\chi^{s/c}_{B\!A,\Lambda}(q)&=&-\frac1{\beta L^2}\sum_p Z^{s/c}_{B,\Lambda,q}(p)\dot D^{ph}_{\Lambda,q}(p)\,Z^{s/c}_{A,\Lambda,q}(p)\label{chiRGfinalsc}\\
\dot Z^{s/c}_{\diamond,\Lambda,q}(p)&=&-\frac1{\beta L^2}\sum_{p'} Z^{s/c}_{\diamond,\Lambda,q}(p')\dot D^{ph}_{\Lambda,q}(p')\,\Gamma^{s/c}_{\Lambda,q}(p',p),\label{ZRGfinalsc}
\ea
where $\diamond=A$ or $B$, $\Gamma^{s}_{\Lambda,q}(p',p)=-X\Gamma_\Lambda(p',p+q,p)$ and $\Gamma^{c}_{\Lambda,q}(p',p)=(2-X)\Gamma_\Lambda(p',p+q,p)$. 

}

\section{Relation to the Wilsonian approach}\label{Wilson}
%+++++++++++++++++++++++++++++++++++++++++++

A key ingredient to the Wilsonian RG is the idea to replace the given problem by a different one with less degrees of freedom but with the same low energy behavior. To achieve this, the effect of the eliminated degrees of freedom is incorporated in a renormalization of the parameters of the effective low-energy theory.

This strategy has been successfully followed using Brillouin-Wigner perturbation theory in the strong coupling limit of various many-body problems (see for example \cite{Emery,Auerbach,Fazekas,Auerbach00}). The best known example is the Hubbard model at half-filling which is represented, in the limit of strong coupling, by a Heisenberg spin Hamiltonian in the limit of strong coupling.

In the weak coupling limit, the application of the Brillouin-Wigner formalism to compute an effective Hamiltonian is less evident and has not been followed to my knowledge. A tractable implementation of the same idea (i.e. elimination of high energy degrees of freedom and renormalization of parameters of the effective theory) is given more easily in the functional integral representation, which is the content of the following section. 

It should nevertheless be mentioned that the idea of effective Hamiltonians on a reduced Hilbert space led to a most powerful numerical tool for one-dimensional systems: the density matrix renormalization group (see \cite{Noack99} for an introduction). An alternative route to effective Hamiltonians was presented recently by Wegner \cite{Wegner94} and applied to the 2D Hubbard model \cite{Grote01} and other correlated systems \cite{Kehrein99,Kehrein01,Heidbrink01}.

\subsection{Functional integral formulation}
%------------------------------------------

The functional integral formulation of quantum many-particle systems is presented in detail in \cite{Negele}. Here we only mention some of the results.

In the functional integral formulation, the annihilation- and creation operators $c^{}_{\v r\sigma}$ and $c^\dagger_{\v r\sigma}$ are replaced by anti-commuting Grassmann fields $\psi_{(\tau,\v r,\sigma)}$ and  $\bar\psi_{(\tau,\v r,\sigma)}$. They depend on  the ``imaginary time'' variable $\tau\in[0,\beta]$ in addition to the labels of the single-particle states and they satisfy anti-periodic boundary conditions in the $\tau$ variable. 

The Hamilton operator is transformed into the action 
\be
S[\psi]=-\int_0^\beta\ud\tau\left(\sum_{\v r\sigma} \bar\psi_{(\tau,\v r,\sigma)}(\partial_\tau-\mu)\psi_{(\tau,\v r,\sigma)}+H[\tau,\psi]\right),
\ee
where $H[\tau,\psi]$ is obtained from the Hamiltonian by the replacements $c_{\v r\sigma}\to\psi_{(\tau,\v r,\sigma)}$ and  $c^\dagger_{\v r\sigma}\to\bar\psi_{(\tau,\v r,\sigma)}$. 

The partition function is given by the functional integral
\be 
Z=\int{\cal D}\psi\, e^{S[\psi]},\label{Z}
\ee
where ${\cal D}\psi$ is a short-hand notation for $\prod_x\ud\psi_x\ud\bar\psi_x$ and the Green's functions are
\ba
G^n(x_1,\ldots,x_n|x_{2n},\ldots,x_{n+1})&=&\frac1Z\int{\cal D}\psi\, e^{S[\psi]}\,\psi^{}_{x_1}\cdots \psi^{}_{x_n}\bar\psi_{x_{n+1}}\cdots \bar\psi_{x_{2n}}\\
&=&\langle\psi^{}_{x_1}\cdots \psi^{}_{x_n}\bar\psi_{x_{n+1}}\cdots \bar\psi_{x_{2n}}\rangle,\label{Gn}
\ea
where we have introduced the notation $\langle\ldots\rangle$ for averaging over Grassmann monomials.

It is useful to perform a Fourier transformation 
\be
\psi_{\sigma k}=(\beta L^2)^{-1/2}\int_0^\beta\ud\tau\, e^{ik_0\tau}\sum_{\v r}e^{-i\v k\v r}\psi_{(\tau,\v r,\sigma)},\label{Fourier}
\ee
where $k=(k_0,\v k)$ contains the Matsubara frequency $k_0$ as in section \ref{Green}. 

The Fourier transformed one- and two-particle Green's functions are 
\be
G(k)=-\langle\psi_{k\sigma}\bar\psi_{k\sigma}\rangle
\ee
\be
G^{\sigma\sigma'}(k_1,k_2,k_3)=\beta L^2\, \langle\psi_{k_1\sigma}\psi_{k_2\sigma'}\bar\psi_{k_3\sigma'}\bar\psi_{k_1+k_2-k_3\sigma}\rangle\label{G2}
\ee

After Fourier transformation, the action reads 
\be
S[\psi]=S_0[\psi]-W[\psi],
\ee
with
\be
 S_0[\psi]=\sum_{\sigma,k}\bar\psi_{\sigma k}(ik_0-\xi_{\v k}) \psi_{\sigma k},
\ee
where $\xi_{\v k}=e_{\v k}-\mu$ and
\be
W[\psi]=\frac12\frac1{\beta L^2}\sum_{k_1,k_2,k_3}g(\v k_1,\v k_2,\v k_3) \sum_{\sigma, \sigma'}\bar\psi_{k_1\sigma} \bar\psi_{k_2\sigma'}\psi_{k_3\sigma'}\psi_{k_1+k_2-k_3\,\sigma}.\label{int}
\ee
The functional integration measure is also readily expressed in terms of the Fourier transformed fields ${\cal D}\psi=\prod_{k\sigma}\ud\psi_{k\sigma}\ud\bar\psi_{k\sigma}$. 

It is convenient to introduce the partition function with source term 
\be 
Z[\eta]=\int d\mu_C[\psi]\,  e^{-W[\psi]+(\bar\eta,\psi)+(\bar\psi,\eta)},
\label{Zeta}
\ee
where we used the short-hand notation 
 $(\bar\chi,\psi):=\sum_{\sigma k} \bar\chi_{\sigma k}\psi_{\sigma k}$ and the normalized
Gaussian measure is  defined by
\be 
d\mu_C[\psi]:=\frac{{\cal D}\psi\ e^{(\bar\psi,C^{-1}\psi)}}{\int{\cal D}\psi\ e^{(\bar\psi,C^{-1}\psi)}}.\label{Gaussian}
\ee
The connected part of the correlation functions are obtained as 
functional derivatives \cite{Negele}
\be
\langle\psi_1\cdots\psi_n\bar\psi_{n+1}\cdots\bar\psi_{2n}\rangle_{\rm c}=
\left.\frac{\delta^{2n}\log Z[\eta]}{\delta\!\eta_{2n}\cdots
\delta\!\eta_{n+1}
\delta\!\bar\eta_{n}\cdots\delta\!\bar\eta_{1}}\right|_{\eta=0},
\ee
where we have written $\psi_i$ instead of $\psi_{k_i\sigma_i}$.

\subsection{Low energy effective action and relation to one-particle irreducible vertices}\label{effaction}
%---------------------------------------------------------------------------------

In the spirit of Section \ref{cutoff},  the bare propagator is now endowed
with an infrared cutoff $\Lambda$ on the band energy  
$C_\Lambda(k)=\Theta(|\xi_{\v k}|-\Lambda)C(k)$. 

One now defines the {\it effective interaction }
\be
{\cal W}_\Lambda[\chi]=-\log  \int d\mu_{C_\Lambda}[\psi]
e^{-W[\psi+\chi]}\label{effint}
\ee
which depends on a Grassmann field $\chi$ (not to be mistaken with the susceptibilities, for which I have chosen the same symbol).
Note that the integration with respect to $d\mu_{C_\Lambda}[\psi]$ is perfectly
 defined, 
although $C_\Lambda^{-1}$ is not. This can be seen most easily in the 
expansion of ${\cal W}_\Lambda[\chi]$ in terms 
of Feynman diagrams. The evaluation of these diagrams involves only 
$C_\Lambda$ and never $C_\Lambda^{-1}$. Whenever  $C_\Lambda^{-1}$ 
appears in an intermediate step of a calculation (see below), it may be 
regularized by replacing the zero in the Heavyside function by an 
infinitesimal number.

${\cal W}_\Lambda[\chi]$ has a twofold interpretation. On the one hand, we 
can restrict the field $\chi$ to the low energy degrees of freedom 
$\psi_{{\rm <}\, k\sigma}=\Theta(\Lambda-|\xi_{\v k}|)\psi_{k\sigma}$. 
The object
\be
S^{\rm eff}_\Lambda[\psi_{\rm <}]=(\bar\psi_{\rm <},C^{-1} \psi_{\rm <})-
{\cal W}_\Lambda
[\psi_{\rm <}]
\ee
corresponds then to Wilson's effective action, which describes the system in 
terms of $\psi^{\rm <}$ only. In fact, for observables (or Green's functions) which depend only on the low energy fields, one shows
\ba
\langle O[\psi^<]\rangle&=&\frac1Z\int{\cal D}\psi^<\int{\cal D}\psi^>\ e^{S[\psi^{<},\psi^{>}]}\ O[\psi^<]\\
&=&\frac1Z\int{\cal D}\psi^<\  e^{S^{\rm eff}_\Lambda[\psi^{<}]}\ O[\psi^<]
\ea
and
\be
Z=\int{\cal D}\psi^<\int{\cal D}\psi^>\ e^{S[\psi^{<},\psi^{>}]}=\int{\cal D}\psi^<\  e^{S^{\rm eff}_\Lambda[\psi^{<}]}.
\ee

On the other hand, ${\cal W}_\Lambda$ is the generating functional of amputated
 connected correlation functions with infrared cutoff $\Lambda$ because of the
 identity  
\be
\log Z_\Lambda[\eta]=-(\bar\eta,C_\Lambda \eta)-{\cal W}_\Lambda[C_\Lambda\eta]
,\label{gen}
\ee
where $Z_\Lambda$ is given by Eq. \eref{Z}, with $C$ replaced by 
$C_\Lambda$. 

\subsubsection{Proof}
To prove Eq. \eref{gen}, we write the Gaussian measure (Eq. \eref{Gaussian}) explicitly
\be
 e^{-{\cal W}_\Lambda[\chi]}=\frac{\int{\cal D}\psi\, e^{(\bar\psi,C_\Lambda^{-1}\psi)-W[\psi+\chi]}}{\int{\cal D}\psi\, e^{(\bar\psi,C_\Lambda^{-1}\psi)}},\nonumber
\ee
then, only in the numerator, we perform a shift of the integration variables $\psi\to \psi-\chi$
\ba
e^{-{\cal W}_\Lambda[\chi]}&=&\frac{\int{\cal D}\psi\, e^{(\bar\psi-\bar\chi,C_\Lambda^{-1}(\psi-\chi))-W[\psi]}}{\int{\cal D}\psi\, e^{(\bar\psi,C_\Lambda^{-1}\psi)}}\nonumber\\
&=&e^{(\bar\chi,C_\Lambda^{-1}\chi)}\,Z_\Lambda[C_\Lambda^{-1}\chi].\nonumber
\ea
Eq. \eref{gen} follows by putting $\chi=C_\Lambda\eta$ and taking the logarithm.\Qed    

\subsubsection{}
As a consequence of Eq. \eref{gen}, the quadratic part of ${\cal W}_\Lambda$ is related
 to the self-energy $\Sigma_\Lambda$ by 
\ba
\left.\frac{\delta^2}{\delta\!\chi_{\sigma k}\delta\!\bar\chi_{\sigma k}}
{\cal W}_\Lambda[\chi]\right|_{\chi=0}&=&-C^{-1}_\Lambda(k)-\langle
\psi_{\sigma k}\bar\psi_{\sigma k}\rangle_{\rm \Lambda}\,C^{-2}_\Lambda(k)\\
&=&\frac{\Sigma_\Lambda(k)}{1-C_\Lambda(k)\Sigma_\Lambda(k)},\label{self}
\ea
where we have used the following identity for the full electron propagator 
$G_\Lambda(k)=-\langle\psi_{\sigma k}\bar\psi_{\sigma k}\rangle_{\rm \Lambda}
=C_\Lambda(k)(1-C_\Lambda(k)\Sigma_\Lambda(k))^{-1}$. 
Therefore in the case $|\xi_{\v k}|<\Lambda$ the right-hand side of Eq. 
\eref{self} simply becomes $\Sigma_\Lambda(k)$. 

Similarly the quartic part of  ${\cal W}_\Lambda$ is related to the one particle 
irreducible vertex $\Gamma_\Lambda$. In fact,
differentiating  Eq. \eref{gen} we find 
\ba
\left.\frac{\delta^{4}{\cal W}_\Lambda[\chi]}
{\delta\!\chi_{\sigma k_4}\delta\!\chi_{\sigma' k_3}\delta\!\bar
\chi_{\sigma' k_2}\delta\!\bar\chi_{\sigma k_1}}\right|_{\chi=0}
&=&-\langle\psi_{\sigma k_1}\psi_{\sigma' k_2}\bar\psi_{\sigma' k_3}\bar
\psi_{\sigma k_4}\rangle_{\rm c,\Lambda}\prod_{i=1}^4C_\Lambda^{-1}(k_i)\\
&=&-\frac{\Gamma_\Lambda^{\sigma\sigma'}(k_1,\ldots,k_4)}{\beta V\prod_{i=1}^4
[1-C_\Lambda(k_i)\Sigma_\Lambda(k_i)]}.\label{vert}
\ea
In the last line we have used Eqs. \eref{vertexdef} and \eref{G2}.

The quartic part of ${\cal W}_\Lambda$ is of the same form as Eq. \eref{int} 
with an effective coupling function $g_\Lambda(k_1,k_2,k_3)$ that now depends on the frequencies as well as the momenta. Taking functional derivatives of Eq. \eref{int}, we find for $|\xi_{\v k_i}|<\Lambda$,  
\be
(1-\delta_{\sigma \sigma'}X)g_\Lambda(k_1,k_2,k_3)=-\Gamma^{\sigma \sigma'}_{\Lambda}(k_1,k_2,k_3)
\ee
and thus

\Rahmen{
\be
g_\Lambda(k_1,k_2,k_3)=-\Gamma_{\Lambda}(k_1,k_2,k_3).\label{correspondence}
\ee
}

$g_\Lambda$ is equal, up to the sign, to a connected amputated 
correlation function if all $|\xi_{\v k_i}|>\Lambda$ and to the 1PI vertex in the opposite case. $g_\Lambda$ is therefore not continuous at $|\xi_{\v k_i}|=\Lambda$.
A formal and non-perturbative proof of these relations was given  by Morris 
\cite{Morris94} for a bosonic field theory. The derivation given above is perturbative, but a generalization of the non-perturbative proof of Morris to fermions appears to be straightforward. Morris has also shown that $\Sigma_\Lambda$ and 
$\Gamma_\Lambda$ are continuous at $|\xi_{\v k_i}|=\Lambda$, in contrast to $g_\Lambda$.

\subsection{Other one-loop RG equations}\label{compare}
%-----------------------------------

The effective interaction satisfies the following exact RG equation for $\partial_\Lambda{\cal W}_\Lambda=\dot{\cal W}_\Lambda$
\be 
\dot{\cal W}_\Lambda[\chi]=\sum_{\sigma,k}\dot C_\Lambda(k)\
\frac{\delta^2{\cal W}_\Lambda[\chi]}{\delta\!\chi_{\sigma k}\delta\!
\bar\chi_{\sigma k}}-\sum_{\sigma,k}\dot C_\Lambda(k)\ \frac{\delta{
\cal W}_\Lambda[\chi]}{\delta\!\chi_{\sigma k}}\frac{\delta{
\cal W}_\Lambda[\chi]}{\delta\!\bar\chi_{\sigma k}}.\label{exact}
\ee 
This equation was first derived by Polchinski in the context of a scalar field theory \cite{Polchinski84}. A proof is given in Appendix \ref{proof}. Zanchi and Schulz \cite{Zanchi,Zanchi00} proposed to develop ${\cal W}_\Lambda$ up to order six in the fermionic variables and to neglect terms of higher order. Terms of order six are not present in the original interaction but they are produced by the RG procedure. Their effect is then to renormalize the effective coupling function $g_\Lambda$. The result is a closed one-loop equation for the coupling function 
$g_\Lambda(k_1,\ldots,k_4)$  where $|\xi_{\v k_i}|<\Lambda$. It is identical, within the correspondence $g_\Lambda=-\Gamma_\Lambda$, to our Eq. \eref{1loop} with one difference. In the RG equation of Zanchi and Schulz, the vertices on the right hand-side are not evaluated at the scale $\Lambda$, but at a higher scale $\tilde\Lambda$, which is given by the band energy of the single-particle propagators, i.e. $\tilde\Lambda=\Max\{|\xi_{\v p}|,|\xi_{\v k-\v p}|\}$ in the p-p term of Eq. \eref{1loop}, $\tilde\Lambda=\Max\{|\xi_{\v p}|,|\xi_{\v p+\v q_{1}}|\}$ in the p-h 1 term and $\tilde\Lambda=\Max\{|\xi_{\v p}|,|\xi_{\v p+\v q_{2}}|\}$ in the p-h 2 term. Their equation is thus non-local in $\Lambda$. It is a flow equation with memory, i.e. the flow at scale $\Lambda$ doesn't only depend on the vertex function $\Gamma_\Lambda$ but also on the history of the flow.

Since this is not very convenient, it was proposed \cite{Halboth00} 
to develop Eq. \eref{exact} into Wick ordered polynomials of the fermionic 
variables instead of monomials as it was done above. Wick ordering with
 respect to the low energy propagator $D_\Lambda=C-C_\Lambda$ 
results in the same one-loop equation as above but now all the couplings are 
evaluated at the actual RG variable $\Lambda$ and $-\ud\left[C_\Lambda(p)
C_\Lambda(q)\right]/{\ud\Lambda}$ has to be replaced by $\ud\left[D_\Lambda(p)
D_\Lambda(q)\right]/{\ud\Lambda}$, i.e., one propagator is at the energy $\Lambda$ and the energy of the second propagator is now restricted to be {\it smaller} than $\Lambda$. This is different from our Eq. \eref{1loop}, where the second propagator is restricted to higher energies. 

We have seen above that the coupling function of the low-energy effective theory equals, up to a sign, the 1PI vertex $\Gamma_\Lambda$. This correspondence can not be generalized to higher order vertices, i.e. it would be completely wrong to say that the sixth order term of ${\cal W}_\Lambda$ is related to the 1PI part of the three particle Green's function and so forth. In addition, the correspondence Eq. \eref{correspondence} relies on the choice of the cutoff which ensures that $C_\Lambda(p)=0$, for $|\xi_{\v p}|<\Lambda$. The correspondence  no longer holds for an alternative scheme, where for example the finite temperature is used to regularize the theory instead of the infrared cutoff. 

The general 1PI vertices are obtained by performing a Legendre transformation on the functional ${\cal W}_\Lambda$. Wetterich has presented a renormalization group scheme for bosonic field theories, working with this Legendre transformed quantity rather that with the effective action defined in Eq. \eref{effint}  \cite{Wetterich93,Tetradis94}. The idea was implemented recently for the many-fermion problem in \cite{Honerkamp,HSFR01,Salmhofer01}. The resulting RG equation is identical to Eq. \eref{1loop} apart from the self-energy (see end of section \ref{floweq}).

The RG equations of the three groups \cite{Zanchi00,Halboth00,HSFR01} differ in the treatment of the diagrams which give not a leading order contribution in the limit $\Lambda\to0$. For example, the (leading order) p-p diagram at total momentum $k=0$ features two internal propagators with exactly opposite momenta. They have exactly the same band energy. So the non-locality of the Zanchi-Schulz equation is not present in this diagram. For the same reason, the energy-constraint of the Wick-ordered scheme can not have any effect in this case. So the three RG schemes treat such diagrams identically. The same is true for the p-h diagrams in the case of perfect nesting. However, as I have argued in section \ref{floweq}, only the leading order terms of these RG equations make sense. The subleading terms are of the same order as others that have been neglected. In fact, a consistent treatment of subleading terms requires going beyond the one-loop approximation.

\begin{figure}  
\centerline{\includegraphics[width=4cm]{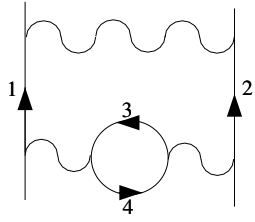}}
\caption{Example of a two loop parquet diagram.}\label{parquetfig2}
\end{figure}

The equivalence of the different RG equations to leading order can also be understood in the following way. By successively integrating the one-loop RG equation \eref{1loop} and expressing the result in terms of  $\Gamma_{\Lambda_0}\approx-g$, one obtains the full series of parquet diagrams. However the structure of Eq. \eref{1loop} introduces a constraint on the energies of internal lines. For example the parquet diagram of Fig. \ref{parquetfig2} is generated by Eq. \eref{1loop} with the following constraint on the propagators 1,2,3 and 4
\be
\Min\{|\xi_1|,|\xi_2|\}\leq\Min\{|\xi_3|,|\xi_4|\}.\label{constraint}
\ee 
Higher order parquet diagrams are generated with similar constraints, i.e. with a certain ``ordering'' of the band energies, when one goes from the inner loops to the exterior loops of a given parquet diagram. It can be checked by introducing this constraint into Eq. \eref{2loop} of Appendix \ref{parquetex}, that the constraint does not change the value of the diagram to leading logarithmic order in $\Lambda$. 

The different RG equations \cite{Zanchi00,Halboth00,HSFR01} all generate the whole series of parquet diagrams, but the constraints are different. In the Wick ordered scheme, the constraint \eref{constraint} is changed into 
$$
\Max\{|\xi_1|,|\xi_2|\}\leq\Max\{|\xi_3|,|\xi_4|\}.
$$
The RG equation of Zanchi and Schulz introduces the most restrictive constraint
$$
\Max\{|\xi_1|,|\xi_2|\}\leq \Min\{|\xi_3|,|\xi_4|\},
$$
i.e. both propagators of the inner loop are higher in energy than both propagators of the exterior loop. All these constraints are irrelevant for the leading logarithmic order in $\Lambda$. I conclude that the RG equation Eq. \eref{1loop} and those of Refs. \cite{Zanchi00,Halboth00,HSFR01} are all equivalent to leading  order in $\Lambda$.

\section{The case of a general Fermi surface revised}\label{revise}
%---------------------------------------------------

In the following, we identify the leading contributions to the one-loop RG equation in the limit of small energies. The main part of the work on the half-filled lattice with nearest-neighbor hopping will be treated in the next chapter. Here I shortly reconsider the case of a general non-nested Fermi surface which has been treated before in Sections \ref{ladder} and \ref{superconductivity}.

\subsection{Low-energy scattering processes}\label{genscattproc}
%...........................................

First one identifies the possible scattering processes which connect four momenta on the Fermi surface because only they will be relevant to leading order. These processes have to satisfy momentum conservation, i.e. we are looking for solutions of the equation 
\be
\v k_1+\v k_2=\v k_3+\v k_4,\label{momentum}
\ee
where $\v k_1,\ldots,\v k_4$ are momenta on the Fermi surface ($\xi_{\v k_i}=0$). Note that this equation is understood modulo a reciprocal lattice vector $2\pi\Z^2$.

One obvious solution is that incoming and outgoing momenta are equal. The corresponding scattering processes are called forward and exchange scattering, respectively with vertices $\Gamma^f$ and $\Gamma^e$ defined by $\Gamma^f(\v k,\v k')=X\Gamma^e(\v k,\v k')=\Gamma(\v k,\v k',\v k')$. To simplify notation I omit the subscript $\Lambda$ in $\Gamma_\Lambda$ from now on. A second general solution corresponds to scattering of particle pairs with zero total momentum. The corresponding vertex is $\Gamma^{BCS}(\v k,\v k')=\Gamma(\v k,-\v k,-\v k')$.

If the Fermi surface is big enough, there are additional Umklapp processes, where the momentum conservation is violated by an integer multiple of $2\pi$ in one or even in both directions. A simple geometric criterion for the existence of low-energy Umklapp processes for the case of a closed and convex Fermi surface is as follows \cite{Kumar97}. Let $FV=\{\v p;\xi_{\v p}<0\}$ be the Fermi volume. Given two momenta $\v k$ and $\v k'$ on the Fermi surface, then the center of mass $\frac12(\v k+\v k')$ is in the Fermi volume, due to convexity. Inversely, for every point $\v p$ of the Fermi volume there exists at least one pair of Fermi momenta  $\v k$ and $\v k'$ such that $\frac12(\v k+\v k')=\v p$. Low-energy Umklapp processes exist, if the Fermi volume has some non-zero intersection with its own translation $\v G/2+FV$, where $ \v G/2$ is half of a reciprocal lattice vector (i.e. $(\pi,0)$, $(0,\pi)$ or $(\pi,\pi)$). Given two Fermi momenta $\v k_1,\v k_2$ such that the center of mass $\frac12(\v k_1+\v k_2)$ is in the intersection $FV\cap(\v G/2+FV)$, then there exists a pair of Fermi momenta $\v k_3,\v k_4$, such that $\v k_1+\v k_2=\v k_3+\v k_4+\v G$. Umklapp processes are continuous two-parameter families like forward-, exchange- or BCS-scattering, not restricted to certain spots on the Fermi surface, nor are they related to some ``Umklapp surface'', as claimed in \cite{HSFR01}\footnote{In \cite{HSFR01}, the dispersion relation $\xi_{\v p}=-2t(\cos p_x+\cos p_y)+4t'\cos p_x \cos p_y-\mu$ of a generalized tight-binding model was considered. In the parameter range $0<t'<0.5$ and $\mu>-4t'$, the Fermi surface consists of four disconnected arcs in the Brillouin zone $BZ=]-\pi,\pi]\times]-\pi,\pi]$. If we choose on the other hand the region $BZ'=[0,2\pi[\times[0,2\pi[$ to represent the momenta $\v p\in \R^2/2\pi Z^2$, then the Fermi surface is closed and convex and the abovementioned criterion applies. But attention: the Brillouin zone BZ' is equivalent to the Brillouin zone BZ only modulo $2\pi Z^2$. Thus some Umklapp processes in BZ' correspond to non-Umklapp processes in BZ. But the lattice model does not really care about which processes are Umklapp and which are not. BZ and BZ' are mathematically equivalent}.

\subsection{Analysis of the bubbles}\label{circ}
%..................................

In order to identify the leading contributions to Eq. \eref{1loop},
 we neglect for the moment the angular dependence of the vertex. The three different contributions p-p, p-h 1 and p-h 2 are proportional to 
\be
\dot B^{ pp/\,ph}(\Lambda,k)=\frac1{\beta L^2}\sum_p\frac{\ud\left[C_\Lambda(p)C_\Lambda(k\mp p)\right]}{\ud\Lambda},\label{bubbles}
\ee
where $k=k_1+k_2$ in the p-p contribution, $k=k_3-k_1$ in the p-h 1 contribution and $k=k_3-k_2$ in the p-h 2 contribution.

We consider the thermodynamic limit and zero 
temperature and therefore replace $1/{\beta V}\sum_k$ by
 $\int \frac{d^{2+1}k}{(2\pi)^{2+1}}$ in the calculations. 
Taking explicitly the derivative with respect to $\Lambda$ and integrating over the frequency $p_0$ (for $k_0=0$) we find 
\be
\dot B^{ pp/\,ph}(\Lambda,\v k)= \mp 2\int\!\frac{d^{2}p}{(2\pi)^{2}}\,\delta(|\xi_{\v p}|-\Lambda)\, \frac{\Theta(|\xi_{\v q}|
-\Lambda)\, \Theta(\pm \xi_{\v p}\xi_{\v q})}{\Lambda+|\xi_{\v q}|}, \label{explicit}
\ee
 where $\v q=\v k\mp\v p$.

Within the Wick ordered scheme \cite{Halboth00}, the first step function in Eq.
 \eref{explicit} would be replaced by $\Theta(\Lambda-|\xi_{\v q}|)$, since the 
second propagator in Eq. \eref{bubbles} is restricted to be in the low energy part. I have verified that this alternative scheme would not change the final results, confirming the conclusion of Section \ref{compare}. 

For general values of $\v k$ one finds that both bubbles $\dot B^{pp,ph}(\Lambda,\v k)$ are proportional to $\log\Lambda$. For the special value $|\v k|=2$, one finds $\dot B^{pp}\sim \dot B^{ph}\sim\Lambda^{-1/2}$. The strongest divergence comes from 
the p-p bubble at $\v k=0$, namely
$\dot B^{pp}(\Lambda,\v 0)\sim\Lambda^{-1}$. Correspondingly, the dominant contribution
in the low-energy regime $\Lambda\to 0$ comes from the p-p channel. As it was argued before in Section \ref{ladder}, the dependence of $\Gamma^{BCS}_{\Lambda}$ on the frequencies $k_0,k'_0$ and the band energies $\xi_{\v k},\xi_{\v k'}$ is irrelevant so that $\Gamma^{BCS}_{\Lambda}$ can be treated as a function of two (angular) variables instead of six, a result that is already known from the more standard scaling analysis used in \cite{Polchinski92,Shankar94,froehlich96}. We therefore write $\Gamma^{BCS}(\v k,\v k')$ instead of $\Gamma^{BCS}(k,k')$, because the frequency dependence has been neglected.

Does it make any formal sense to take into account the non-leading p-h terms of Eq. \eref{1loop} in the absence of nesting? One has to compare the non leading terms $\sim\Gamma_\Lambda^2\log\!\Lambda$  with other contributions, that have been neglected in the derivation of Eq. \eref{1loop}. According to a careful analysis by Salmhofer and Honerkamp \cite{Salmhofer01}, the first neglected terms are or the order $\sim\Gamma^3_\Lambda\log^2\!\Lambda$ (see Eq. 108 of \cite{Salmhofer01}). The contributions to the exact RG flow can be classified into three classes: 1. leading terms of the one-loop equation, 2. subleading terms of the one-loop equation and 3. terms which are neglected in the one-loop equation. Schematically,
\be
\dot\Gamma_\Lambda=\underbrace{\mbox{``leading terms''}}_{\sim\Gamma_\Lambda^2\Lambda^{-1}}+ \underbrace{\mbox{``subleading terms''}}_{\sim\Gamma_\Lambda^2\log\!\Lambda}+\underbrace{\mbox{``neglected terms''}}_{\sim\Gamma^3_\Lambda\log^2\!\Lambda}.
\ee

The RG flow is split into three different energy regimes. 
\begin{enumerate}
\item {\bf High energies}, in which $\Lambda$ is not small. In this regime, the subleading terms of the one-loop equation dominate the neglected two-loop terms, provided $\Gamma_\Lambda\log\!\Lambda$ is small. Thus Eq. \eref{1loop} can in principle be used to to calculate $\Gamma_\Lambda$, starting with a cutoff equal to the bandwidth, where the vertex is given by the bare coupling. In this regime, there is no small parameter to further simplify the functional RG equation \eref{1loop}, so using it is technically very difficult. But in the high energy regime ($g\log\Lambda\ll1$) naive perturbation theory provides a controlled and much more feasible method of calculation than the RG. 
\item {\bf Small energies}, where $\Lambda$ is small such that $(\Lambda\log\Lambda)^{-1}\gg\Gamma_\Lambda\log\!\Lambda\gtrsim1$. Naive perturbation theory breaks down in this regime, but the leading terms of equation \eref{1loop} are superior to the neglected terms even if $\Gamma$ is not necessarily small. This is where the one-loop RG equation is most useful. Note, that the subleading terms of Eq. \eref{1loop} are no longer superior to the neglected terms in this regime. They have to be neglected to be consistent.
\item {\bf The critical regime}, where $\Lambda$ is close to the critical energy scale $\Lambda_c$ at which $\Gamma_\Lambda$ is diverging. If $\Lambda$ is too close to $\Lambda_c$, the neglected terms are no longer negligible and the one-loop approximation is no longer accurate. Note however that the weaker the initial interaction is, the more $\Lambda$ can approach $\Lambda_c$ before the one-loop approximation breaks down. 
\end{enumerate}

It is worthwhile to discuss the behavior of the bubbles for small but finite values of $\v k$. To be specific, I consider the example of very low filling,
 where the single-electron spectrum is approximately parabolic $\xi_{\v p}=\v 
p^2-1$ (all energies are given in units of the Fermi energy and all momenta in 
units of $k_F$). The energy shell $|\xi_{\v p}|=\Lambda$ consists of two circles with radius $\sqrt{1\pm\Lambda}$. For the p-p bubble one finds in the limit $|\v k|,\Lambda\ll 1$
\be 
\dot B^{pp}(\Lambda,\v k)=\frac{-1}{2\pi^2\sqrt{|\v k^2-\Lambda^2|}}\left\{\begin{array}{ll}
2\arctan{\sqrt{\frac{\Lambda-|\v k|}{\Lambda+|\v k|}}} & 
\mbox{if }|\v k|<\Lambda,\\
\log{\frac{|\v k|+\sqrt{\v k^2-\Lambda^2}}{\Lambda}} & \mbox{if }|\v k|>\Lambda.
\end{array}\right.
\ee
If we renormalize a vertex with a small total momentum $\v k=\v k_1+\v k_2$ 
the p-p contribution to its flow is approximately independent of $\v k$ as long as $\Lambda\gg|\v k|$. In this case we replace $\Gamma_\Lambda(k_1,k_2,k_3)$ by 
$\Gamma^{BCS}_\Lambda(k_1,-k_3)$ even if the total momentum is not exactly 
zero. This replacement is no longer justified when $\Lambda$ is of order 
$|\v k|$. The flow then depends strongly on $\v k$ and cannot be controlled. 
Nevertheless, there is no danger because a few renormalization steps later, if 
$\Lambda\ll|\v k|$, the flow is suppressed and the coupling under consideration no longer contributes.

We can do the same type of analysis for the p-h bubble with a small momentum 
transfer, which renormalizes couplings that are close to the forward- or 
exchange scattering $g^f_\Lambda(\v k,\v k'):=Xg^e_\Lambda(\v k,\v k'):=g_\Lambda(
\v k,\v k',\v k',\v k)$. We obtain 
\be 
\dot B^{ph}(\Lambda,\v k)=\left\{\begin{array}{ll}
0 & \mbox{if }|\v k|<\Lambda,\\
\frac1{2\pi^2|\v k|}\log{\frac{|\v k|+\sqrt{\v k^2-\Lambda^2}}{\Lambda}} & 
\mbox{if }|\v k|>\Lambda.
\end{array}\right.
\ee
It gives a big $\v k$-dependent contribution if $\v k$ is of order 
$\Lambda$. But this flow is again suppressed if $\Lambda$ is further reduced. 

The results presented above for the isotropic case should remain valid as long as the Fermi 
surface is both far away from van Hove singularities and not nested. The 
presence of Umklapp scattering does not change the result, since the contribution of Umklapp processes to the flow is not of leading order.

It is not excluded that Umklapp processes influence the flow in an important way before one enters the low-energy regime \cite{HSFR01}. But these are most likely non perturbative effects and I doubt that they can reliably be accounted for within a one-loop approach.

\section{Nesting}\label{nestingsection}
%---------------

I now consider the case of the half-filled $t_x-t_y$ model, presented in Section \ref{Nesting}. 

\subsection{Particle-hole symmetry}\label{phnesting}
%.................................

It is interesting to note that the nesting property \eref{dispnesting} and thus also the relation of the p-p and p-h bubbles Eq. \eref{bubblenesting} are consequences of the particle-hole symmetry of the model. 

For a general single-band model on the square lattice, where the hopping is restricted to nearest neighbors, the one-body Hamiltonian $H_0$ is invariant under the particle-hole transformation 
\be
c^{}_{\v k\sigma}\to c^\dagger_{\v k+\v Q\sigma}.\label{phtrans}
\ee
The interaction given by Eq. \eref{HI} does in general not satisfy this symmetry. However the term 
\be
\tilde H_I=H_I-H'\label{phsymmH}
\ee
where
\be
H'=\frac12\sum_{\sigma,\v k}\left(\frac1{L^2}\sum_{\v p}(2-X)g(\v k,\v p,\v p)\right) n_{\sigma \v k},
\ee
 is particle-hole symmetric, provided 
\be
g(\v k_1,\v k_2,\v k_3)=g(\v k_1+\v Q,\v k_2+\v Q,\v k_3+\v Q). \label{phsym}
\ee

For example, the nearest-neighbor interaction Eq. \eref{gUVJWK} satisfies the condition \eref{phsym}, except for the Coulomb assisted hopping term $K$. If $K=0$, the term $H'$ is proportional to the particle number and can be absorbed in the chemical potential. 

The transformation \eref{phtrans} interchanges occupied and empty single-particle states. The particle hole symmetry therefore relates the system at particle density $n$ to the system at density $2-n$. At half filling ($n=1$), the particle hole symmetry implies the following exact identities
\be
G(k_0,\v k)=-G(-k_0,\v k+\v Q).\label{nestingpropagator}
\ee
and (using time reversal invariance, spin rotation invariance and parity)
\be
\Gamma(k_1,k_2,k_3)=\Gamma(Q-k_1,Q-k_2,Q-k_3).
\ee

As a consequence of Eq. \eref{nestingpropagator}, self energy corrections might change the details of the Fermi surface, but not the nesting property. In fact the dispersion $\tilde\xi_{\v p}=\xi_{\v p}+\Sigma(0,\v p)$ satisfies Eq. \eref{dispnesting} and the relation \eref{bubblenesting} remains true if one replaces the bare propagators $C$ in the definition of the bubbles by exact Green's functions $G$. 

\subsection{Low-energy scattering processes}
%...........................................

\begin{figure}  
\centerline{\includegraphics[width=15cm]{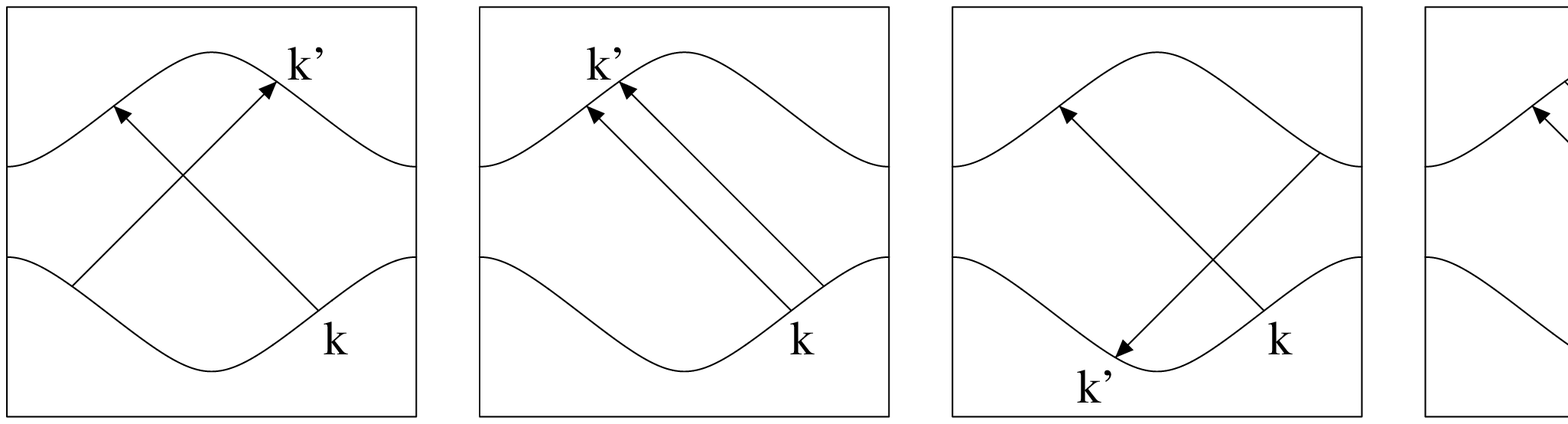}}
\caption{Examples of scattering processes $\Gamma^d(k,k')$ for different choices of $\v k$ and $\v k'$.}\label{Gdirect}
\end{figure}

\begin{figure}  
\centerline{\includegraphics[width=15cm]{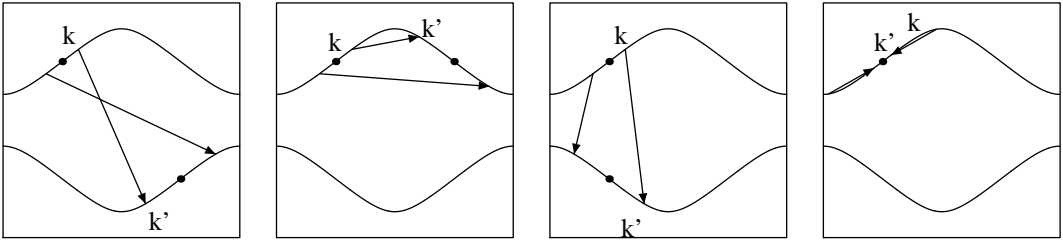}}
\caption{Examples of scattering processes $\Gamma^\eta(k,k')$ for different choices of $\v k$ and $\v k'$. The ``center of mass'' of the two incoming and outgoing particles which are are fixed to $(\pm\pi/2,\pm\pi/2)$, are indicated by points.}\label{Geta}
\end{figure}

While the generic forward, exchange and BCS scattering processes are connected to zero momentum transfer and zero total momentum, the peculiar (particle-hole symmetric) geometry of the present Fermi surface allows for low-energy scattering processes associated with the momentum $\v Q=(\pi,\pi)$. The processes with direct or exchanged momentum transfer $\v Q$ are $\Gamma^d(\v k,\v k')=X\Gamma^x(\v k,\v k')=\Gamma(\v k,\v Q+\v k',\v k')$ and those with a total momentum $\v Q$ are $\Gamma^\eta(\v k,\v k')=\Gamma(\v k,\v Q-\v k,\v Q-\v k')$. These are low-energy scattering processes for $\v k$ and $\v k'$ chosen freely on the Fermi surface, since $\xi_{\v Q\pm\v k}=-\xi_{\v k}$. Some examples are shown in Figs. \ref{Gdirect} and \ref{Geta}. 

 
This list of the possible low energy processes is complete but the
classification into $\Gamma^f$, $\Gamma^e$, $\Gamma^{BCS}$, (defined in Section \ref{genscattproc}) and   $\Gamma^d$, $\Gamma^x$, $\Gamma^\eta$ is not unique. In fact, two two-parameter families intersect in a one-parameter family as $\Gamma^{BCS}(\v k,\v k+\v Q)=\Gamma^{d}(\v k,\v Q-\v k)$,  $\Gamma^{BCS}(\v k,\v Q-\v k)=\Gamma^{x}(\v k,\v Q-\v k)$, etc.

\subsection{RG flow in the p-p and p-h channels}\label{threechannelflow}
%-----------------------------------------------

As a consequence of the particle-hole symmetry, the p-p and p-h bubbles are related by Eq. \eref{bubblenesting}. Thus in addition to $\Gamma^{BCS}$ the vertices $\Gamma^x$ and $\Gamma^d$ have to be renormalized. 

However the renormalization of a general BCS vertex is the same as without nesting. For example we have 
$$
\dot \Gamma^{BCS}(\v k,\v k')\ =\dot R^{pp}_\Lambda+\dot R^{ph1}_\Lambda+\dot R^{ph2}_\Lambda,
$$
where $\dot R^{pp}_\Lambda\sim \dot B^{pp}(\Lambda,0)\sim\Lambda^{-1}$,  $\dot R^{ph1}_\Lambda\sim \dot B^{ph}(\Lambda,\v k+\v k')$ and  $\dot R^{ph2}_\Lambda\sim \dot B^{ph}(\Lambda,\v k-\v k')$. We may neglect $\dot R^{ph1}$ and $\dot R^{ph2}$ for small values of the cutoff, except when $\v k\pm\v k'=\v Q+O(\Lambda)$. Similarly, for  generic $\v k$ and $\v k'$, $\Gamma^x(\v k,\v k')$ is renormalized exclusively by the p-h 1 term and $\Gamma^d(\v k,\v k')$ only by p-h 2. The flow equations \eref{1loop} written in terms of the functions 
$\Gamma^{BCS}$, $\Gamma^s=-\Gamma^x$ and $\Gamma^c=2\Gamma^d-\Gamma^x$ have a simple form. They read 
\ba
\dot \Gamma^{BCS}(\v k,\v k')&=&\dot R^{pp}(\v k,-\v k,-\v k')\\
\dot \Gamma^{s/c}(\v k,\v k')&=&\dot R^{s/c}(\v k,\v k'+\v Q,\v k').
\ea
The notation $\Gamma^{s/c}$ is the same as in Section \ref{corr}, except that the transfered momentum is understood to be $\v Q$. The vertices $\Gamma^{s/c}$ are thus relevant for spin- and charge susceptibilities of momentum $\v Q$.

Because of the similarity of Eq. \eref{rg2} with Eq. \eref{rg1}, one obtains three closed RG equations  
\ba
\dot\Gamma^{BCS}(\v k,\v k')&=&\frac1{\beta L^2}\sum_p\Gamma^{BCS}(\v k,\v p)\dot D^{pp}_{k=0}(p)\Gamma^{BCS}(\v p,\v k')\label{bcseq1}\\
\dot\Gamma^{s/c}(\v k,\v k')&=&-\frac1{\beta L^2}\sum_p\Gamma^{s/c}(\v k,\v p)\dot D^{ph}_{q=(0,\v Q)}(p)\Gamma^{s/c}(\v p,\v k').\label{sceq1}
\ea
Note that we have systematically replaced $\Gamma(k,k')$ by $\Gamma(\v k,\v k')$, because the dependence on the frequencies is irrelevant. Due to particle-hole symmetry, Eqs. \eref{bcseq1} and \eref{sceq1} have the identical form
\be
\dot\Gamma^{\diamond}(\v k,\v k')=\frac1{\beta L^2}\sum_p\Gamma^{\diamond}(\v k,\v p)\dot D^{pp}_{k=0}(p)\Gamma^{\diamond}(\v p,\v k'),
\ee
where  $\diamond$ stands for $BCS$, $s$ or $c$. After integrating over the frequency $p_0$, one finally obtains
\be
-2\Lambda\,\dot \Gamma^{\diamond}(\v k,\v k')=\frac1{L^2}\sum_{\v p}
\delta(|\xi_{\v p}|-\Lambda)\,\Gamma^{\diamond}(\v k,\v p) \Gamma^{\diamond}
(\v p,\v k').\label{decrg}
\ee

In spite of the apparent decoupling of $\Gamma^{BCS}$, $\Gamma^{s}$ and  $\Gamma^{c}$, there is in principle some coupling by hybrid vertices such as $\Gamma(\v p,-\v p,-\v p+\v Q)$. However the weight of such terms in the sum on the right hand-side of Eq. \eref{decrg} is negligible. For example, the ``mixing'' vertices $\Gamma^{BCS}(\v k,\v p)$, where $\v p=\v Q\pm\v k+O(\Lambda)$, give merely a $O(\Lambda)$ contribution to the right hand side of Eq. \eref{decrg}  (see also \cite{Shankar94}). The situation is more difficult if the Fermi surface contains a van Hove singularity, i.e. in the case $t_x=t_y$ considered in Chapter \ref{chapter3}. The non-uniform density of states increases the weight of certain points in the $\v p$-integration of Eq. \eref{decrg} and thus weakens the phase space argument used here. 

We conclude that if $t_x\neq t_y$, the charge, spin and superconducting instabilities do not influence each other in leading order. Moreover the charge and spin instabilities can be studied in exactly the same manner as it was done for superconductivity in Section \ref{superconductivity}.  

For the repulsive Hubbard interaction, the dominant instability will be a spin-density wave, since $\Gamma^s_{\Lambda_0}=-\Gamma^x_{\Lambda_0}\approx U$ is positive, whereas $\Gamma^c_{\Lambda_0}$ and $\Gamma^{BCS}_{\Lambda_0}$ are negative. One expects thus a Slater insulator with a spin-density wave at weak coupling and, as $U$ is increased,  a crossover to the antiferromagnetic Mott insulator.

\section{Van Hove singularities}\label{vHsection}
%------------------------------

We now consider the case where the Fermi surface passes through a van Hove singularity, i.e. a saddle point of the dispersion relation.
The density of states is  logarithmically diverging at the 
saddle points and leads to a more singular behavior of the p-p bubble at zero momentum. Differentiating Eq. \eref{zeropBubble} with respect to $\Lambda$, one gets
\be
\dot B^{pp}(\Lambda,\v 0)=-\frac{\nu(\Lambda)+\nu(-\Lambda)}{2\Lambda}\label{vHoveBubble}
\ee
Assuming a logarithmic behavior of the density of states, it follows that $\dot B^{pp}(\Lambda,\v 0)\sim-\Lambda^{-1}\log\Lambda$ and thus $B^{pp}(\Lambda,\v 0)\sim\log^2\Lambda$. 

\begin{figure}
        \centerline{\includegraphics[width=4cm]{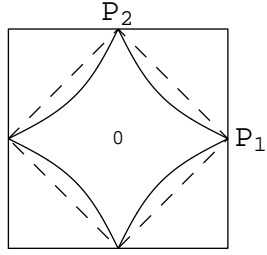}}
\caption{The Fermi surface at van Hove filling: $t'=0.3$ (solid line) and $t'=0$ (dashed line).}
\label{ttprimeFS}
\end{figure}

 To be specific, consider the dispersion relation of a generalized tight-binding model: $\xi_{\v k}=-2(\cos k_x+\cos k_y)+4t'(\cos k_x\cos k_y+1)$. The unit of energy is given  by the hopping amplitude between nearest neighbors. A finite electron 
hopping $0<t'<1/2$ between next nearest neighbors has been included 
and the chemical potential is fine-tuned such that the Fermi surface contains 
the saddle points  at $\v P_1=(\pm\pi,0)$ and $\v P_2=(0,\pm\pi)$ (see Fig. \ref{ttprimeFS}).

If $t'\neq0$, the leading ($\log^2\Lambda$) terms arise only in the p-p channel at zero momentum. In addition, there are divergences in $B^{pp}(\v Q)$, $B^{ph}(\v Q)$ and in $B^{pp}(\v q)$ for small $\v q$, but these divergences are only $\sim\log\Lambda$ and are thus not of the leading logarithmic order. 
If one restricts the RG flow to the leading terms, only the BCS 
coupling function $\Gamma^{BCS}(\v k,\v k')$ is renormalized and the only possible instabilities are of the superconducting type.
The flow of  $\Gamma^{BCS}(\v k,\v k')$ is similar to the one explained in detail in the next section, but without
 competing charge and spin instabilities. 

For repulsive interactions, where the BCS flow is towards weak coupling, the 
approximation is clearly not sufficient and non-leading terms might play an important role. Many efforts to include non-leading terms in the RG flow  have been made \cite{Lederer87,Dzyaloshinskii96,Furukawa98,IKK01,IK01,HSPRL01}, but a consistent treatment has not yet been attained. In the following, I mention some of the problems which would have to be solved to obtain a consistent treatment of the first sub-leading terms (i.e. terms of the order $\Lambda^{-1}$).

Let us first study the behavior of the p-p bubble at small but finite $|\v k|$.  
It is instructive to focus on the contributions of a small patch surrounding a saddle point, say the region  
$$P_1=\left\{\v p\ ;\sqrt{1-2t'}\,|p_x-\pi|+\sqrt{1+2t'}\,|p_y|\leq 2\rho\right\}.$$ 
The parameter $\rho$ is small enough so that 
$\xi_{\v p}$ can be replaced by its limiting quadratic form close to the saddle point $\v P_1=(\pi,0)$. I have computed the bubbles $\dot B^{pp}_{P_1}(\Lambda,\v k)$ and $\dot B^{ph}_{P_1}(\Lambda,\v k)$  defined by Eq. \eref{explicit}, with the summation restricted to $\v p\in P_1$. The values of  $\dot B^{pp}_{P_1}(\Lambda,\v k)$ and $\dot B^{ph}_{P_1}(\Lambda,\v k)$ depend sensitively on $\xi_{\v P_1+\v k}$. Both bubbles are  negligible if $|\xi_{\v P_1+\v k}|\gg \Lambda$, but for $|\xi_{\v P_1+\v k}|\ll \Lambda$ (and $\Lambda,\v k^2\ll\rho^2$) we get
\be
\dot B^{pp}_{P_1}(\Lambda,\v k)=\frac{-1}{(2\pi)^2\sqrt{1-4t'^2}\,\Lambda}\ \log\frac{4\Lambda\rho^2}{(\Lambda+\rho\, k_+)(\Lambda+\rho\, k_-)}.\label{vHpp}
\ee
and
\be
\dot B^{ph}_{P_1}(\Lambda,\v k)=\frac{1}{(2\pi)^2\sqrt{1-4t'^2}\,\Lambda}\left[\Theta(\rho\,k_+-\Lambda)\frac{\rho\,k_+-\Lambda}{\rho\,k_+}+\Theta(\rho\,k_--\Lambda)\frac{\rho\,k_--\Lambda}{\rho\,k_-}\right],\label{phvH}
\ee
where $k_\pm:=|\sqrt{1-2t'}\,k_x\pm\sqrt{1+2t'}\,k_y|$. Note that $k_+\,k_-=|\xi_{\v P_1+\v k}|$.

The behavior is most interesting in the regime $k_+k_-\ll\Lambda\ll\rho\,k_\pm$.  The p-h bubble, zero for 
sufficiently small $|\v k|$, gives a contribution $\sim\Lambda^{-1}$ there. Thus 
for a given small momentum transfer $\v k$, the p-h contributions are considerable over many RG iterations, in contrast to the case without van Hove singularities.
Even more strikingly, the first subleading (i.e. $\sim\Lambda^{-1}$) part of the p-p diagram is very sensitive with respect to both the size and the direction of the total momentum $\v k$ in this regime. Thus for a given small momentum $\v k$, the p-p contribution has a complicated behavior on $\v k$ over many RG iterations, in contrast to the case of Sections \ref{circ} without van Hove singularities.

We conclude that the RG equation to order $\Lambda^{-1}$ for the vertex $\Gamma_\Lambda(\v k_1,\v k_2,\v k_3)$ depends in a delicate and non-trivial manner on the momenta.

In contrast to the case of the circular Fermi surface, self-energy corrections are not negligible in the present situation. In fact, they are expected to change the shape of the Fermi surface.

For the corrections to the single particle dispersion and the quasi particle weight, one finds in second order perturbation theory (see for example Eq. 7 of \cite{Dzyaloshinskii96})
\ba
\partial_{p_0}\Sigma_\Lambda(p)&\sim&g^2\log^2\Lambda\label{SEcorr}\\
\nabla_{\v p}\Sigma_\Lambda(p)&\sim&g^2\log^2\Lambda\,\cdot\,\nabla_{\v p}\xi_{\v p}.\nonumber
\ea
These corrections are negligible within the leading-order approximation, where we assume $g\log^2\Lambda\sim1$. An attractive coupling diverges at this scale. However, in the case of repulsive interactions the RG flow can be followed to smaller energies such that $g\log\Lambda\sim1$. At these energy scales, self energy corrections have to be taken seriously. In other words, a consistent calculation to the order $\Lambda^{-1}$ requires to follow simultaneously the RG flow of the self energy and of the vertex.

Due to the additional logarithm in the p-p bubble the theory is not
renormalizable in the usual sense of field theory, i. e. it is
not possible to send the bare momentum cutoff to infinity while
keeping some physical correlation functions finite (even after the
introduction of counter-terms in the microscopic Hamiltonian and of
wave-function renormalization). Gonzalez, Guinea and 
Vozmediano \cite{Gonzalez96,Guinea97,Gonzalez99,Gonzalez00} proposed a field theoretical RG scheme where the coupling
 constants for forward and exchange scattering are renormalized only 
by the p-h diagrams. In this approach the p-p channel is treated
separately in connection with a renormalized chemical potential. It is
however not clear from our approach that the RG equations 
will not mix p-p and p-h diagrams, if all contributions 
$\sim\Lambda^{-1}$ are taken into account.

%auto-ignore

\chapter{Instabilities of the half-filled lattice with nearest-neighbor hopping}\markboth{CHAPTER 4. INSTABILITIES OF THE HALF-FILLED LATTICE...}{}\label{chapter3}
%=============================================================================

\section{Motivation}\label{motivation}
%------------------

This chapter is dedicated to the nearest-neighbor tight-binding model (i.e. $t_x=t_y=1$ and $t'=0$) at half filling where the Fermi surface is a square (see Fig. \ref{ttprimeFS}). This special case has common features with the two cases considered before. There is exact nesting associated to particle-hole symmetry and the Fermi level is at a van Hove singularity.  This case is more difficult, than the anisotropic model of Section \ref{nestingsection}, but it leads to nontrivial results already to leading (one-loop) order, unlike the case of Section \ref{vHsection} without nesting. 

\begin{figure} 
        \centerline{\includegraphics[width=6cm]{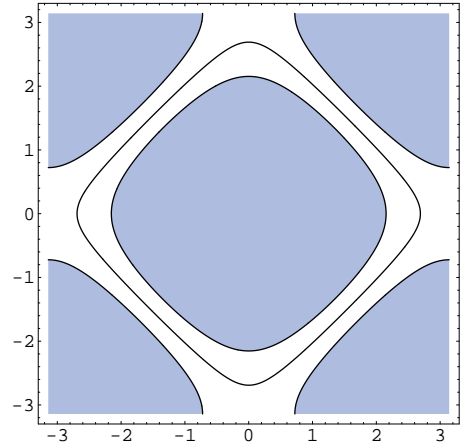}}
\caption{The Fermi surface of the nearest-neighbor tight-binding model slightly below half-filling. The grey region corresponds to high energies, $|\xi_{\v k}|>\Lambda$. The case shown here is in the parquet regime $\Lambda>\mu$, where RG equations are not yet susceptible for the doping away from half-filling.}\label{parquetBZ} 
\end{figure}

As a motivation, I would like to relate the subsequent analysis to the partially numerical investigation of the  Hubbard model  by Zanchi and Schulz \cite{Zanchi00}.  According to them,  close to half filling the flow RG separates into two regimes. For not too low energies $\Lambda>\mu$, where $\mu$ is the chemical potential, the RG flow is not yet susceptible  of the deviation from half-filling (see Fig. \ref{parquetBZ}). Consequently, the RG flow is complicated by the simultaneous appearance of divergences in the p-p and p-h channels and by the presence of van Hove singularities. Zanchi and Schulz call this the parquet regime, since the interplay of fluctuations of different kinds is the essence of the parquet method. 

\begin{figure} 
        \centerline{\includegraphics[width=6cm]{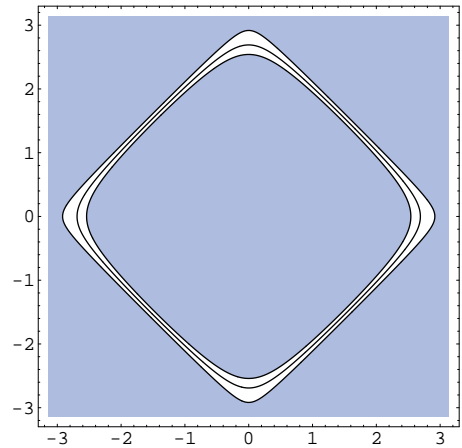}}
\caption{Same as Fig. \ref{parquetBZ}, but for a lower cutoff $\Lambda<\mu$. This is the BCS regime, where the saddle points have already been integrated out and the curvature of the Fermi surface becomes relevant.}\label{BCSBZ} 
\end{figure}

For lower energies $\Lambda<\mu$, the distance of the Fermi level from the van Hove singularity as well as the absence of nesting becomes relevant (Fig. \ref{BCSBZ}). There, the RG flow is entirely in the p-p channel as in Chapter \ref{chapter1}. This is called the BCS regime.

\begin{figure} 
        \centerline{\includegraphics[width=6cm]{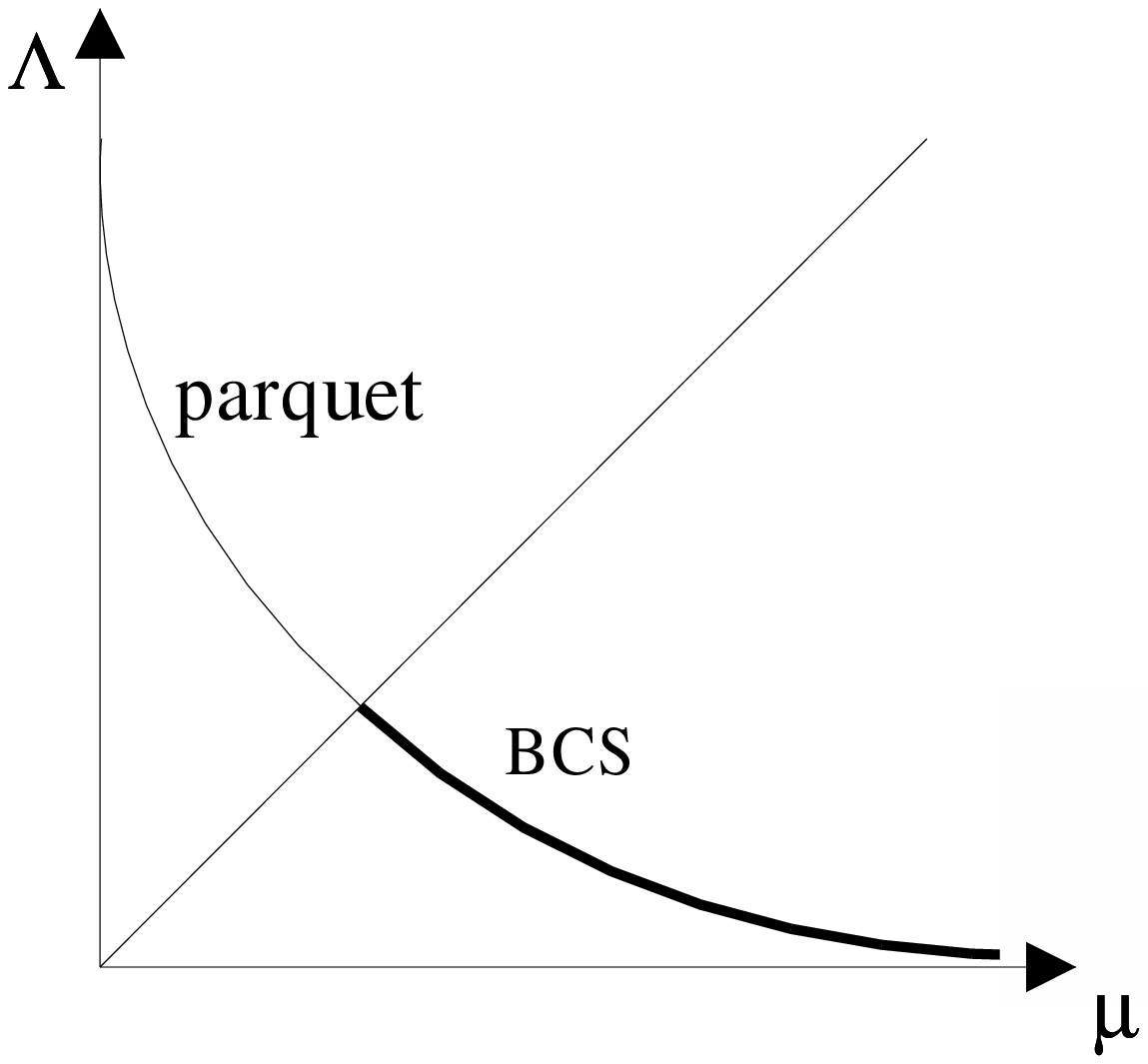}}
\caption{The schematic phase diagram according to Zanchi and Schulz \cite{Zanchi00}.}\label{Drazen} 
\end{figure}

Zanchi and Schulz have found that the critical energy scale $\Lambda_c$, where the RG flow diverges, depends strongly on the electron density (i.e. on the chemical potential $\mu$). It is largest at half filling and becomes exponentially small upon doping. One can now draw the schematic phase diagram shown in Fig. \ref{Drazen}. The instabilities at $\Lambda_c$ are qualitatively different in the two regimes. In the BCS regime, the instability is entirely in the p-p channel and the diverging quantity is the vertex function $\Gamma^{BCS}$, i.e. the scattering of particle pairs with a vanishing total momentum. As a consequence, the superconducting susceptibilities diverge, but not the density-wave or flux-phase susceptibilities which depend on nesting. The interpretation of this instability as the onset of superconductivity is therefore quite obvious, although there is no proof.

The situation is much less clear if the instability takes place in the parquet regime. There the numerical investigations find many diverging vertices of different kinds and simultaneously diverging susceptibilities of the SDW- and of the superconducting type, although the SDW susceptibilities are clearly dominating at half filling.
 How should one interpret this kind of instability? 

 I consider the half-filled case as an important key to a better understanding of the instability in the parquet regime because it allows us to take the limit of weak interactions and thus very low energies $\Lambda$ without leaving ultimately the parquet regime.

The remaining part of the chapter is organized as follows. In Section \ref{squareFS}, the renormalized couplings are classified according to both the location of momenta with respect to the van Hove points and the channels characterizing the different instabilities. It is argued that to leading order there is no mixing between superconducting, charge and spin instabilities except from momenta very close to the van Hove points. A simple way of disentangling this special behavior at the van Hove points and the generic behavior elsewhere is presented in Section \ref{2patch} and contrasted to an earlier approach where the momentum dependence was altogether neglected. The asymptotic behavior of the RG flow  allows to draw a phase diagram including superconductivity, density waves and flux phases, depending on the values of the bare couplings. This phase diagram agrees with symmetry considerations linking the various order  parameters, as shown in Section \ref{symmsection}. The content of of Sections \ref{squareFS}-\ref{symmsection} was published in \cite{Binz02}. Finally in Section \ref{uniform}, I investigate the behavior of some additional physical quantities in the vicinity of each instability. These quantities are the uniform spin susceptibility, the charge compressibility, but also more exotic susceptibilities which measure the tendency towards $\eta$-pairing (pairing of Cooper pairs with momentum $(\pi,\pi)$) or the tendency towards a deformation of the Fermi surface (Pomeranchuk instability).

\section{RG flow of general vertices and susceptibilities} \label{squareFS}
%--------------------------------------------------------

\subsection{Low energy scattering processes}
%...........................................

\begin{figure} 
        \centerline{\includegraphics[width=10cm]{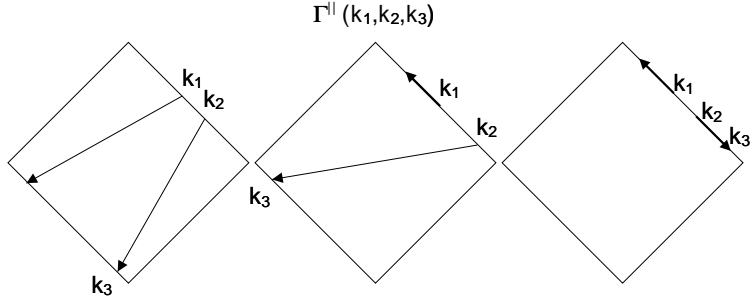}}
\caption{Examples of processes described by $\Gamma^\parallel$. The momenta $\v k_1$, $\v k_2$ and $\v k_3$ can be chosen freely on two parallel sides of the square Fermi surface. }
\label{channels}
\end{figure}

I first identify the possible scattering processes which connect four momenta on the Fermi surface and satisfy momentum conservation modulo a reciprocal lattice vector. In addition to the six classes of processes discussed in Section \ref{nestingsection}, there are now further processes, which are related to the fact that the Fermi surface consists of straight lines.  If three points $\v k_1,\v k_2,\v k_3$ are 
chosen freely on two parallel sides of the square, the resulting $\v k_4=\v
 k_1+\v k_2-\v k_3$ lies automatically on the Fermi surface, giving rise to
 a three-parameter family $\Gamma^\parallel(\v k_1,\v k_2,\v k_3)$. 
Some examples are shown in Fig. \ref{channels}.  

As in Section \ref{nestingsection}, the list of the possible low energy processes is complete but the
classification into $\Gamma^f$, $\Gamma^e$, $\Gamma^{BCS}$, $\Gamma^d$, $\Gamma^x$, $\Gamma^\eta$ and $\Gamma^\parallel$ is not unique. For example if $\v k$ and $\v k'$
belong to the same pair of parallel sides of the square, the
two-parameter families $\Gamma^f(\v k,\v k')$, $\Gamma^e(\v k,\v k')$, $\Gamma^{BCS}(\v k,\v k')$, $\Gamma^d(\v k,\v k')$, $\Gamma^x(\v k,\v k')$ and
$\Gamma^\eta(\v k,\v k')$ belong to the larger three parameter family $\Gamma^\parallel$. Furthermore, two two-parameter families intersect in a one-parameter family as $\Gamma^{BCS}(\v k,\v k+\v Q)=\Gamma^{d}(\v k,\v Q-\v k)$,  $\Gamma^{BCS}(\v k,\v Q-\v k)=\Gamma^{x}(\v k,\v Q-\v k)$, etc., where $\v Q=(\pi,\pi)$. Finally three two-parameter families can intersect in scatterings between the two saddle points as $\Gamma^{BCS}(\v P_1,\v P_2)=\Gamma^d(\v P_1,\v P_2)=\Gamma^x(\v P_1,\v P_2)$. Where $\v P_1=(\pi,0)$ and $\v P_2=(0,\pi)$ are the two saddle points.

\subsection{Self-energy effects}
%..............................

As it has been explained in Section \ref{phnesting}, the nearest-neighbor tight-binding model is particle-hole symmetric. We further assume that the interaction also respects this symmetry. The self energy then satisfies the exact relation $\Sigma_\Lambda(p_0,\v p)=-\Sigma_\Lambda(-p_0,\v p+\v Q)$. Imposing in addition the point symmetries of the square lattice, one can conclude that the self-energy vanishes on the Fermi surface ($p_0=\xi_{\v p}=0$). This means that the Fermi surface is not modified by self-energy effects.

For the corrections to the single particle dispersion and the quasi particle weight, one can derive relations similar to Eq. \eref{SEcorr}, namely (in second order perturbation theory)
\ba
\partial_{p_0}\Sigma_\Lambda(p)&<&\mbox{const.}\,\cdot\,g^2\log^3\Lambda\label{omegaSigma}\\
\nabla_{\v p}\Sigma_\Lambda(p)&<&\mbox{const.}\,\cdot\,g^2\log^3\Lambda\,\cdot\,\nabla_{\v p}\xi_{\v p}.\label{kSigma}
\ea
 The calculations are given in Appendix \ref{selfenergy}. In the absence of perfect nesting (van Hove filling, but $t'\neq0$), it is known that the bounds Eq. \eref{kSigma} and  Eq. \eref{omegaSigma} overestimate the self-energy terms by one power of the logarithm (see Eq. \eref{SEcorr}). I do not know whether this is also true if $t'=0$. In any case the bound given above is sufficient to exclude a leading order correction $\sim g^2\log^4\Lambda$. I conclude that self-energy corrections are negligible within our approximation, where $g\log^2\Lambda\sim1$. 

In spite of the fact that self-energy effects are negligible within the leading order (or one-loop) approximation, it cannot be excluded that they influence the physics considerably. In fact, due to properties of the self-energy, the half-filled 2D Hubbard model is not a Fermi liquid above the critical temperature for the SDW \cite{Rivasseau01}, in contrast to the 2D jellium model which is a Fermi liquid above the superconducting critical temperature \cite{Disertori00}. But quite generally, a consistent treatment of self-energy effects requires to go beyond the one-loop approximation, i.e. it would be inconsistent to include self-energy effects at our level of approximation.

\subsection{RG flow of general vertices}\label{fscouplings}
%.......................................

We now investigate the one-loop corrections to the coupling constants by an analysis of the bubbles \eref{explicit}. As a consequence of particle-hole symmetry, the p-p and p-h bubbles are related by Eq. \eref{bubblenesting}.

\begin{figure} 
        \centerline{\includegraphics[width=10cm]{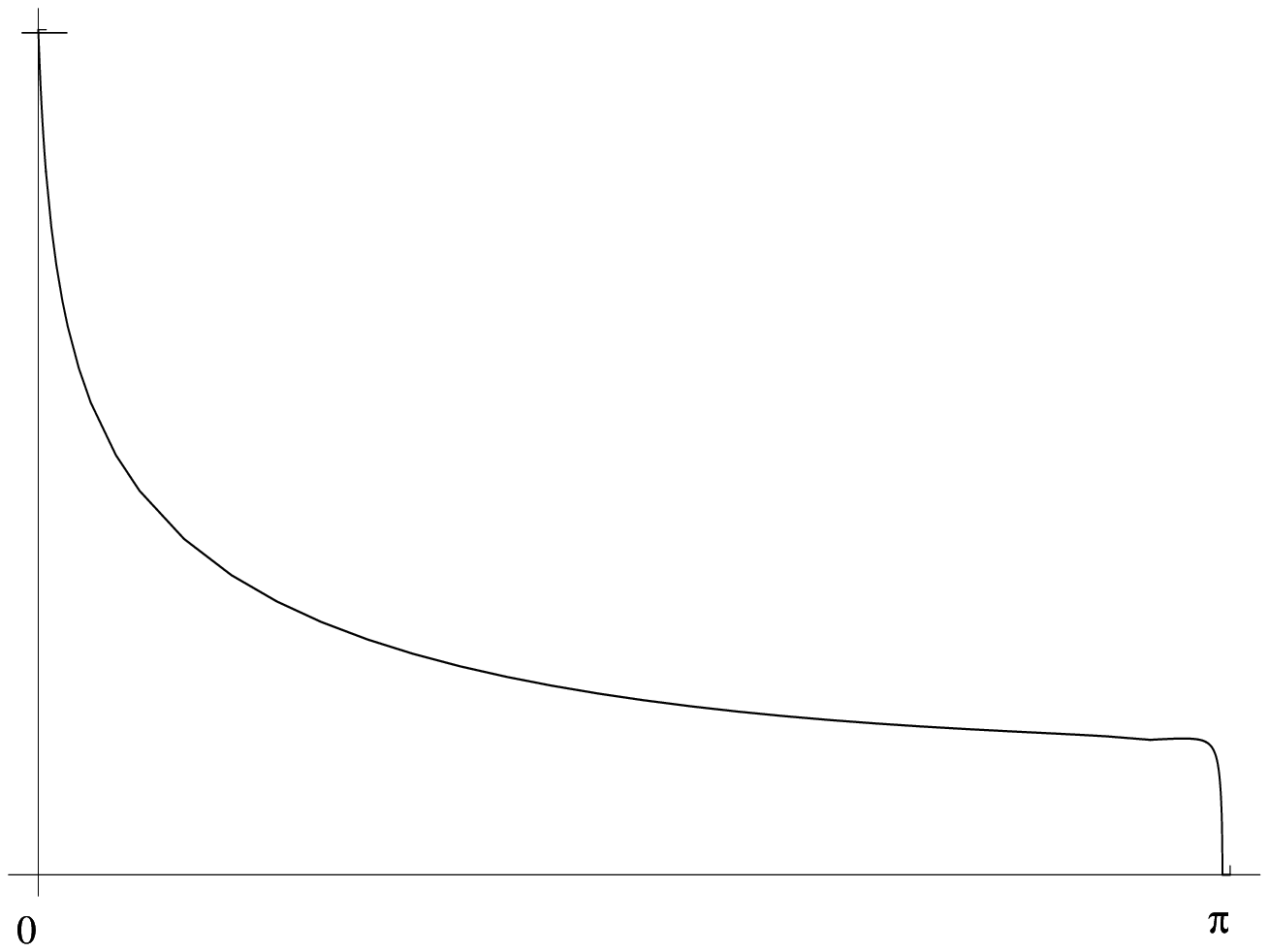}}\vspace{-11pt}
\centerline{\hspace{-3.6cm}\raisebox{5.0cm}[0pt][0pt]{$-\dot B^{pp}(\Lambda,\kappa,\kappa)$}\hspace{6.6cm}\raisebox{0.9cm}[0pt][0pt]{$\kappa$}}
\caption{Plot of the p-p bubble for $\v k=(\kappa,\kappa)$, parallel to the square Fermi surface and $\Lambda=0.01$. The small mark on the vertical axes indicates the maximal value $-\dot B^{pp}(\Lambda,0,0)$.}
\label{Bpp}
\end{figure}

Both the p-p and p-h bubbles are of order  $\Lambda^{-1}$  whenever $\v k=n(\pi,\pi)+\kappa(1,\pm1)$ ($n\in\Z,\kappa\in\R$). The reason is that the Fermi surface
 consists of straight lines. If $\v k$ is parallel to  such a line, it satisfies the conditions of Section \ref{Nesting} both for p-p and p-h nesting.  It follows that the three parameter vertex function $\Gamma^\parallel $ is renormalized by 
contributions of order $\Lambda^{-1}$ from every term $\dot R^{pp}$,  $\dot R^{ph1}$, and $\dot R^{ph2}$  in Eq. \eref{1loop}.

 In Fig. \ref{Bpp} we show a plot of $\dot B^{pp}(\Lambda,\v k)$ for $\v k=(\kappa,\kappa)$. The analytic expression is given in Appendix \ref{Bppapp}. Here I only mention the asymptotic result for small $\kappa$
\be
\dot B^{pp}(\Lambda,\kappa,\kappa)\ \MapRight{\kappa\ll1}\ \frac{-1}{2\pi^{2}\Lambda}\,\log\frac{16}{\Lambda+4\kappa}.\label{bubbleofk}
\ee

By restricting ourselves to the logarithmically dominant
 terms of order $\Lambda^{-1}\log\Lambda$ we can go a long way
 using an analytical approach, as will be shown now. We consider the two-parameter vertex functions $\Gamma^{BCS}$, $\Gamma^x$ and $\Gamma^d$ and use the same argumentation  as in Section \ref{threechannelflow}. 
For  generic $\v k$ and $\v k'$, the vertex $\Gamma^{BCS}(\v k,\v k')$ is renormalized only by the p-p term,    $\Gamma^x(\v k,\v k')$
 only by the p-h 1 term and $\Gamma^d(\v k,\v k')$ only 
by p-h 2. 

As a consequence, one obtains the RG equations for the functions 
$\Gamma^{BCS}$, $\Gamma^s=-\Gamma^x$ and $\Gamma^c=2\Gamma^d-\Gamma^x$ as in Section \ref{threechannelflow}, 
\be
2\Lambda\frac\ud{\ud\Lambda}\Gamma^{\diamond}(\v k,\v k')=-\frac1{L^2}\sum_{\v p}
\delta(|\xi_{\v p}|-\Lambda)\,\Gamma^{\diamond}(\v k,\v p) \Gamma^{\diamond}
(\v p,\v k')\label{rg},
\ee
where $\diamond$ stands for $BCS$, $s$ or $c$. 

Eq. \eref{rg} is valid for momenta $\v k$ and $\v k'$, such that $\v k\pm\v k'$ is not too close to $\v Q$. It was argued in Section \ref{threechannelflow}, that the contribution of these special momentum configurations to the right hand-side of the RG equation is negligible. We will see shortly that this is no longer true here because of the non-uniform density of states in momentum space.

\begin{figure} 
        \centerline{\includegraphics[width=5cm]{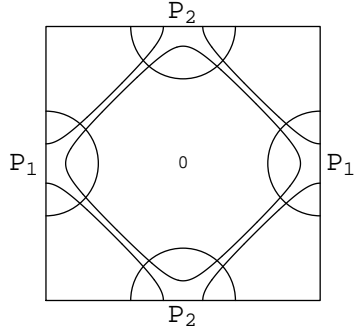}}
\caption{The Brillouin zone with the energy shell $|\xi_{\v p}|=\Lambda$. It is
 separated into two patches $P_1$ and $P_2$ and the remaining part.}\label{bz}
\end{figure}

A large contribution to the integration over $\v p$ in Eq. \eref{rg} comes from a small neighborhood of the saddle-points, due to the diverging density of states. To identify the logarithmically diverging contribution to Eq. \eref{rg} we consider the patches $P_1=\left\{\v p\,;\,|p_x-\pi|+|p_y|<2\rho\right\}$ and $P_2=\left\{\v p\,;\,|p_x|+|p_y-\pi|<2\rho\right\}$ of a size
 $\rho\ll\pi/2$ around the two van Hove points and separate the integral 
$\sum_{\v p}$ into  $\sum_{\v p\in P_1}+\sum_{\v p\in P_2}+\sum_{\v p\in 
B.Z.-P_1-P_2}$, where $B.Z.$ is the whole Brillouin zone (see Fig. \ref{bz}).
 We compute the weight of the patch and of the remaining part of the Brillouin 
zone, assuming that  $\Gamma^\diamond(\v k,\v p)$ is of the same order of magnitude 
for every value of $\v p$. Comparing the values
\be 
\frac1{L^2}\sum_{\v p\in P_1}\delta(|\xi_{\v p}|-\Lambda)=\frac1{2\pi^2} \log(\frac{4\rho^2}{\Lambda}),\label{p1}
\ee
\be 
\frac1{L^2}\sum_{\v p\in B.Z.-P_1-P_2}\!\delta(|\xi_{\v p}|-\Lambda)=\frac1{\pi^2}
\log(\frac4{\rho^2}),\label{p2}
\ee
 we conclude that the patch contribution dominates the remaining part 
if $\Lambda\ll\rho^4$. Under the hypothesis that the functions $\Gamma^\diamond(\v k,\v k')$ are slowly varying, it is then consistent to replace Eq. \eref{rg} by
\be
\frac\ud{{\ud} l}\Gamma^\diamond(\v k,\v k')=\sum_{i=1,2}\,\Gamma^\diamond_i(\v k) 
\Gamma^\diamond_i(\v k'),\label{prg}
\ee
where we have set $\Gamma^\diamond(\v k,\v p)\approx \Gamma^\diamond_i(\v k)$ for 
$\v p\in P_i$ and 
\be
l=B^{pp}_P(\Lambda,\v 0)=\frac1{8\pi^2}\log^2(\frac{4\rho^2}{\Lambda})\label{ldef}
\ee
as the new RG parameter. $B^{pp}_P(\Lambda,\v 0)$ is the value of the p-p bubble, where the loop momentum is restricted to the saddle point patch.
 
Correspondingly, setting $\Gamma^\diamond_i(\v k')\approx \Gamma^\diamond_{ij}$
 for  $\v k'\in P_j$ in \eref{prg} leads to 
\be
\frac\ud{{\ud}l}\Gamma^\diamond_j(\v k)=\sum_{i=1,2} \,\Gamma^\diamond_i(\v k) \,
\Gamma^\diamond_{ij}.\label{prg2}
\ee

Eqs. \eref{prg} and \eref{prg2} are only correct under the hypothesis that $\Gamma^\diamond(\v k,\v k')$ and $\Gamma^\diamond_i(\v k)$ are slowly varying functions.  This hypothesis is certainly justified in the beginning of the RG flow, when $\Gamma\approx-g$. It will be shown now that, starting from this hypothesis, we obtain an RG flow where the abovementioned functions remain well behaved. The hypothesis is thus consistent with the final result.

We are now left with the problem of renormalizing the vertices
$\Gamma^\diamond_{ij}$, which describe scattering of particles near the
saddle points. Eq. \eref{rg} cannot be used here because it is not
valid for $\v k\pm\v k'\approx\v Q$, i.e. if
$\v k\in P_1$ and $\v k'\in P_2$. The RG flow of the vertices
$\Gamma^\diamond_{ij}$ will be discussed in detail in Section \ref{2patch}.

Once the evolution of the $\Gamma^\diamond_{ij}$ as a function of $l$ is 
known, one can integrate Eqs. \eref{prg2} and \eref{prg}. 
Introducing $\Gamma^\diamond_\pm(\v k)=\Gamma^\diamond_1(\v k)\pm \Gamma^\diamond_2(\v k)$
 we get
\be
\Gamma^\diamond_\pm(\v k) =\left.\Gamma^\diamond_\pm(\v k)\right|_{l=l_0}\,\cdot\,
\exp\left[\int_{l_0}^l{\ud} l\,\left(\Gamma^\diamond_{11}\pm 
\Gamma^\diamond_{12}\right)\right]\label{rgpp2}
\ee
and
\be
\Gamma^\diamond(\v k,\v k')=\left.\Gamma^\diamond(\v k,\v k')\right|_{l=l_0}+
\frac12\int_{l_0}^l{\ud} l\,\left[\Gamma^\diamond_+(\v k)\Gamma^\diamond_+(\v k')
+\Gamma^\diamond_-(\v k)\Gamma^\diamond_-(\v k')\right].\label{rgkk}
\ee

We will see that the RG equations of $\Gamma^\diamond_{ij}$ as a function of $l$ yield diverging solutions at a finite value $l=l_c$. Near this critical value they behave asymptotically like 
\be 
\Gamma^\diamond_{ij}(l)\approx\frac{\tilde \Gamma^\diamond_{ij}}{l_c-l}+O(l_c-l)^\alpha,\label{divergentg}
\ee
where the constant $\tilde \Gamma^\diamond_{ij}$ can be determined from the RG equation and $\alpha>-1$. Eqs. \eref{rgpp2} and \eref{rgkk} then give
\be
\Gamma^\diamond_\pm(\v k)\approx A_\pm\cdot\left.\Gamma^\diamond_\pm(\v k)\right|_{l=l_0}\,\cdot\,
\left[(l_c-l)^{-\tilde \Gamma^\diamond_{\pm}}+O(l_c-l)^{-\tilde \Gamma^\diamond_{\pm}+\alpha+1}\right]
\ee
and
\be
\Gamma^\diamond(\v k,\v k')\approx\left.\Gamma^\diamond(\v k,\v k')\right|_{l=l_0}+\sum_{\nu=\pm}B_\nu\left.\Gamma^\diamond_\nu
(\v k)\Gamma^\diamond_\nu(\v k')\right|_{l=l_0}
\left[(l_c-l)^{1-2\tilde \Gamma^\diamond_{\nu}}+O(l_c-l)^{\Min{\{0,2-2\tilde \Gamma^\diamond_{\nu}+\alpha\}}}\right],\label{gkksol}
\ee
where $\tilde \Gamma^\diamond_{\pm}:=\tilde \Gamma^\diamond_{11}\pm\tilde \Gamma^\diamond_{12}$ and $A_\pm$, $B_\pm$ are positive constants.
The vertex function 
$\Gamma^\diamond(\v k,\v k')$ is diverging if $\tilde \Gamma^\diamond_{+}$ or $\tilde \Gamma^\diamond_{-}\geq1/2$. 

The functions $\Gamma^\diamond_+(\v k)$ have $s$-wave symmetry, i.e. they
respect all the point symmetries of the square lattice. On the other hand we see that
$\Gamma^\diamond_-(\v k)$ is of the $d_{x^2-y^2}$-wave type 
[$\Gamma^\diamond_-(k_x,k_y)=\Gamma^\diamond_-(k_x,-k_y)=-\Gamma^\diamond_-(k_y,k_x)$].
 Thus the diverging part of the vertex
function has $s$ or $d_{x^2-y^2}$-wave symmetry.

In the preceding calculation we have distinguished strictly between points far from the saddle points and those close to the saddle points. The scale which distinguishes between ``far'' and ``close'' is the patch size $\rho$, which was introduced by hand. 

The behavior of the overall vertex function $\Gamma^\diamond(\v k,\v k')$
near the critical point depends on the biggest positive value of the constants $\tilde \Gamma^\diamond_\pm$. It will be shown in Section
\ref{our}  that 
$\Max\{\tilde \Gamma^\diamond_\pm\}=1$. In this situation the function $\Gamma^\diamond(\v k,\v
k')$ diverges everywhere with the same power $(l-l_c)^{-1}$ and  
$\Gamma^\diamond(\v k,\v k')$ remains a smooth function upon
renormalization even if $\v k$ or $\v k'$ (or both) approach the
saddle points. 
This justifies a posteriori the estimation of the p-p, p-h 1 and p-h 2 terms of Eq. \eref{1loop} by the bubbles \eref{bubbles} as well as the 
approximations of Eqs. \eref{prg} and \eref{prg2}, namely that $\Gamma^\diamond(\v k,\v k')$ approaches continuously $\Gamma^\diamond_i(\v k)$ as $\v k'\to \v P_i$ and $\Gamma^\diamond_i(\v k)$ approaches $\Gamma^\diamond_{ij}$ as  $\v k \to \v P_j$.
If instead we would find $0<\Max\{\tilde \Gamma^\diamond_\pm\}<1$, the scattering of particles
near the van Hove points would diverge more rapidly than that of particles
at the remaining Fermi surface. This would mean that different regions
in the Brillouin zone behave differently. The region around the
saddle points would become strongly interacting while the remaining
Fermi surface would remain weakly interacting.

\subsection{RG flow of the susceptibilities}
%-------------------------------------------

In order to calculate the susceptibilities we consider Eqs. \eref{chiRGfinal} - \eref{ZRGfinalsc}. We only consider the leading logarithmic terms, i.e. the pairing susceptibilities (Eq. \eref{chiRGfinal} and {\eref{ZRGfinal}}) at a total frequency-momentum $k=0$ and the susceptibilities for density waves and flux phases (Eq. \eref{chiRGfinalsc} and {\eref{ZRGfinalsc}}) at a transfered frequency-momentum $q=(0,\v Q)$. 

In the low energy  regime $\Lambda\to0$, we assume that the frequency
dependence of $Z$ is not important and replace $Z(p)$ by $Z(\v p)$. We can then explicitly perform the frequency integral and obtain, because of the exact nesting, identical equations for the charge, spin and pairing susceptibilities
\ba
2\Lambda\dot\chi^\diamond&=&-\frac1{L^2}\sum_{\v p}\delta(|\xi_{\v p}|-\Lambda)\,Z_B^{\diamond}(\v p)\,Z_A^{\diamond}(\v p),\nonumber\\
2\Lambda\dot Z_{A/B}^{\diamond}(\v p)&=&-\frac1{L^2}\sum_{\v p'}\delta(|\xi_{\v p'}|-\Lambda)\,Z_{A/B}^\diamond(\v p')\,\Gamma^\diamond(\v p',\v p),\label{response}
\ea
where $\diamond=c,s,BCS$. The index $\Lambda$ in the symbols $\chi$, $Z$ and $\Gamma$ have been omitted to simplify the notation.

We treat Eq. \eref{response} in the same way as Eq. \eref{rg}, i.e. we take the leading contribution from the patches $P_1,P_2$ around the two saddle points and assume $Z_{A/B}^\diamond(\v p)\approx Z_{A/B,\,i}^\diamond$ for $\v p\in P_i$. 
It is sufficient to consider complex conjugate form factors $f_A(\v p)=\overline{f_B(\v p)}=f(\v p)$ and thus $Z^\diamond_{A,i}= \overline{Z^\diamond_{B,i}}=Z^\diamond_i$. The `` more general'' angle-dependent susceptibilities \eref{anglesusceptbcs} are in fact not more general in this case\footnote{The reason is as follows. The angle-dependent susceptibility $\chi(\theta,\theta')$ is replaced by a matrix $\chi_{ij}$, where $i,j\in\{1,2\}$ run over the saddle points. The eigenvectors of this two by two matrix (i.e. the natural form factors) are completely determined by the symmetry with respect to the exchange of the two saddle points. One eigenvector is odd, and the other is even with respect to the permutation of the two saddle points.}. We get
\ba
\frac\ud{\ud l}\,\chi^{\diamond}&=&\sum_{j=1}^2|Z^{\diamond}_j|^2,\nonumber\\
\frac\ud{\ud l}Z^{\diamond}_i&=&\sum_{j=1}^2Z^{\diamond}_j\Gamma^\diamond_{ji},\label{presponse1}
\ea
which can be written as
\ba
\frac\ud{\ud l}\,\chi^\diamond_\pm&=&\frac12|Z^{\diamond}_\pm|^2,\nonumber\\
\frac\ud{\ud l}Z^{\diamond}_\pm&=&Z^\diamond_\pm\Gamma^\diamond_{\pm} ,\label{presponse}
\ea
where $Z^\diamond_\pm:=Z^\diamond_1\pm Z^\diamond_2$ and $\chi^\diamond=:\chi^\diamond_++\chi^\diamond_-$. 

If $\Gamma^\diamond_\pm$ is diverging asymptotically like $\tilde
\Gamma^\diamond_\pm/(l_c-l)$ and if $\tilde
\Gamma^\diamond_\pm>=1/2$, the corresponding susceptibility diverges with a critical exponent $1-2\tilde \Gamma^\diamond_\pm$:
\be
\chi^\diamond_\pm\sim(l_c-l)^{1-2\tilde \Gamma^\diamond_\pm}\sim(\Lambda-\Lambda_c)^{1-2\tilde \Gamma^\diamond_\pm}.\label{crit}
\ee
On the other hand $\tilde \Gamma^\diamond_\pm\leq1/2$ leads to a finite
value of the susceptibility.

We thus naturally identify six possible instabilities. In each of the
charge, spin or pairing sectors the form factor can be either even or
odd under the exchange of the two van Hove points (i.e. $Z_1=Z_2$ or
$Z_1=-Z_2$). In the pairing sector, $\chi^{BCS}_+$ clearly corresponds
to $s$-wave superconductivity (sSC)  and $\chi^{BCS}_-$ to
$d_{x^2-y^2}$-wave  superconductivity (dSC). The charge- and spin-density waves (CDW and SDW) are related to the susceptibilities 
$\chi_+^c$ and $\chi_+^s$, respectively. The two remaining
susceptibilities $\chi^{c,s}_-$ correspond to a form factor with
$d_{x^2-y^2}$-wave symmetry in the charge and spin sectors. They
describe the tendency towards the formation of charge or spin flux-phases (CF and SF, see Section \ref{corr}).

These six instabilities of a system with two van Hove 
singularities have been discussed long ago by H. J. Schulz \cite{Schulz89}. 
Here it was shown that they appear naturally in the leading order renormalization group.

\section{Flow of the vertices between saddle points.}\label{2patch}
%==================================================================

We now return to the renormalization of vertices for scattering processes both within and between saddle point patches in the nearest-neighbor tight-binding model ($t'=0$).
The most simple approach is to treat the saddle points in close analogy to the Fermi points of a one-dimensional system where there are just four types of scattering processes, one restricted to the region of a single Fermi point, the other three involving both right and left movers (forward, backward and Umklapp scattering).
This one-dimensional scenario with four effective coupling constants $g_1,\ldots,g_4$ (and thus four vertices given by $\Gamma_i=-g_i$) implicitly assumes that the more detailed wave vector dependence in this region is irrelevant. While this appears to be true for one-dimensional Fermi systems, where going away from a Fermi point means leaving the Fermi surface, we will argue that in the present case the functional dependence of the vertex is relevant in the neighborhood of the saddle points.

\subsection{``One-dimensional'' solution}\label{historical}
%----------------------------------------

\begin{figure} 
        \centerline{\includegraphics[width=7cm]{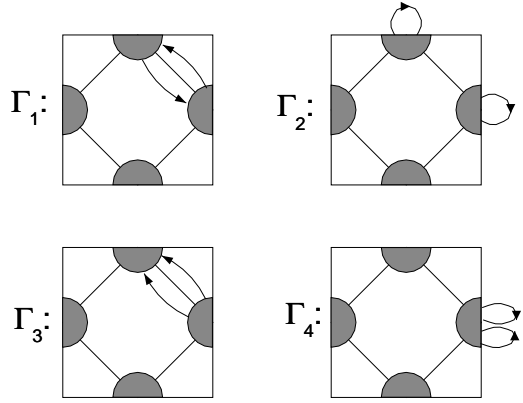}}
\caption{Classification of the vertex function into $\Gamma_1,\ldots,\Gamma_4$.}\label{g14fig}
\end{figure}

In early contributions to this subject \cite{Schulz87,Dzyaloshinskii87}, it was assumed that the vertex function $\Gamma(\v k_1,\ldots,\v k_4)$ takes only four different values
according to how the momenta $\v k_1,\v k_2,\v k_3$ and $\v k_4=\v k_1+\v k_2-\v k_3$ are distributed over the two patches, namely 
\be
\Gamma(\v k_1,\v k_2,\v k_3)\equiv\left\{\begin{array}{ll} 
\Gamma_1\;;\v k_1,\v k_3\in P_1\mbox{ and }\v k_2,\v k_4\in P_2\\ 
\Gamma_2\;;\v k_1,\v k_4\in P_1\mbox{ and }\v k_2,\v k_3\in P_2\\
\Gamma_3\;;\v k_1,\v k_2\in P_1\mbox{ and }\v k_3,\v k_4\in P_2\\ 
\Gamma_4\;;\v k_1,\ldots,\v k_4\in P_1
\end{array}\right.,\label{g14}
\ee
or symmetry-related configurations of the external momenta. The four corresponding scattering processes are illustrated Fig. \ref{g14fig}. The parameters $\Gamma^\diamond_\pm=\Gamma^\diamond_{11}\pm \Gamma^\diamond_{12}$, which control the various instabilities ($\diamond=s,c,BCS$), are readily expressed in terms of the vertices $\Gamma_1,\ldots,\Gamma_4$,
\ba
SDW/SF:&\ \Gamma_\pm^s&=-\Gamma_2\mp \Gamma_3\nonumber\\
CDW/CF:&\ \Gamma_\pm^c&=2\Gamma_1-\Gamma_2\pm \Gamma_3\nonumber\\
sSC/dSC:&\ \Gamma_\pm^{BCS}&=\Gamma_4\pm \Gamma_3.\label{instab}
\ea

$\Gamma_1$ has to be renormalized by the diagrams p-h 2, because the direct momentum transfer $\v k_3-\v k_2$ is close to the nesting vector $\v Q$, but the other contributions coming from p-h 1 and p-p are negligible. Similarly $\Gamma_2$ is renormalized only by p-h 1 and $\Gamma_4$ by p-p. The remaining vertex $\Gamma_3$ on the other hand gets leading contributions from all three channels, p-p, p-h 1 and p-h 2.  

The RG equation is obtained by locating the external momenta $\v k_1,\ldots,\v k_4$ in Eq. \eref{1loop} exactly at the saddle points $\v P_1$ and $\v P_2$ and restricting the sum over $\v p$ to the two patches $P_1,P_2$. The result is 
\ba
\frac\ud{\ud l}\Gamma_1&=&2\Gamma_1(\Gamma_1-\Gamma_2)\nonumber\\
\frac\ud{\ud l}\Gamma_2&=&-\Gamma_2^2-\Gamma_3^2\nonumber\\
\frac\ud{\ud l}\Gamma_3&=&2\Gamma_3(\Gamma_1-2\Gamma_2+\Gamma_4)\nonumber\\
\frac\ud{\ud l}\Gamma_4&=&\Gamma_3^2+\Gamma_4^2.\label{differential}
\ea    

For most initial conditions the numerical solution of these equations diverge asymptotically like in Eq. \eref{divergentg} with coefficients $\tilde \Gamma_i$ satisfying
\ba
\tilde \Gamma_1&=&2\tilde \Gamma_1(\tilde \Gamma_1-\tilde \Gamma_2)\nonumber\\
\tilde \Gamma_2&=&-\tilde \Gamma_2^2-\tilde \Gamma_3^2\nonumber\\
\tilde \Gamma_3&=&2\tilde \Gamma_3(\tilde \Gamma_1-2\tilde \Gamma_2+\tilde \Gamma_4)\nonumber\\
\tilde \Gamma_4&=&\tilde \Gamma_3^2+\tilde \Gamma_4^2.\label{algebra}
\ea   
Eq. \eref{algebra} has many solutions, but the ones which are relevant for the divergences of Eq. \eref{differential} are $\tilde \Gamma_1=0$, $\tilde \Gamma_2=-\tilde \Gamma_4=-1/6$ and $\tilde \Gamma_3=\pm\sqrt5/6$, depending on whether the initial value of $\Gamma_3$ is positive or negative (note that $\Gamma_3$ cannot change its sign). The special feature of these two solutions is that in view of Eqs. \eref{crit} and \eref{instab} three out of the six dominant susceptibilities are diverging with the same critical exponent. Namely
\ba
\chi_{SDW}\sim\chi_{dSC}\sim\chi_{CF}&\sim(\Lambda-\Lambda_c)^{-\gamma}\quad\mbox{ if }\quad \Gamma_3<0,\nonumber\\
\chi_{CDW}\sim\chi_{sSC}\sim\chi_{SF}&\sim(\Lambda-\Lambda_c)^{-\gamma}\quad\mbox{ if }\quad \Gamma_3>0,\label{g3discr}
\ea
where $\gamma=(\sqrt5-2)/3\approx0.08$.  Note, that $\Gamma_3<0$ corresponds to a repulsive coupling $g_3>0$. The divergence of the susceptibilities is thus weak compared to the mean field behavior $\chi\sim(T-T_c)^{-1}$. 

However, for some initial conditions the solutions of Eq. \eref{differential} are not diverging but flow towards the trivial fixed point $\Gamma_1=\Gamma_2=\Gamma_3=\Gamma_4=0$. This means that the RG flow of the vertices between the two saddle point patches does not develop an instability. In this case, the flow can be followed down to small energies such that $\Gamma\log\Lambda\sim1$ and then, the restriction to the saddle-point patches is no longer valid. It means that the low energy behavior can be controlled in this case by subleading terms such as the part of the Fermi surface which is far away from the saddle points, self-energy effects or others.

\subsection{Towards functional renormalization}\label{our}
%----------------------------------------------

The hypothesis of constant vertices $\Gamma_1,\ldots,\Gamma_4$ neglects the fact that in reality all these parameters are functions of incoming and outgoing momenta $\v k_1,\ldots,\v k_4$, three of which can move freely on the Fermi surface within the saddle point patches. Unfortunately, a true functional renormalization is presently beyond the reach of an analytical approach. I therefore search for an appropriate simplified characterization of the momentum dependence of the vertex.

The p-p bubble, which renormalizes the vertex varies most rapidly near $\v k=0$, as it can be seen in Fig. \ref{Bpp}. It is thus natural to admit that the vertex $\Gamma(\v k_1,\v k_2,\v k_3)$ depends strongly on the total momentum $\v k_1+\v k_2$ in the vicinity of $\v k_1+\v k_2=\v 0$ and similarly on the two transfered momenta $\v k_3-\v k_1$ and  $\v k_3-\v k_2$, if one of them is close to $\v Q$. A simple way of modeling this behavior is the following. The vertex is assumed approximately constant except in the vicinity of one of the three special configurations ($\v k_1+\v k_2=\v 0$, $\v k_3-\v k_1=\v Q$ or $\v k_3-\v k_2=\v Q$).  

\begin{figure} 
        \centerline{\includegraphics[width=6cm]{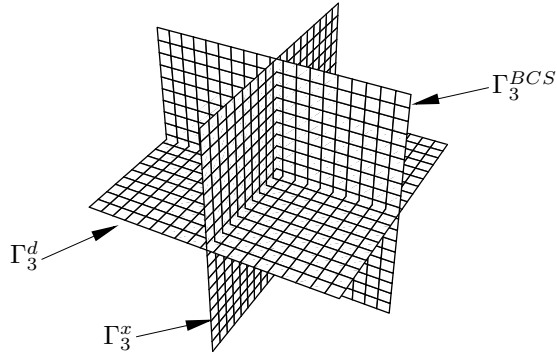}}\vspace{-11pt}
\centerline{\hspace{-3.9cm}\raisebox{1cm}[0pt][0pt]{$\Gamma_3^{x}$}}\vspace{-11pt}
\centerline{\hspace{-6.4cm}\raisebox{2.1cm}[0pt][0pt]{$\Gamma_3^{d}$}}\vspace{-11pt}
\centerline{\hspace{6.9cm}\raisebox{4.4cm}[0pt][0pt]{$\Gamma_3^{BCS}$}}
\caption{Schematic view of the space of three momenta $\v k_1, \v k_2$ and $\v k_3$ moving along the Fermi surface, where the function $\Gamma_3$ is defined. The three planes correspond to the special relations $\v k_1+\v k_2=\v 0$, $\v k_3-\v k_1=\v Q$ and $\v k_3-\v k_2=\v Q$. The point of intersection of the three planes corresponds to the configuration, where the momenta are exactly located at the saddle points. In the approximation proposed here, the vertex takes a different constant value ($\Gamma_3^d$, $\Gamma_3^x$ or $\Gamma_3^{BCS}$) on each of the three planes and yet another constant value $\Gamma_3$ in the remaining phase space.}\label{planes}
\end{figure}

We therefore mimic the true momentum dependence by introducing, in addition to the constants $\Gamma_1,\ldots,\Gamma_4$ representing general values of the momenta (in the patches), other vertices corresponding to specific combinations of momenta. For example, we allow $\Gamma_3(\v k_1,\v k_2,\v k_3)$ to take four different values:
\be 
\Gamma_3(\v k_1,\v k_2,\v k_3)\approx\left\{\begin{array}{ll}
\Gamma_3&\mbox{if }\ |\v k_3-\v k_2- \v Q|,\ |\v k_3-\v k_1 -\v Q|,\ |\v k_1+\v k_2|>O(\sqrt{\Lambda})\\
{ \Gamma_3^{BCS}}&\mbox{if }\ \v k_1+\v k_2= \v 0\\
{ \Gamma_3^x}&\mbox{if }\ \v k_3-\v k_1= \v Q\\
{ \Gamma_3^d}&\mbox{if }\ \v k_3-\v k_2= \v Q
\end{array}\right.\label{specialcouplings}
\ee
The approximation is illustrated schematically in Fig. \ref{planes}.

Similarly, $\Gamma_1(\v k_1,\v k_2,\v k_3)$ takes the values  {$\Gamma_1^d$} or
 $\Gamma_1$,  $\Gamma_2(\v k_1,\v k_2,\v k_3)={ \Gamma_2^x}$ or $\Gamma_2$ and  $\Gamma_4(\v k_1,\v k_2,\v k_3)={ \Gamma_4^{BCS}}$ or $\Gamma_4$. We thus separate the vertices $\Gamma^{BCS}_3,\Gamma^{BCS}_4$ with zero total momentum, $\Gamma^d_1$, $\Gamma^d_3$ with a direct momentum transfer equal to $\v Q$ and
 $\Gamma^x_2,\Gamma^x_3$ with an exchanged momentum transfer of $\v Q$ from the
 general ones ($\Gamma_1,\ldots,\Gamma_4$), where none of these special
 relations among the in- and outgoing momenta applies\footnote{Other special configurations of the momenta will be considered later in Section \ref{uniform}. An overview of all the different vertices and the notations is given in Appendix \ref{notation}.}. In the transition domain (where for example $0<|\v k_1+\v k_2|<\sqrt\Lambda$) the function is unknown, but, as I will argue shortly, its knowledge is not essential. 

The special vertices are related to the vertices $\Gamma^\diamond_\pm$
 introduced in Section \ref{fscouplings} by
\be
\begin{array}{rlll}
SDW/SF:&\ \Gamma_\pm^s&=\Gamma^s_{11}\pm\Gamma^s_{12}&=-\Gamma^x_2\mp \Gamma^x_3\nonumber\\
CDW/CF:&\ \Gamma_\pm^c&=\Gamma^c_{11}\pm\Gamma^c_{12}&=2\Gamma^d_1-\Gamma^x_2\pm (2\Gamma^d_3-\Gamma^x_3)\nonumber\\
sSC/dSC:&\ \Gamma_\pm^{BCS}&=\Gamma^{BCS}_{11}\pm \Gamma^{BCS}_{12}&=\Gamma^{BCS}_4\pm \Gamma^{BCS}_3.
\end{array}\label{ourinstab}
\ee

Since $\Gamma^{BCS}, \Gamma^x$ and $\Gamma^d$ get the strongest contributions from the p-p, p-h 1 and p-h 2 channels, respectively, we say that $\Gamma^{BCS}$ is {\it 
resonant} in the p-p channel, $\Gamma^x$ in the p-h 1 channel and $\Gamma^d$ in the p-h 2 channel. Furthermore the term with the largest contribution, say $\dot R^{pp}$ for a BCS vertex, again only includes BCS vertices and does not mix with non-BCS processes. 

\begin{figure} 
        \centerline{\includegraphics[width=8cm]{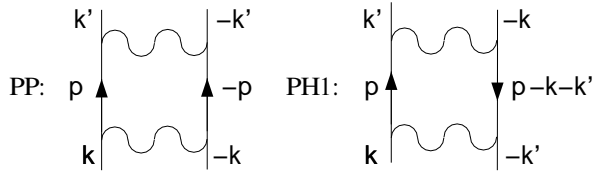}}
\caption{Two diagrams involved in Eq. \eref{ex}. $\v k$ and $\v k'$ are typical vectors belonging to the patch $P_1$ and $P_2$, respectively; $\v p$ is the integration variable}\label{exdiags}\end{figure}

For example, let $\v k\in P_1$ and $\v k'\in P_2$ such that $\Gamma(\v k,-\v k,-\v k')\approx \Gamma^{BCS}_3$ and consider the RG equation for this process:
\be 
\frac\ud{\ud\Lambda}\Gamma^{BCS}_3=\dot R^{pp}+\dot R^{ph1}+\dot R^{ph2}.\label{ex}
\ee
The dominant p-p diagram, shown in Fig. \ref{exdiags}, involves the vertices $\Gamma^{BCS}(\v k,\v p)$ and $\Gamma^{BCS}(\v p,\v k')$. Within our approximation they are replaced by the constants $\Gamma^{BCS}_3$ or $\Gamma^{BCS}_4$, respectively. The p-p contribution  to Eq. \eref{ex} becomes
\be
\dot R^{pp}=2\dot B^{pp}(\Lambda,\v 0)\ \Gamma^{BCS}_3\Gamma^{BCS}_4,
\ee
where $\dot B^{pp}_P(\Lambda,\v k)$ is the p-p bubble restricted to a saddle point patch of size $\rho$, given by Eq. \eref{vHpp} for $t'=0$. 

By contrast, the non-resonant diagram p-h 1 (also shown in Fig. \ref{exdiags}) involves  vertices like $\Gamma(\v k,\v p-\v k-\v k',-\v k')$ which for almost every value of $\v p$ do not satisfy one of the special relations $\v k_1+\v k_2= O(\sqrt\Lambda)$, $\v k_3-\v k_2=\v Q+ O(\sqrt\Lambda)$ or $\v k_3-\v k_1=\v Q+ O(\sqrt\Lambda)$. This diagram therefore includes only the general vertices $\Gamma_2$ and $\Gamma_3$ and no special  vertices $\Gamma^{BCS}, \Gamma^d$ or $\Gamma^x$. Thus  
\be
\dot R^{ph1}=2\dot B^{ph}_P(\Lambda,\v k+\v k')\ \Gamma_2 \Gamma_3,\label{phasespace}
\ee
where $\dot B^{ph}_P(\Lambda,\v k)=-\dot B^{pp}_P(\Lambda,\v k-\v Q)$. It follows that the RG flow depends on the ratio $\dot B^{ph}_P(\Lambda,\v k+\v k')/\dot B^{pp}_P(\Lambda,\v 0)$. In order to obtain a closed set of equations we replace this ratio by a constant. 

Let us first define 
\be
\alpha(\Lambda, \v k+ \v k'):=\frac{\dot B^{pp}_P(\Lambda,\v k+\v k'-\v Q)}{\dot B^{pp}_P(\Lambda,\v 0)},
\ee
which varies in principle between $0$ and $1$. Its value is $1$ if $\v k+\v k'=\v Q$, i.e. if the process to be renormalized is at the same time a $\Gamma^{BCS}$ and a $\Gamma^x$. Such processes exist of course, but they will not influence the RG equations in a relevant way. According to Eq. \eref{p1} a region in the Brillouin zone of width $\sim\sqrt\Lambda$ or smaller can be safely ignored within logarithmic precision. We thus assume that $\v k+\v k'-\v Q\sim\sqrt\Lambda$ or bigger. The biggest values are obtained if $\v k+\v k'$ is parallel to the Fermi surface $\v k+\v k'-\v Q=(\kappa,\kappa)$. For $\kappa\geq\sqrt\Lambda$ we get from Eq. \eref{vHpp} 
\be 
\alpha(\Lambda,\v k+\v k')\leq\frac{\log{\frac{4\rho^2}{\Lambda+2\rho\sqrt\Lambda}}}{\log{\frac{4\rho^2}{\Lambda}}}\ \MapRight{\Lambda\ll\rho^2}\ \frac12.
\ee
In the following we replace $\alpha(\Lambda,\v k+\v k')$ by a constant $\alpha\leq1/2$.  All the nearly resonant diagrams are treated in the same way, i.e. they include general  vertices $\Gamma_1,\ldots \Gamma_4$ only and their amplitude is reduced with respect to the resonant ones by a factor $\alpha\leq1/2$.  

Our approximation scheme leads to a set of RG equations for the special vertices $\Gamma_1^d$, $\Gamma_2^x$, $\Gamma_3^{BCS}$, $\Gamma_3^d$, $\Gamma_3^x$ and $\Gamma_4^{BCS}$
\ba
\frac\ud{\ud l}\Gamma_1^d&=&2\Gamma_1^d(\Gamma_1^d-\Gamma_2^x)+2\Gamma_3^d(\Gamma_3^d-\Gamma_3^x)\nonumber\\
\frac\ud{\ud l}\Gamma_2^x&=&-\left(\Gamma_2^x\right)^2-\left(\Gamma_3^x\right)^2\nonumber\\
\frac\ud{\ud l}\Gamma_3^{BCS}&=&2\Gamma_3^{BCS}\Gamma_4^{BCS}+2\alpha\, \Gamma_3(\Gamma_1-2\Gamma_2)\nonumber\\
\frac\ud{\ud l}\Gamma_3^{x}&=&-2\Gamma_2^{x}\Gamma_3^{x}+2\alpha\, \Gamma_3(\Gamma_1-\Gamma_2+\Gamma_4)\label{1stapp}\\
\frac\ud{\ud l}\Gamma_3^{d}&=&2(2\Gamma_1^{d}\Gamma_3^{d}-\Gamma_2^{x}\Gamma_3^{d}-\Gamma_1^{d}\Gamma_3^{x})+2\alpha\, \Gamma_3(\Gamma_4-\Gamma_2)\nonumber\\
\frac\ud{\ud l}\Gamma_4^{BCS}&=&\left(\Gamma_3^{BCS}\right)^2+\left(\Gamma_4^{BCS}\right)^2.\nonumber
\ea

The general  vertices $\Gamma_1,\ldots,\Gamma_4$ are resonant in none of the three channels. The RG flow of $\Gamma_1,\ldots,\Gamma_4$ is thus given by Eqs. \eref{differential}, with the right hand side multiplied by $\alpha$.

Eqs. \eref{1stapp} can be rewritten in terms of the  vertices which are associated with the dominant instabilities (see Eq. \eref{ourinstab})
\ba 
\frac\ud{\ud l}\Gamma^s_\pm&=&\left(\Gamma_\pm^s\right)^2\mp2\alpha\, \Gamma_3(\Gamma_1-\Gamma_2+\Gamma_4)\nonumber\\
\frac\ud{\ud l}\Gamma^c_\pm&=&\left(\Gamma_\pm^c\right)^2\mp2\alpha\, \Gamma_3(\Gamma_1+\Gamma_2-\Gamma_4)\label{1stapp2}\\
\frac\ud{\ud l}\Gamma^{BCS}_\pm&=&\left(\Gamma_\pm^{BCS}\right)^2\mp2\alpha\, \Gamma_3(2\Gamma_2-\Gamma_1).\nonumber
\ea
For $\alpha=0$, this is a set of six independent equations, one for each instability. In fact, if the non-resonant diagrams are completely neglected, the RG becomes equivalent to the summation of ladder diagrams. Even for $0<\alpha<1$ the special  vertices associated with the different instabilities still do not influence each other, but each RG equation has a source term coming from the general vertices $\Gamma_1,\ldots,\Gamma_4$. For $\alpha=1$ and initial conditions  $\Gamma_1^d=\Gamma_1$, $\Gamma_2^x=\Gamma_2$, $\Gamma_3^d=\Gamma_3^x=\Gamma_3^{BCS}=\Gamma_3$, $\Gamma_4^{BCS}=\Gamma_4$, Eq. \eref{differential} is perfectly recovered (since these conditions are then conserved by the RG flow).

\subsection{Results}
%------------------

One can search for asymptotic solutions of the form $\Gamma(l)=\tilde \Gamma\cdot(l_c-l)^{-1}$ of Eq. \eref{1stapp2} by solving the resulting algebraic equations for the $\tilde \Gamma$. We first consider the possibility of diverging general vertices $\Gamma_1,\ldots,\Gamma_4$. In this case it follows from our analysis of Eq. \eref{differential} that the asymptotic behavior of the general vertices is given by $\tilde \Gamma_1=0,\ \tilde \Gamma_2=-\tilde \Gamma_4=-1/(6\alpha)$ and $\tilde \Gamma_3=\pm\sqrt5/(6\alpha)$, depending on the sign of $\Gamma_3$. By inserting this behavior into Eq. \eref{1stapp2} it is easily seen that a real solution for $\tilde \Gamma^s_\pm,\tilde \Gamma^c_\pm$ and $\tilde \Gamma^{BCS}_\pm$ requires $\alpha\geq\sqrt{80/81}\approx0.994$. But as we argued above, the appropriate values of $\alpha$ are $\leq1/2$.

It follows that for acceptable values of $\alpha$ a special vertex can only diverge if $\tilde \Gamma_1=\ldots=\tilde \Gamma_4=0$. This means that this special vertex diverges at a higher energy scale than the general vertices. 

\begin{figure} 
        \centerline{\includegraphics[width=9cm]{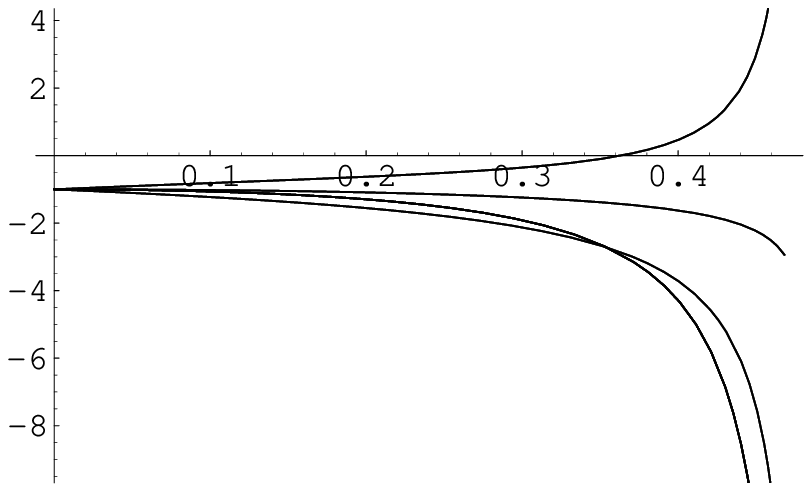}}\vspace{-11pt}
\centerline{\hspace{9.4cm}\raisebox{2.8cm}[0pt][0pt]{$\Gamma^\diamond_1$}\hspace{-0.7cm}\raisebox{1cm}[0pt][0pt]{$\Gamma_2^\diamond$}\hspace{-1.7cm}\raisebox{0.5cm}[0pt][0pt]{$\Gamma_3^\diamond$}\hspace{0.8cm}\raisebox{5cm}[0pt][0pt]{$\Gamma_4^\diamond$}\hspace{0cm}\raisebox{3.8cm}[0pt][0pt]{$l$}}\vspace{-11pt}
\centerline{\hspace{-3cm}\raisebox{5cm}[0pt][0pt]{\fbox{$\alpha=1$}}}\vspace{-11pt}
\centerline{\hspace{-11cm}\raisebox{2.5cm}[0pt][0pt]{\LARGE a)}}
\vspace{6pt}
\centerline{\includegraphics[width=9cm]{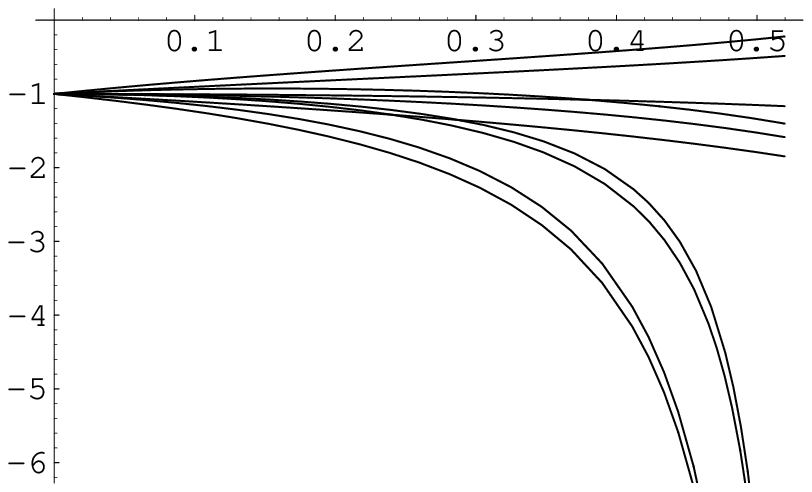}}\vspace{-11pt}
\centerline{\hspace{7.3cm}\raisebox{0.8cm}[0pt][0pt]{$\Gamma_2^x$} \hspace{0.2cm}\raisebox{1cm}[0pt][0pt]{$\Gamma_3^x$} \hspace{-0.8cm}\raisebox{2cm}[0pt][0pt]{$\Gamma_3^d$} \hspace{0.2cm}\raisebox{2.3cm}[0pt][0pt]{$\Gamma_1^d$} \hspace{0.7cm}\raisebox{5.3cm}[0pt][0pt]{$l$}}\vspace{-11pt}
\centerline{\hspace{-3cm}\raisebox{2cm}[0pt][0pt]{\fbox{$\alpha=1/2$}}}\vspace{-11pt}
\centerline{\hspace{-11cm}\raisebox{2.5cm}[0pt][0pt]{\LARGE b)}}
\vspace{6pt}
\centerline{\includegraphics[width=9cm]{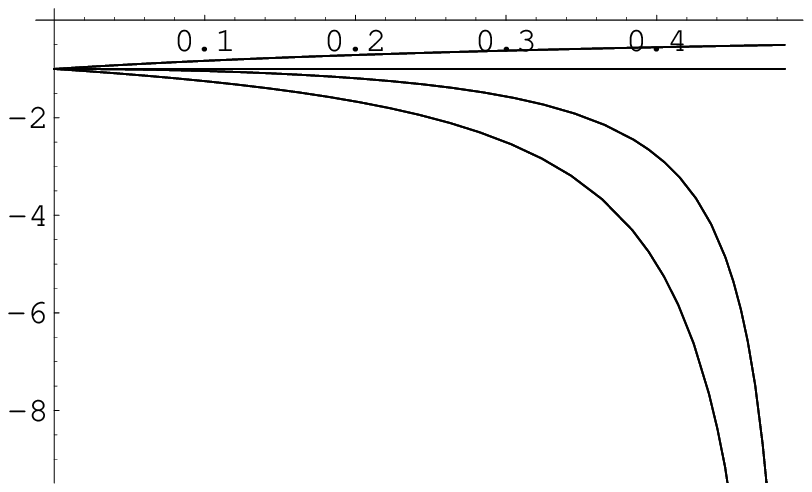}}\vspace{-11pt}
\centerline{\hspace{9.3cm}\raisebox{5.5cm}[0pt][0pt]{$l$}}\vspace{-11pt}
\centerline{\hspace{-3cm}\raisebox{2cm}[0pt][0pt]{\fbox{$\alpha=0$}}}\vspace{-11pt}
\centerline{\hspace{4.5cm}\raisebox{1.5cm}[0pt][0pt]{$\Gamma^x_2=\Gamma^x_3$}}\vspace{-11pt}
\centerline{\hspace{7.9cm}\raisebox{3.8cm}[0pt][0pt]{$\Gamma^d_1=\Gamma^d_3$}}\vspace{-11pt}
\centerline{\hspace{-11cm}\raisebox{2.5cm}[0pt][0pt]{\LARGE c)}}\vspace{-11pt}
\centerline{\hspace{11cm}\raisebox{5cm}[0pt][0pt]{$\begin{array}{l}\Gamma_3^{BCS}=\Gamma^{BCS}_4\\ \Gamma_1=\cdots=\Gamma_4\end{array}$}}
\caption{Three plots of the ten vertices $\Gamma_1,\Gamma_2,\Gamma_3,\Gamma_4,\Gamma_1^d,\Gamma_2^x,\Gamma_3^d,\Gamma_3^x,\Gamma_3^{BCS}$ and $\Gamma_4^{BCS}$ as a function of the RG variable $l$, for different values of the parameter $\alpha$. All the vertices are initially chosen equal to $-1$. The six non diverging vertices of Figure b) are, from top to bottom, $\Gamma_4^{BCS}, \Gamma_4, \Gamma_1, \Gamma_3^{BCS}, \Gamma_3$, and $\Gamma_2$.}\label{tencflow}
\end{figure}

To illustrate the behavior of the RG equations for different values of the parameter $\alpha$, I have plotted some numerical solutions of Eq. \eref{1stapp} in Fig. \ref{tencflow}. The initial conditions have been chosen such that all the vertices are initially equal to $-1$, i.e. the bare value of the repulsive Hubbard model at $U=t=1$. The RG flow is quite similar for $\alpha=1/2$ and for $\alpha=0$ (Fig. \ref{tencflow} b) and c)). Both RG flows are governed by the divergence of the $\Gamma^x$ and $\Gamma^d$ vertices, while other vertices stay small\footnote{The case $\alpha=0$ features the exact degeneracies $\Gamma_1=\Gamma_2=\Gamma_3=\Gamma_4$, $\Gamma_3^{BCS}=\Gamma_4^{BCS}$, $\Gamma_3^{d}=\Gamma_1^{d}$ and $\Gamma_3^{x}=\Gamma_2^{x}$. This is an artifact of both, the special initial conditions and of $\alpha=0$.}. I conclude that the RG flow depends little on the parameter $\alpha$ in the range $0<\alpha<1/2$. 

In contrast the case $\alpha=1$ (Fig. \ref{tencflow} a)), which is equivalent to the ``one-dimensional'' approach Eq. \eref{differential}, shows a completely different flow. If one would choose $\alpha$ close to $1$ but still smaller than $\sqrt{80/81}$, the flow would first resemble the one of Fig. \ref{tencflow} a) and the dominance of the $\Gamma^x$ and $\Gamma^d$ vertices would start only close to the critical scale $l_c$. But for acceptable values $0<\alpha<1/2$ the flow differs from Fig. \ref{tencflow} a) already long before the critical scale. 

The most striking difference to the ``one-dimensional'' solution is that some of the six special vertices can diverge, while the others remain finite. This occurs here because the mixing of the flow for these vertices has been neglected on the basis of a phase space argument (see the discussion  before Eq. \eref{phasespace}). This argument is certainly valid as long as the vertex function is slowly varying, but it may be questioned close to the instability, where the vertex function gets peaked.  It is argued however in Section \ref{consistency} that the non leading vertices can nevertheless stay finite in the case of an instability.

One important feature however is shared by Figs. \ref{tencflow} a) and b), in contrast to Fig. \ref{tencflow} c). Namely the RG flow for $\alpha>0$ leads to a positive value of the quantity $\Gamma^{BCS}_4-\Gamma^{BCS}_3$. This would lead to $d$-wave superconductivity if there was no competing SDW. The main conclusion of Schulz's original paper \cite{Schulz87} that the Hubbard model with an electron density slightly lower than one becomes a $d$-wave superconductor remains valid in our more elaborate scheme\footnote{The argument for the occurrence of superconductivity was as follows. In a doped system with an electron density slightly lower than one, the RG flow will first be similar to the half-filled case and $\Gamma^{BCS}_4-\Gamma^{BCS}_3$ develops a positive value. Finally for low enough energies, the competing density-wave and flux-phase instabilities are suppressed because of the absence of nesting and $d$-wave superconductivity can develop unhamperedly.}. 

\begin{figure} 
        \centerline{\includegraphics[width=9cm]{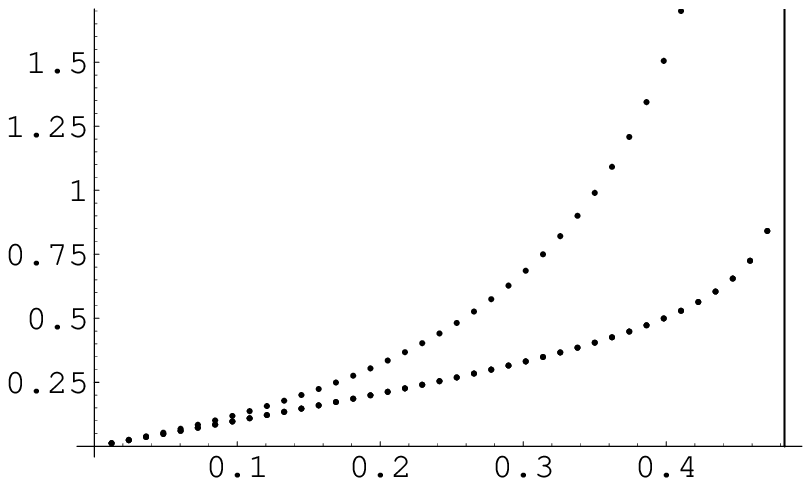}}\vspace{-11pt}
\centerline{\hspace{8.cm}\raisebox{0.5cm}[0pt][0pt]{$l$}\hspace{-5.8cm}\raisebox{2.3cm}[0pt][0pt]{$\chi_{SDW}$}\hspace{1cm}\raisebox{1.2cm}[0pt][0pt]{$\chi_{dSC}=\chi_{CF}$}}\vspace{-11pt}
\centerline{\hspace{-3cm}\raisebox{4cm}[0pt][0pt]{\fbox{$\alpha=1$}}}
\vspace{6pt}
\centerline{\includegraphics[width=9cm]{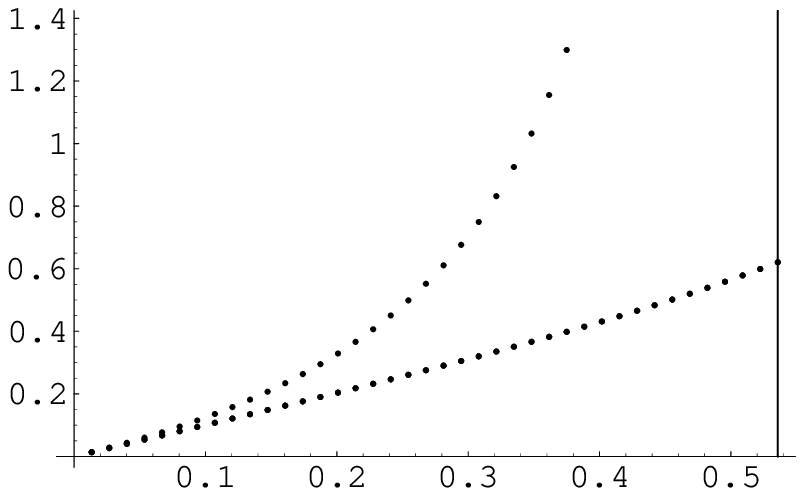}}\vspace{-11pt}
\centerline{\hspace{7.2cm}\raisebox{0.5cm}[0pt][0pt]{$l$}\hspace{-5.8cm}\raisebox{2.3cm}[0pt][0pt]{$\chi_{SDW}$}\hspace{0.8cm}\raisebox{1.2cm}[0pt][0pt]{$\chi_{dSC}=\chi_{CF}$}}\vspace{-11pt}
\centerline{\hspace{-3cm}\raisebox{4cm}[0pt][0pt]{\fbox{$\alpha=1/2$}}}
\caption{RG flow of the susceptibilities $\chi_{SDW}$, $\chi_{dSC}$ and $\chi_{CF}$. Two of them, $\chi_{dSC}$ and $\chi_{CF}$, are exactly degenerate for the initial conditions at hand. This is due to a special symmetry as explained in Section \ref{symmsection}. The location of $l_c$ is indicated by a vertical line.}\label{tencsusc}
\end{figure}

\begin{figure} 
        \centerline{\includegraphics[width=9cm]{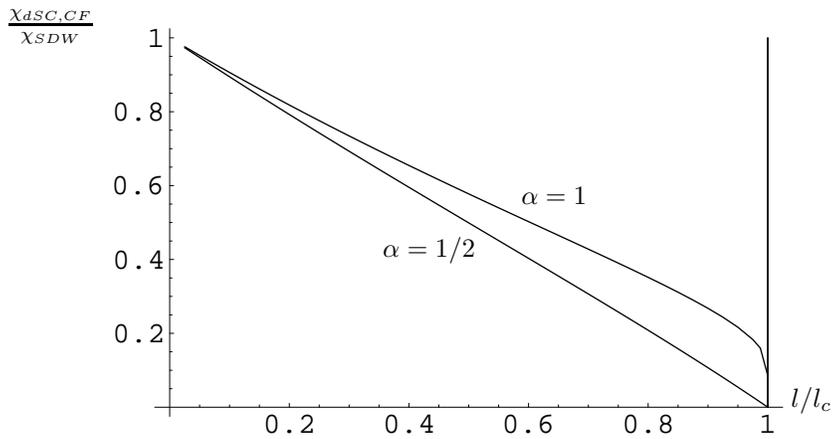}}\vspace{-11pt}
\centerline{\hspace{-0.7cm}\raisebox{5.5cm}[0pt][0pt]{$\frac{\chi_{dSC,CF}}{\chi_{SDW}}$} \hspace{3.7cm}\raisebox{2.5cm}[0pt][0pt]{$\alpha=1/2$}\hspace{0.5cm} \raisebox{3.2cm}[0pt][0pt]{$\alpha=1$}
\hspace{2.6cm}\raisebox{0.5cm}[0pt][0pt]{$l/l_c$}}
\caption{Plot of the ratio between the non-leading and leading susceptibilities as a function of $l/l_c$. It shows a qualitatively different behavior for $\alpha=1$ and $\alpha=1/2$.}\label{suscvergleich} 
\end{figure}

The divergence of the leading vertices is characterized by $\tilde \Gamma^\diamond_\pm=1$. Eq. \eref{crit} then implies
the asymptotic behavior $\chi\sim(\Lambda-\Lambda_c)^{-1}$,
corresponding to a mean field exponent. Fig. \ref{tencsusc} shows the RG flow of the susceptibilities $\chi_{SDW},\chi_{dSC}$ and $\chi_{CF}$ in the case of Hubbard-like initial conditions (all the vertices equal to $-1$). It has been obtained by solving Eq. \eref{presponse}. The SDW susceptibility dominates the two others in both cases $\alpha=1$ and $\alpha=1/2$. But whereas all three susceptibilities diverge with the same asymptotic behavior $\sim(l_c-l)^{(2-\sqrt5)/3}$ in the ``one-dimensional'' solution  $\alpha=1$, the non-dominant ones saturate in the more realistic case $\alpha=1/2$. The difference between  $\alpha=1$ and $\alpha=1/2$ is less pronounced for the susceptibilities than for the vertices, but it is clearly visible. For a better comparison, see Fig. \ref{suscvergleich} where the ratio between the non-leading and leading susceptibilities is shown as a function of $l/l_c$.
 
Finally, in view of 
Eq. \eref{gkksol} the coupling function $\Gamma^\diamond(\v k,\v k')$
diverges as $(\Lambda-\Lambda_c)^{-1}$ everywhere on the Fermi
surface and remains a smooth function upon renormalization, as anticipated in Section \ref{fscouplings}. 

In conclusion, the RG scheme presented in Section \ref{our} interpolates between the ``one dimensional'' solution and the generalized ladder approximation (or generalized RPA), and shares features of both of them. Like in the ``one dimensional'' solution, our RG flow generates enhanced superconducting fluctuations in addition to the leading SDW and strongly suggests a SDW-dSC transition upon doping in the repulsive Hubbard model. But like the RPA, our RG flow leads to only one diverging susceptibility and to the mean-field critical exponent $\chi\sim(T-T_c)^{-1}$.

\subsection{Phase diagram}\label{phasediagram}
%------------------------

For a curved Fermi surface away from the van Hove singularity, the dominant instability is superconductivity, as discussed in Chapter \ref{chapter1}. For the square Fermi surface the flow equations \eref{1stapp2} show that there are several possible instabilities, $s$- and $d$-wave superconductivity, charge- and spin-density waves, charge and spin flux-phases. The values of the initial vertices will determine which of these instabilities, if any, occurs first, i.e. at the largest energy scale. 

The initial conditions for the flow equations are given by the vertices $\Gamma_1,\ldots,\Gamma_4$ at a cutoff $\Lambda_0\ll\rho^4$, when the flow enters the asymptotic regime. They are calculated by naive perturbation theory, i.e. to lowest order $\Gamma_\Lambda=-g$, where $g$ is the microscopic coupling function. This choice is a good approximation if the interaction is small enough such that the vertices vary little before entering the asymptotic regime. 

To be specific I consider the interaction given by Eq. \eref{UVJWK} consisting of on-site and nearest-neighbor terms. It yields the following starting values of the vertices 
\ba
 \Gamma_1=\Gamma^d_1=-U+4V+J+4W,&\qquad \Gamma_2=\Gamma_2^x=-U-4V-J+4W,\label{UVJ}\\
 \Gamma_3=\Gamma^d_3=\Gamma^x_3=\Gamma^{BCS}_3=-U+4V-3J-4W,&\qquad   \Gamma_4=\Gamma^{BCS}_4=-U-4V+3J-4W.\nonumber
\ea
The Coulomb-assisted hopping term $K$ does not appear in Eq. \eref{UVJ}, because it vanishes exactly at the Fermi surface of the half-filled nearest-neighbor tight-binding band. 

\begin{figure} 
        \centerline{\includegraphics[width=8cm]{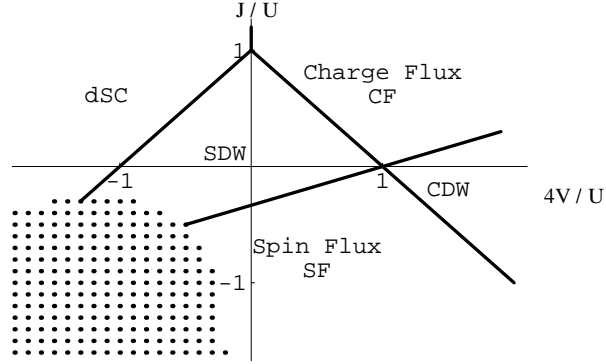}}
\caption{The phase diagram for $U>0$ and $\alpha=1/2$. In the dotted region all the vertices flow to zero. The size of the dotted region depends on the parameter $\alpha$, the other features are $\alpha$-independent}\label{upos}
\end{figure}
\begin{figure} 
        \centerline{\includegraphics[width=8cm]{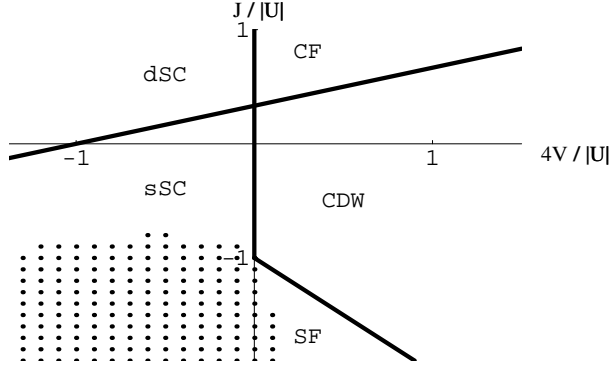}}
\caption{The phase diagram for $U<0$ and $\alpha=1/2$.}\label{uneg}  
\end{figure}

I have solved Eq. \eref{1stapp2} numerically for various initial conditions and obtained the phase diagram as a function of $U$, $V$ and $J$. The result is shown in Figs. \ref{upos} and \ref{uneg} for, respectively, positive and negative values of $U$. $W$ has been put to zero. The predicted phase for the repulsive Hubbard model ($U>V=J=0$) is a
 spin-density wave (SDW) as expected and quite strong nearest neighbor
 terms are needed to establish a flux phase or (only for attractive 
$V$) a $d$-wave superconductor. An unexpected feature of Fig. 
\ref{upos} is that the SDW can be destabilized by {\it 
positive} values of $J$.

In contrast to the ``old'' RG equations \eref{differential} our more elaborate scheme Eq. \eref{1stapp} produces only one diverging susceptibility while
 the others remain finite. Only at the phase boundaries, where the two
 neighboring phases are degenerate, both susceptibilities diverge.

In a certain parameter range (the dotted region in Figs \ref{upos} and \ref{uneg}) all the vertices flow to zero. In this case the behavior is not necessarily dominated by the saddle points and subleading contributions to the RG flow such as self energy corrections can play an important role.

It is interesting to note that most features of the phase diagram are independent of the parameter $\alpha$. In fact, only the size of the dotted regime depends on it, but neither of the transition lines between two neighboring phases. The reason is that these transition lines are entirely defined by some symmetries, a point which will be further explained in Section \ref{symmsection}. 

Since the phase diagram is independent of $\alpha$, we can choose $\alpha=0$. There it is clear, that the leading instability is given by the biggest positive value of the initial vertices $\Gamma_\pm^\diamond$, namely
\ba
sSC:\qquad&\frac12\Gamma^{BCS}_+&=-U-4W\nonumber\\ 
dSC:\qquad&\frac12\Gamma^{BCS}_-&=-4V+3J\nonumber\\
SDW:\qquad&\frac12\Gamma^s_+&=U+2J\label{instvertices}\\
SF:\qquad&\frac12\Gamma^s_-&=4V-J-4W\nonumber\\
CDW:\qquad&\frac12\Gamma^c_+&=-U+8V\nonumber\\
CF:\qquad&\frac12\Gamma^c_+&=4V+3J+4W.\nonumber
\ea
A numerical integration of Eq. \eref{1stapp2} is only required to check whether the vertices diverge or flow to zero.

Let me conclude this section with a remark about superconductivity in a slightly doped system, where the density-wave and flux-phase instabilities are suppressed.   As we have seen in Chapter \ref{chapter1}, a nearest-neighbor repulsion $V>0$ is efficient to suppress $d$-wave superconductivity. This is also evident in Eq. \eref{instvertices}. However, it is easy to see from Eqs. \eref{1stapp2} and \eref{differential}, that $\dot\Gamma^{BCS}_->0$, if initially $\Gamma_3<0$ and $2\Gamma_2-\Gamma_1<0$.  This means that the $d$-wave superconducting fluctuations are enhanced by the RG flow even in the presence of nearest-neighbor repulsions $4V<U$ (assuming $J=W=0$ and $U>0$). I have checked that for $V<0.18 U$, the enhancement is sufficient to turn the initially negative value of $\Gamma^{BCS}_-$ into a positive one before the SDW instability takes place. In this parameter regime, $d$-wave superconductivity is the most favorable instability for the doped system.

\subsection{A consistency test}\label{consistency}
%------------------------------

The approximation scheme presented in section \ref{our} leads to a complete decoupling of the p-p and p-h channels for $\alpha=0$. In this case, the solution of the RG equation is given by the summation of the p-p or p-h ladders Figs. \ref{ppladders}, \ref{ph1ladders} and \ref{ph2ladders}. 

We have seen that, although Eq. \eref{1stapp} for $\alpha>0$ goes beyond the RPA, its solutions are asymptotically the same as for $\alpha=0$. 
The reason is that in our approximate treatment of the non-resonant diagrams only the general  vertices $\Gamma_1,\ldots,\Gamma_4$ intervene and not the (diverging) special vertices $\Gamma_1^d$, $\Gamma_2^x$ etc. This approximation is certainly justified in the beginning of the RG flow, i.e. as long as the vertex does not vary much as a function of the momenta. However, in our solution  we obtain diverging special vertices and finite general vertices, i.e. the vertex functions get a strong peak close to $\v k_1+\v k_2=\v 0$, $\v k_3-\v k_2=\v Q$ or $\v k_3-\v k_1=\v Q$, depending on the nature of the instability. Using the schematic picture of Fig. \ref{planes}, the vertex becomes peaked at one (or several) of the three planes. It is not a priori clear whether our approximation is justified, once the vertex is strongly peaked. 

For example the (non-resonant) contribution of the diagram p-h 1 of Fig. \ref{exdiags} to the renormalization of $\Gamma_3^{BCS}$ involves an integration over the vertex function $\Gamma(\v k,\v p-\v k-\v k',-\v k',\v p)$, where $\v p$ is the integration variable moving along the one-dimensional energy shell $|\xi_{\v p}|=\Lambda$. It is an integral over a one dimensional curve in the space of momenta $(\v k_1,\v k_2,\v k_3)$. For special values of $\v p$, the curve crosses one of the planes specified by $\v k_1+\v k_2=\v 0$, $\v k_3-\v k_2=\v Q$ or $\v k_3-\v k_1=\v Q$. For example for $\v p=\v k'$ the vertex is equal to $\Gamma_3^{BCS}$. The question is whether one can neglect the contribution close to this point even when $\Gamma_3^{BCS}$ diverges.

\begin{figure}  
\centerline{\includegraphics[width=7cm]{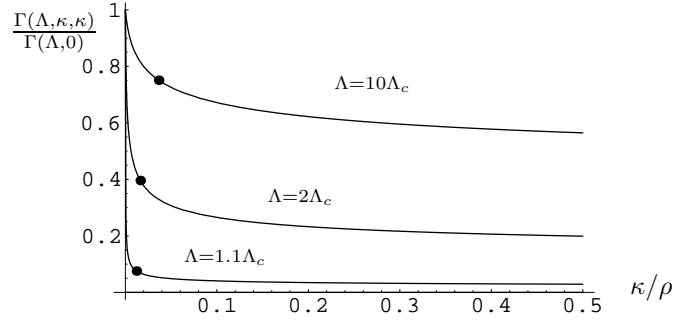}}\vspace{-11pt}
\centerline{\hspace{-5.3cm}\raisebox{0.7cm}[0pt][0pt]{$ ^{\Lambda=1.1\Lambda_c}$}\raisebox{1.5cm}[0pt][0pt]{$ ^{\Lambda=2\Lambda_c}$}\raisebox{3.0cm}[0pt][0pt]{$ ^{\Lambda=10\Lambda_c}$}\hspace{2.9cm}\raisebox{0.4cm}[0pt][0pt]{$\kappa/\rho$}\hspace{-8.8cm}\raisebox{3.8cm}[0pt][0pt]{$\frac{\Gamma(\Lambda,\kappa,\kappa)}{\Gamma(\Lambda,0)}$}}
\caption{Plot of the vertex $\Gamma(\Lambda,\kappa,\kappa)$ normalized by its maximal value at $\kappa=0$, for the choice of parameters $\Lambda_0=0.1\rho^2$, $g=-1$ and different values of $\Lambda$. The vertex depends relatively weakly on $\kappa$ for $\Lambda$ far above $\Lambda_c$, whereas the peak at $\kappa=0$ gets more and more pronounced and narrow as $\Lambda_c$ is approached. Fig. \ref{Gk} is in nice accordance with our approximation, where the vertex function is replaced by a constant for $\sqrt{\Lambda/2}<\kappa<\rho$. The thick points indicate $\kappa=\sqrt{\Lambda/2}$.}\label{Gk}
\end{figure}

In order to answer this question we have to know the value of the vertex
for small but finite total momentum. Within the ladder approximation one obtains
\be
\Gamma_\pm^{BCS}(\Lambda,\v q)=\frac{-g_{\v q,\pm}^{BCS}}{1+g_{\v q,\pm}^{BCS}B_P^{pp}(\Lambda,\v q)},\label{gq}
\ee  
where $\v q$ is the (small) total momentum\footnote{$\v q=\v p-\v k'$ in the example of section \ref{our}.} and $g_{\v q,\pm}^{BCS}=g(\v P_1,\v q-\v P_1,\v q-\v P_1)\pm g(\v P_1,\v q-\v P_1,\v q-\v P_2)$. Eq. \eref{gq} can be obtained either by solving the RG equation \eref{1stapp2} for $\alpha=0$, where the RG variable $l$  is replaced by $B_P^{pp}(\Lambda,\v q)$, or by explicitly summing the ladder series \eref{seriespp3} where the integral over the angle $\theta$ is replaced by a sum over the two saddle points.

 A similar expression is obtained for $\Gamma^s$ and $\Gamma^c$ as functions of the deviation of the momentum transfer from $\v Q=(\pi,\pi)$. The couplings $g_{\v q,\pm}^{\diamond}$ depend weakly on $\v q$ and can thus be approximated by their value at $\v q=0$. I will therefore study the function 
\be
\Gamma(\Lambda,\v q)=\frac{-g}{1+g\, B_P^{pp}(\Lambda,\v q)},\label{gq1}
\ee  

For negative values of $g$, the vertex diverges at a
critical scale $\Lambda_c$. For $\Lambda>\Lambda_c$,  $\Gamma(\Lambda,\v q)$ has a maximum at $\v q=0$, which diverges for $\Lambda\to\Lambda_c$. Fig. \ref{Gk} illustrates the behavior of the vertex for $\v q=(\kappa,\kappa)$ parallel to the Fermi surface where, from Eq. \eref{vHpp}, the p-p bubble is given by
\be
B^{pp}_P(\Lambda,\kappa,\kappa)=\int_{\Lambda}^{\Lambda_0}\!\ud\tilde\Lambda\,\frac1{(2\pi)^2\tilde\Lambda}\log\frac{4\rho^2}{\tilde\Lambda+2\rho\kappa}.
\ee

\begin{figure} 
        \centerline{\includegraphics[width=7cm]{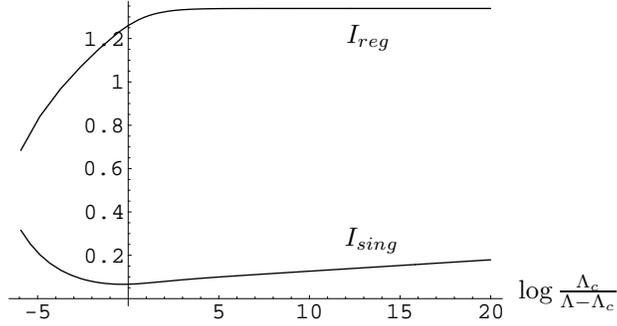}}\vspace{-11pt}
\centerline{\hspace{6.cm}\raisebox{1.1cm}[0pt][0pt]{$I_{sing}$}\hspace{-0.7cm}\raisebox{3.8cm}[0pt][0pt]{$I_{reg}$}\hspace{1.7cm}\raisebox{0.4cm}[0pt][0pt]{$\log\frac{\Lambda_c}{\Lambda-\Lambda_c}$}}
\caption{Numerical evaluation of $I_{{sing}}$ and $I_{{reg}}$ for $L_0=0.1\rho^2$ and $g=-1$ as a function of $\log\frac{\Lambda_c}{\Lambda-\Lambda_c}$. (Taking smaller values of $\Lambda_0$ or $g_0$ reduces the weight of $I_{{sing}}$ with respect to $I_{{reg}}$.)}\label{IL}
\end{figure}

We estimate the contribution of the peak in $\Gamma(\Lambda,\v q)$ to the non-resonant diagram by integrating this function over a curve in $\v q$-space. The biggest contribution is obtained from $\v q=(\kappa,\kappa)$ (parallel to the Fermi surface). Within this worst case scenario, we compare the contribution from the vicinity of the singular peak
\be 
I_{{sing}}(\Lambda)=\int_0^{\sqrt{\Lambda}}\ud\kappa\,\Gamma(\Lambda,\kappa,\kappa)
\ee 
with the regular remaining contribution 
\be 
I_{{reg}}(\Lambda)=\int_{\sqrt{\Lambda}}^{\rho}\ud\kappa\,\Gamma(\Lambda,\kappa,\kappa).
\ee
The non uniform density of states is not taken into account for the moment. The integrals can be calculated numerically and are shown in Fig. \ref{IL}. $I_{{sing}}$ is suppressed with respect to $I_{{reg}}$, because the integration range  is much smaller. Finally for $\Lambda\to\Lambda_c$, $I_{{sing}}$ is diverging like $\log{(\Lambda-\Lambda_c)}$ (note the positive slope in Fig. \ref{IL}) whereas $I_{{reg}}$ saturates. But the divergence of $I_{sing}$ manifests itself only exponentially close to $\Lambda_c$ and is in any case negligible compared to the resonant diagrams $\sim{(\Lambda-\Lambda_c)}^{-2}$. I conclude that the approximation presented in section \ref{our}, namely evaluating the non-resonant diagrams with the constant general values $\Gamma_1,\ldots,\Gamma_4$, is consistent.

In view of the preceding analysis, the RG equation of a non-leading vertex is affected by the singularity through a logarithmically diverging term, i.e. $\dot\Gamma_{i}\sim\log(\Lambda-\Lambda_c)$. Since $\int_{\Lambda_0}^{\Lambda_c}\ud\Lambda\log(\Lambda-\Lambda_c)<\infty$, this confirms the result stated above in Section \ref{our} that non leading vertices remain finite at $\Lambda_c$. The same applies to non-leading susceptibilities.\footnote{The result that $I_{sing}$ diverges as $\log(\Lambda-\Lambda_c)$ can also be obtained analytically by noticing that the vertex function behaves close to the peak like 
$$\Gamma(\Lambda,\kappa,\kappa)\approx\frac a{\Lambda-\Lambda_c+b\,\left|\frac\kappa\Lambda\right|},$$ 
where $a$ and $b$ are positive constants. This behavior is an artifact of our zero-temperature calculation. If the same calculation would be done for $\Lambda=0$ but a finite temperature $T$, the vertex would be analytic in $\v q$ and behave like
$$\Gamma_T(\Lambda=0,\v q)\approx\frac a{T-T_c+c\,\left(\frac{\v q}T\right)^2},$$
where $c$ is another positive constant. But in a finite temperature calculation, the right-hand side of the RG equation is not given by one-dimensional integrals over energy shells $|\xi_{\v p}|=\Lambda$, but over a region $|\xi_{\v p}|<T$ with a finite width. As a consequence, the contribution of the peak in $\Gamma_T(\v q)$ is given by a two-dimensional integral 
$$I_{sing}=\int\!\ud^2\v q\, \Gamma_T(\v q)\MapRight{T\to T_c}\sim\log(T-T_c).
$$
It shows the same logarithmic behavior close to the instability as in the zero temperature case with a cutoff.} 

Note that this behavior can only be obtained because the momenta are allowed to move continuously on the Fermi surface. If instead we would discretize the Fermi surface and replace the continuous vertex function by a finite set of constants, a divergence of one vertex $\Gamma_c$ at a scale $\Lambda_c$ would imply the divergence, at the same scale $\Lambda_c$, of all these vertices that have $\Gamma_c$ appearing in the right hand side of the RG equation. 

Recently, the discretized RG has been studied in detail for a simpler model without van Hove singularities and without Umklapp scattering  \cite{dusuel01}.
It was found that there is a factor $1/N$ between the biggest non-dominant vertices  and the dominant ones, where $N$ is the number of patches. This is consistent with our result in the continuous case ($N\to\infty$), that non-dominant vertices  stay finite at $\Lambda_c$ while the dominant  vertices diverge.
Similar results were also found in the large-$N$ limit of half-filled $N$-leg ladders \cite{Ledermann01}. 

\begin{figure} 
        \centerline{\includegraphics[width=7cm]{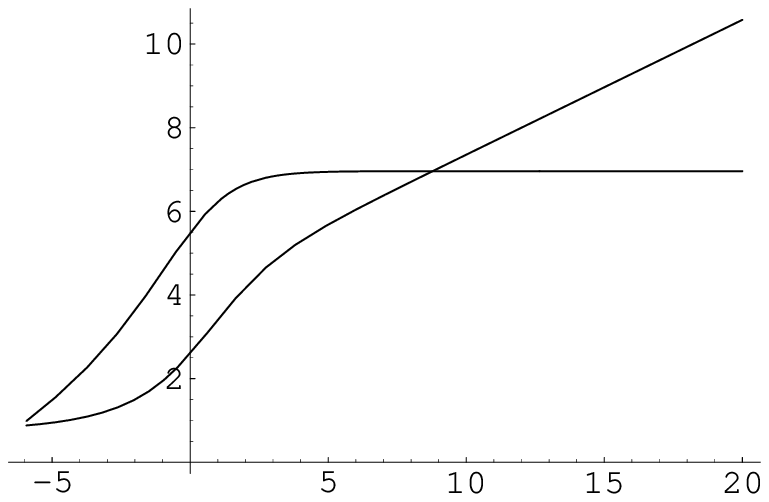}}\vspace{-11pt}
\centerline{\hspace{6.cm}\raisebox{3.9cm}[0pt][0pt]{$\tilde I_{sing}$}\hspace{-0.7cm}\raisebox{2.5cm}[0pt][0pt]{$\tilde I_{reg}$}\hspace{1.7cm}\raisebox{0.4cm}[0pt][0pt]{$\log\frac{\Lambda_c}{\Lambda-\Lambda_c}$}}
\caption{Numerical evaluation of $\tilde I_{{sing}}$ and $\tilde I_{{reg}}$  for $L_0=0.1\rho^2$ and $g=-1$}\label{IIL}
\end{figure}

Taking into account the non uniform density of states could increase the contribution of the peak, if $\kappa=0$ corresponds to the maximal density of states. But this implies $\v k'=(0,\pi)$ in the example of Fig. \ref{exdiags} and is typically not the case. In order to test the really worst case, I have evaluated the following integrals
\be 
\tilde I_{{sing}}(\Lambda)=\int_0^{\sqrt{\Lambda}}\frac{\ud\kappa}{\sqrt{\Lambda+\kappa^2}}\,\Gamma(\Lambda,\kappa,\kappa)\label{tildeIsing}
\ee 
and
\be 
\tilde I_{{reg}}(\Lambda)=\int_{\sqrt{\Lambda}}^{\rho}\frac{\ud\kappa}{\sqrt{\Lambda+\kappa^2}}\,\Gamma(\Lambda,\kappa,\kappa).\label{tildeIreg}
\ee
The denominator in the integrand mimics the situation where the path of integration $|\xi_{\v p}|=\Lambda$ gets close to the van Hove singularity at the same time as $\v q=\v p-\v P_i$ approaches zero. The result is shown in Fig. \ref{IIL}. The singular part is no longer small compared to the regular part, but the divergence of $I_{sing}$ is of course  still not stronger than $\sim\log{(\Lambda-\Lambda_c)}$. As I mentioned before, Fig. \ref{IL} corresponds to a more typical situation then Fig. \ref{IIL}. 

\section{Special symmetries}\label{symmsection}
%=========================

I will now discuss the phase diagram at half filling (Fig. \ref{upos} and
\ref{uneg}) in terms of special symmetries on the lines separating two different phases. For that purpose I consider the p-h symmetric Hamiltonian of Eq. \eref{phsymmH}.

The symmetry group of the noninteracting Hamiltonian $  H_0$ is
extremely large. From any function $d_{\v k}$ we can build an operator 
$N_d=1/2\sum_{\sigma,\v k}d_{\v k}\,c^{\dagger}_{\sigma \v k}c^{}_{\sigma \v k}$
that commutes with $  H_0$ and thus generates a continuous group of symmetry
transformations $\exp(i\alpha N_d)$. We find that $N_d$ commutes with the complete Hamiltonian $H$ if and only if
\be 
\left(d_{\v k_1}+d_{\v k_2}-d_{\v k_3}-d_{\v k_4}\right)\, g(\v k_1,\v k_2,\v k_3)=0\quad \forall\ \v k_1,\ldots \v k_4,\label{condNd}
\ee
where here and in the following $\v k_4=\v k_1+\v k_2-\v k_3$. 

Let the order parameter for superconductivity with a form factor $f$ be given by 
\be
\Delta_f=\frac1{L^2}\sum_{\v p}f(\v p)\,c_{\v p\down}c_{-\v p\up}.
\ee
The symmetry $N_d$ relates superconducting order parameters with form factors $f$ and $f\cdot d$ by $[N_d,\Delta_f]=-2\Delta_{d\cdot f}$, where $d\cdot f$ means the point-wise multiplication of the two functions. This relation can be iterated to $[N_d,\Delta_{d\cdot f}]=-2\Delta_{d^2\cdot f}$. If $d$ is chosen as a sign function, such that  $d^2(\v k)=1$, then the two order parameters $\Delta_f$ and $\Delta_{d\cdot f}$ are related by the symmetry transformation\be
\Delta_{d\cdot f}=\exp{(i\pi N_d/4)}i\Delta_{f}\exp{(-i\pi N_d/4)}.\label{symmop}
\ee
A proof is most easily given by solving a second-order differential equation for the operator $O(\phi)=\exp{(i\phi N_d)}\Delta_{f}\exp{(-i\phi N_d)}$.

Clearly, Eq. \eref{symmop} transforms an $s$-wave superconductor into a $d-$wave superconductor and vice-versa. As a consequence the susceptibilities for $s$- and $d$-wave superconductivity must be exactly equal, provided condition \eref{condNd} holds. The symmetry $N_d$ relates in a similar way spin- or charge-density waves to the corresponding flux phases. It might
therefore control the transition lines SDW/spin flux-phase, CDW/charge flux-phase and sSC/dSC. 

Similarly the operators 
${\vec S_d}=1/2\sum_{\sigma,\sigma',\v k}d_{\v k}\, 
c^\dagger_{\sigma \v k}\vec \sigma_{\sigma\sigma'}  c_{\sigma' \v k}$
relate the spin-density wave to the charge flux-phase and the charge-density wave to the spin flux-phase. They commute with $  H$ if the following two conditions hold 
\be 
\left\{\begin{array}{l}
(d_{\v k_1}- d_{\v k_2}+d_{\v k_3}-d_{\v k_4})\, g(\v k_1,\ldots,\v
k_4)=0\quad \forall\ \v k_1,\ldots \v k_4\\
(d_{\v k_1}+ d_{\v k_2}-d_{\v k_3}-d_{\v k_4})\, (1-X)g(\v k_1,\ldots,\v
k_4)=0\quad \forall\ \v k_1,\ldots \v k_4.
\end{array}\right.\label{condSd}
\ee

Another symmetry introduced by Lieb \cite{Lieb89} and then further
 investigated by Yang and Zhang \cite{Yang90,Zhang90} is generated by the 
pseudo spin operator $\eta_s=\sum_{\v k}s_{\v
k}\,  c_{\up\v Q-\v k}  c_{\down\v k}$, where the function
 $s_{\v k}$ satisfies $s_{\v Q-\v
k}=s_{\v k}$. It turns a $s$-wave  superconductor
into a charge-density wave and  a $d$-wave superconductor into a charge flux-phase (and vice
versa). $\eta_s$ commutes with $H_0$ because of the exact nesting
($\xi_{\v Q-\v k}+\xi_{\v k}=0$) and it commutes with the full
Hamiltonian provided
\be 
\left\{\begin{array}{l}
\sum_{\v p}s_{\v p}\,g(\v p,\bar\v p,\bar\v k,\v k)= s_{\v k}\sum_{\v
p}(2-X)g(\v k,\v p,\v p,\v k)\quad \forall\ \v k\\
s_{\v k_1}(1-X)g(\v k_1,\ldots,\v k_4)-s_{\v k_3}g(\bar\v k_3,\v
k_2,\v k_4,\bar\v k_1)+s_{\v k_4}g(\bar\v k_4,\v k_2,\v k_3,\bar\v
k_1)=0
\end{array}\right.\label{condeta}
\ee
$\forall\ \v k_1,\ldots \v k_4$, where $\bar\v k:=\v Q-\v k$.

Finally Zhang \cite{Zhang97}  considered the operators $\vec \Pi_d=1/2
\sum_{\sigma,\sigma',\v k}d_{\v k}\, 
c_{\sigma \v Q-\v k} \left(\vec\tau\tau^y\right)_{\sigma\sigma'} 
c_{\sigma' \v k}$ connecting a spin-density wave to a $d$-wave superconductor and a spin flux-phase to a $s$-wave superconductor. The symmetry condition
is of the same form as \eref{condeta} but with
$s_{\v k_1}$ replaced by a function $d_{\v k_1}$ that satisfies  $d_{\v Q-\v
k}=-d_{\v k}$. 

It is in general difficult to satisfy the conditions \eref{condNd} to \eref{condeta}. For example they do not hold for the $U-V-J-W$ interaction Eq. \eref{UVJWK} (the Coulomb assisted hopping $K$ is supposed zero, because it breaks the p-h symmetry). The only exception is the pseudo spin
symmetry $\eta_s$ which is exact for $V=W=0$ and $s_{\v k}=1$. 

However the restriction
of the model to the two saddle point patches has more chance of being
symmetric. We take $s_{\v k}=1$ everywhere whereas $d(\v k)=1$ for $\v k
\in P_1$ and $d(\v k)=-1$
for $\v k\in P_2$. For this simple choice the symmetry
generators $N_d$, $\vec S_d$, $\eta_s$ and $\vec \Pi_d$ together with the total spin- and charge operators form a $so(6)\oplus so(2)$ Lie algebra. The commutation relations of the symmetry generators and the relevant order parameters
are listed in reference \cite{markiewicz98}. 

We now assume that the coupling function $g(\v
k_1,\ldots,\v k_4)$ takes only four different values $g_1,\ldots g_4$ as indicated in Fig. \ref{g14fig}. The symmetry conditions are then $g_3=0$ for  $N_d$,  $ g_1=0$ for
$\vec S_d$,  $ g_2+ g_4=2 g_1$ for  $\eta_s$ and   $ g_2+ g_4=0$ for  $\vec
\Pi_d$. These hyper-planes in our four dimensional coupling space define exactly the transition planes of the phase diagram (shown in Figs. \ref{upos} and \ref{uneg} for $ g_1,\ldots, g_4$ parametrized by $U,V$ and $J$).

We have thus shown that the
transition planes of the phase diagram are fixed by exact symmetries of the $ g_1,\ldots,g_4$- model. This is a strong indication that the phase 
diagram shown in Figs. \ref{upos} and \ref{uneg} is the correct one at sufficiently weak coupling. Such a determination of an exact phase diagram by simple 
symmetry considerations was also possible for a one-dimensional
 system \cite{baeriswyl80}. 

Let us now look at the symmetries of the effective interaction $g_\Lambda(\v k_1,\v k_2,\v k_3)=-\Gamma_\Lambda(\v k_1,\v k_2,\v k_3)$ (see Section \ref{effaction}). In  Section \ref{our}, we have parametrized this function in terms of ten instead of
four constants. The symmetry conditions for the effective interaction read
\ba
\Gamma_3^d=\Gamma_3^x=\Gamma_3^{BCS}=\Gamma_3=0&\mbox{ for }& N_d,\nonumber\\
\Gamma_1=\Gamma_1^d=\Gamma_3^d-\Gamma_3^x=0&\mbox{ for }&\vec S_d, \nonumber\\
2\Gamma_1-\Gamma_2-\Gamma_4=2\Gamma_1^d-\Gamma_2^x-\Gamma^{BCS}_4=2\Gamma_3^d-\Gamma_3^x-\Gamma^{BCS}_3=0&\mbox{ for }&\eta_s, \\
\Gamma_2+\Gamma_4=\Gamma_2^x+\Gamma^{BCS}_4=\Gamma_3^x-\Gamma^{BCS}_3=0&\mbox{ for
}&\vec \Pi_d. \nonumber
\ea
 These
symmetries are respected by our approximate RG equation \eref{1stapp}
for every value of the parameter $\alpha$. In fact it is
easy to show that if one of these conditions is satisfied at the initial scale $l_0$ it remains to
be so at any scale $l$. This is not completely trivial since an
arbitrary approximation scheme might violate the symmetries of the
model.

\markright{UNIFORM SUSCEPTIBILITIES...}\section{Uniform susceptibilities, Fermi surface deformations and $\eta$-pairing}\markright{UNIFORM SUSCEPTIBILITIES...}\label{uniform}
%-------------------------------------------------------------------------------------
 
\subsection{More conventional and generalized susceptibilities}\label{uniform1}
%--------------------------------------------------------------

So far we have identified six leading instabilities of  the half-filled nearest-neighbor tight-binding band:  spin- and charge-density waves (SDW and CDW),  spin and charge flux-phases (SF and CF) and finally $s$-wave and $d$-wave superconductivity (sSC and dSC). Each of these instabilities is now further characterized by looking at some additional physical quantities. The framework is linear response theory, which was presented in Section \ref{corr}. In the preceding sections we have focused on the spin- and charge susceptibilities at momentum $\v Q=(\pi,\pi)$ and pairing susceptibilities at zero total momentum. The quantities addressed here are obtained from them by a momentum shift of $\v Q$, i.e. we consider spin and charge susceptibilities at small momenta $\v q\to\v 0$ and pairing with a total momentum of the pairs close to $\v Q$.

The uniform spin susceptibility $\chi_S$ is defined by
\be 
\chi_S=\lim_{\v q\to\v 0}\chi^s(q_0=0,\v q),
\ee
where $\chi^s(q)$ is the spin susceptibility of Section \ref{corr} with a trivial form factor $f_A(\v p)=f_B(\v p)=1$. The small momentum limit needs to be specified here, because the p-h bubble $B^{ph}(q)$ has no well defined limit $q\to0$. From Eqs. \eref{uniformBph} and \eref{qlimitbubble}, one obtains 
\be
\lim_{\v q\to\v 0}B^{ph}(q_0=0,\v q)\MapRight{T\to 0}-\nu(T), \label{qlimbub}
\ee
whereas $B^{ph}(q_0,\v q=\v 0)=0$. It is important that there is no infrared cutoff $\Lambda$ in Eq. \eref{qlimbub}, because otherwise, the result would be zero rather than $-\nu(T)$. Therefore, in the remaining part of this section, we assume a finite temperature $T$ but $\Lambda=0$.   

An enhancement of the uniform spin susceptibility indicates ferromagnetic tendencies, and a suppression of this quantity  signals the absence of low energy spin excitations (spin gap).

The uniform charge susceptibility is the analogous quantity in the charge sector. The external field then couples to the total particle number $N$ and corresponds to a change in the chemical potential. The  charge susceptibility is thus defined as 
\be
\chi_N=\lim_{\v q\to\v 0}\chi^c(q_0=0,\v q)=\frac1{L^2}\frac{\partial\langle N\rangle}{\partial\mu},\label{chargesusc}
\ee
where the partial derivative is taken at constant temperature and system volume. This quantity is often called charge compressibility. The relation between Eq. \eref{chargesusc} and the compressibility is explained in Appendix \ref{comprapp}. A suppressed charge compressibility at low temperature signals a charge gap and is symptomatic of insulating systems (both band- or Mott insulators). A diverging compressibility would indicate phase separation (see Appendix \ref{comprapp}).
 
We also investigate the linear response with respect to the operators $N_d$ and $\vec S_d$, which have been introduced in the preceding section. These operators are obtained by introducing a $d_{x^2-y^2}$-wave form factor into the particle number and total spin operators, respectively, in a similar way as one obtains the charge and spin flux-phases by introducing such a form factor into the corresponding density waves. The susceptibilities $\chi_{N_d}$ and $\chi_{S_d}$ have the same definition as $\chi_N$ and $\chi_S$, but with $f_A(\v p)=f_B(\v p)=\cos p_x-\cos p_y$ (or another smooth function with the same $d$-wave symmetry). They have the  following interpretation. 

If $\chi_{N_d}$ is diverging, the system is likely to establish a non-zero value of $\langle N_d\rangle\approx\langle N_1\rangle-\langle N_2\rangle$, where $N_i$ is the occupation number of the saddle point patch $P_i$. If one patch, say $P_1$, is more occupied than the other, it means that the electrons are moving more easily in the $x$-direction than in the $y$-direction. I interpret this as a spontaneous deformation of the square Fermi surface towards an open Fermi surface as shown in Fig. \ref{nestedFS}, breaking the discrete $D_4$ lattice symmetry. Such an instability (called Pomeranchuk instability) has been predicted recently in \cite{Halboth01}. This instability is likely to move the Fermi level away from the van Hove singularity and it destroys the flatness of the Fermi surface, but it does {\it not} destroy perfect nesting. In fact, as it was shown in Section \ref{nestingsection}, perfect p-h nesting will persist unless the particle-hole symmetry is spontaneously broken.  Since the flat Fermi surface and the van Hove singularity both increase the mixing of p-p and p-h diagrams\footnote{In fact, a flat Fermi surface has an infinity of nesting vectors. This increases the value of non-ladder parquet diagrams. The non-uniform density of states at the van Hove singularity focuses on special scattering processes, like $\Gamma_3$, which are simultaneously resonant in more than one channel. For a uniform density of states, these processes would be negligible.}, a Pomeranchuk instability would increase the asymptotic decoupling of the p-p and p-h channels (the main result of Section \ref{2patch}). 

A diverging susceptibility $\chi_{S_d}$ would indicate that there are two different Fermi surfaces for spin up and spin down electrons, respectively, with opposite deformations relative to the square Fermi surface of the bare system.

From the particle-hole symmetry of the model, the  pairing susceptibilities with a momentum close to $\v Q$ can be treated in the same way as the  p-h susceptibilities at a small momentum $\v q\to\v 0$. We therefore investigate the linear response with respect to the operators $\eta_s$ (singlet $s$-wave Cooper pairing with a total momentum $\v Q$) and $\Pi^y_d$ (triplet $d$-wave Cooper pairing with a total momentum $\v Q$). These operators have been introduced in the preceding section. The susceptibilities $\chi_{\eta_s}$ and $\chi_{\Pi_d}$ correspond to $\lim_{\v k\to\v Q}\chi^{BCS}(0,\v k)$, with $s$-wave and $d$-wave form factors, respectively. 

Unfortunately, the present investigation has to stay rather qualitative. The reason is that we are addressing quantities, which do not feature a leading logarithmic divergence in the perturbation theory. For this reason it is not straightforward to calculate them within a one-loop theory. The one-loop RG is nevertheless able to give a qualitative estimate of the six susceptibilities $\chi_S$, $\chi_N$, $\chi_{S_d}$, $\chi_{N_d}$, $\chi_{\eta_s}$ and $\chi_{\Pi_d}$ in the vicinity of each of the six instabilities SDW, CDW, SF, CF, sSC and dSC. In the following section, we only present and discuss the results. The calculations are shown in Appendix \ref{unifapp}.

\subsection{Results}\label{uniform2}
%-----------------------------------

 The results which are established in Appendix \ref{unifapp} (Eqs. \eref{gS}-\eref{divergingpeak}, \eref{Eq}, \eref{e2} and \eref{e3}) are summarized in  Table \ref{tabsusc}. It turns out that the result does not depend on the form factor ($s$- or $d$-wave) of the instability, i.e. it is the same for SDW as for SF, etc. 

\begin{table}[h]
\center{\begin{tabular}{c|c|c|c}
& SDW/SF & sSC/dSC & CDW/CF \\
\hline
$\chi_N$&$0$&?&$0$\\
$\chi_S$&?&$0$&$0$\\
$\chi_{N_d}$&$\infty$&$\infty$&$\infty$\\
$\chi_{S_d}$&$0$&$0$&$\infty$\\
$\chi_{\eta_s}$&$0$&-&$\infty$\\
$\chi_{\Pi_d}$&$0$&-&$0$\\
\end{tabular}}
\caption{Behavior of various susceptibilities in the vicinity of spin, superconducting or charge instabilities, according to the results of Appendix \ref{unifapp}.  ``$0$'' means that the susceptibility is suppressed and ``$\infty$'' indicates a divergence. In the two occasions, where ``?'' appears, there is a competition between two terms of opposite sign, which can not be decided within our approach. The sign ``-'' means that from our approach no dramatic behavior is expected at the instability.}\label{tabsusc}
\end{table}

In the cases where a susceptibility is suppressed, I obtain a vanishing susceptibility at the critical energy scale, corresponding to a behavior $\chi\sim|\log(T-T_c)|^{-1}$ (see Appendix \ref{unifapp}). This can of course not be true, since no susceptibility is strictly zero at finite temperature. The real behavior is more likely of the form $\chi(T)\sim\exp(-T_c/T)$. This is probably how one should interpret the RG result $\chi\MapRight{T\to T_c}0$, namely that $\chi$ is suppressed at an energy scale given by $T_c$.

Quite remarkably, the susceptibility $\chi_{N_d}$ diverges for any of the six instabilities. Whereas the suppression of susceptibilities above $T_c$ is very slow (only logarithmic), $\chi_{N_d}$ diverges even at a higher energy scale than $T_c$ according to the results of Appendix \ref{unifapp}. This indicates that the square Fermi surface is generally unstable with respect to a deformation towards the open Fermi surface of Fig. \ref{nestedFS} (Pomeranchuk instability \cite{Halboth01}). 
 As it was argued above, this deformation does not imply the destruction of perfect nesting, i.e. it does not hinder the nesting related instabilities SDW,CF, etc. In view of Section \ref{nestingsection}, I suspect that the deformed Fermi surface will increase the decoupling of the different p-p and p-h channels, which was anticipated in Section \ref{2patch}. 

Apart from that, we see that the spin instabilities SDW and SF are characterized by a suppressed  charge compressibility $\chi_N$, indicating the opening of a charge gap. This is expected from the mean field description of the SDW and SF phases. But it is not trivial that a weak coupling RG theory, where no quasi-particle gap is implemented, can reproduce this result. For the uniform spin susceptibility, there is a competition between two effects, which presumably cancel each other. In fact, the SDW ground state allows for low energy spin excitations and there is no reason for a vanishing or diverging spin susceptibility. The numerical works \cite{Halboth00} and \cite{HSFR01} both obtained a finite spin susceptibility at the SDW transition.

In the case of $s$-wave or $d$-wave superconductivity, there is a suppressed spin susceptibility $\chi_S$. Within BCS theory, this is due to the fact that the electrons get paired into singlet Cooper pairs.  To polarize the spins, one has to break these pairs and this needs energy (spin gap). It is remarkable that a RG theory of the (unpaired) normal state already yields the suppression of $\chi_S$. For the charge compressibility, I get a competition between two effects. Although the present calculation is not precise enough to give a conclusive answer, I suspect that the  compressibility is neither zero nor diverging at the superconducting transition. In the numerical calculations away from half filling \cite{Halboth00,HSFR01}, a strong upturn of the charge compressibility was observed, but only very close to the superconducting transition.

The charge instabilities CDW and CF have vanishing susceptibilities $\chi_N$ and $\chi_S$. This indicates the opening of a charge- and spin gap. More unexpected is the divergence of the susceptibilities $\chi_{S_d}$ and $\chi_{\eta}$. This suggests that the formation of a charge flux-phase (or $d$-density wave as proposed in \cite{Chakravarty01}) could be accompanied by even more exotic phenomena such as $\eta$-pairing. The diverging $\chi_{S_d}$ indicates that there could be two Fermi surfaces, one for spin up and one for spin down electrons. These points certainly require further investigations before definite conclusions can be drawn.

%auto-ignore

\chapter{Conclusion}\label{conclusion}
%------------------

The present thesis provides a systematic analysis of the possible instabilities of weakly interacting electrons on a 2D square lattice. This is done using a one-loop  Wilsonian RG approach, i e. to the leading logarithmic order in the energy-scale variable $\Lambda$.

In the general case where the Fermi surface is neither nested nor at a van Hove singularity, the one-loop RG equation becomes equivalent to the ladder approximation and leads to Kohn-Luttinger superconductivity at very low energies. Other instabilities do not occur in the weak coupling limit. In particular, the inclusion of particle-hole contributions and of certain self-energy effects is not consistent within a one-loop calculation. The superconducting instabilities of the extended Hubbard model, including a nearest-neighbor interaction, have been studied in Chapter \ref{chapter1}. 

If the Fermi surface passes through a van Hove singularity without being nested, the one-loop approximation is probably not sufficient, even for weak coupling. An appropriate calculation for this case has still to be invented. It has to include self-energy effects as well as higher-order (two-loop) diagrams. Furthermore it may be necessary to use explicitly a finite temperature to regularize the theory instead of the band cutoff $\Lambda$, since the two are equivalent only within leading logarithmic precision. A step in this direction has been made in \cite{HS01}. 

Most results have been established for the half-filled nearest-neighbor tight-binding model with its square Fermi surface.
Besides $s$- and $d$-wave superconductivity I have identified
 commensurate density waves and flux phases in both  the charge 
and spin sectors as the dominant instabilities. The transition lines of the phase diagram as a function of some parameters of the interaction are fixed by exact symmetries and therefore robust for various approximation schemes. 

In addition, the results of Section \ref{uniform} show that the square Fermi surface of the half-filled nearest-neighbor tight-binding model is generally unstable with respect to a deformation towards an open Fermi surface. This result is in line with the partly numerical work of Halboth and Metzner \cite{Halboth01}. It is argued that such a deformation, although it breaks the $D_4$ symmetry of the square lattice, cannot destroy perfect nesting (unless the particle-hole symmetry is also spontaneously broken). The deformed Fermi surface would then be well described by the anisotropic half-filled nearest-neighbor $t_x$-$t_y$ model. The deformed Fermi surface will be shifted away from the saddle points (the latter are pinned to $(0,\pi)$ and $(\pi,0)$ by symmetry) and it is no more a straight line. Both features, curvature of the Fermi surface and detuning from the van Hove singularity, contribute to an enhanced decoupling of spin, charge and superconducting instabilities. If the deformation is weak, such that the saddle point regions still contain a major part of the density of states, the instabilities and also the phase diagram will be exactly the same as for the square Fermi surface.

In the case of a charge-density wave or a charge flux-phase (or $d$-density wave \cite{Chakravarty01}), I have detected additional diverging susceptibilities, which signal a tendency towards $\eta$-pairing (pairing into Cooper pairs with momentum $(\pi,\pi)$) and  spin-dependent Fermi-surface deformations. Further studies would be required to give a definite interpretation of these features.

On a technical level, I have found that the dominant RG flow of the 
two-particle vertex function in the limit of small energies is controlled
 by scattering processes with momenta close to van Hove points.
 Nevertheless, the low energy effective action contains relevant 
couplings between electrons everywhere near the Fermi surface. The
couplings far away from the saddle points diverge at the same critical
energy and with the same power $\sim(\Lambda-\Lambda_c)^{-1}$ as the
couplings at the saddle points. From this point of view, there is no
sign of a scenario with strong effective couplings at the
saddle points and weak couplings on the remaining Fermi surface.

It was shown that  the RG equations for different superconducting and density-wave instabilities, although strongly coupled at the initial stage, become decoupled in the asymptotic limit of small energies.
Thus  the asymptotic result turns out to be similar to 
that of a generalized random phase approximation (RPA). The
 decoupling arises because the effective coupling function is strongly
 enhanced only for special configurations of the external momenta,
 i.e. in a small region of $\v k$-space. 
In fact according to our analysis, the diverging part of the effective interaction
 restricted to the two patches $P_1$
and $P_2$ (formulated in terms of creation and annihilation operators) writes 
\ba
&&\sum_{i,j=1}^2g^{BCS}_{ij}\sum_{\v k\in P_i,\v k'\in P_j,\v
q}f_\Lambda(\v q)\quad c_{\up\v k}^\dagger c^\dagger_{\down-\v k+\v
q}c_{\down-\v k'+\v q}c_{\up\v k'}\nonumber\\
&&+\frac12\sum_{i,j=1}^2g^{d}_{ij}\sum_{\v k\in P_i,\v k'\in P_j,\v
q}\!\!\!f_\Lambda(\v q)\sum_{\sigma\sigma'} c^\dagger_{\sigma\v
k} c^\dagger_{\sigma'\v k'+\v Q+\v
q} c_{\sigma'\v k'} c_{\sigma\v k+\v Q+\v q}\label{effectivetheory}\\
&&+\frac12\sum_{i,j=1}^2g^{x}_{ij}\sum_{\v k\in P_i,\v k'\in P_j,\v
q}\!\!\!f_\Lambda(\v q)\sum_{\sigma\sigma'} c^\dagger_{\sigma\v
k}c^\dagger_{\sigma'\v k'+\v Q+\v
q}c_{\sigma'\v k+\v Q+\v q}c_{\sigma\v k'},\nonumber
\ea
where $f_\Lambda(\v q)$ is a strongly peaked function such that $|\v
q|<\sqrt\Lambda$, but $\v k$ and $\v k'$ are within patches of size
$\rho$. (The effective $\Lambda$-dependent coupling constants are related to the vertices introduced in Section \ref{fscouplings} by $g^\diamond_{ij}=-\Gamma^\diamond_{ij}$.) The strongly peaked nature of $f_\Lambda(\v q)$ includes that
the effective interaction gets long ranged in real space. If instead we 
would insist on a local interaction, $f_\Lambda(\v q)$ would be a constant
 and Eq. \eref{effectivetheory} is simplified to the model considered by H.
 J. Schulz \cite{Schulz87,Schulz89}. The non-locality emerges here from the
 unbiased treatment of the most diverging terms in the RG equations. 
 
In a one-dimensional system one never generates non local effective 
interactions. The reason is that the low energy excitations of a one-dimensional electron gas are constrained to two privileged points in $\v k$-space: the Fermi 
points. This special geometry allows for a strong mixing between the
various interaction channels, because reducing the arguments of the $g_\Lambda(k_1,k_2,k_3)$
 function to the Fermi points fixes at once the total momentum and momentum
 transfer. Consequently, superconducting and density wave instabilities remain coupled in one dimension. It is however not unusual to encounter non-locality in the vertex 
function $\Gamma_\Lambda(k_1,k_2,k_3)$ where $\xi(\v k_i)>\Lambda$ is allowed. 

 The saddle points of the two-dimensional dispersion 
have a different status than the Fermi points of the one-dimensional electron gas. They are privileged only due to the diverging density of states, but the low energy excitations exist on the whole Fermi surface. As a consequence, the effective interaction ($g_\Lambda(k_1,k_2,k_3)=-\Gamma_\Lambda(k_1,k_2,k_3)$, where $\xi(\v k_i)<\Lambda$) depends on  $\v k$ points which can move continuously on the Fermi surface. The high sensitivity of
 the RG equations on the external momenta leads to a non local effective
 interaction at very low energies. 

In spite of the asymptotic decoupling, our RG flow develops enhanced correlations in the $d$-wave superconducting sector, although they are dominated by the SDW at half filling. This supports the by now well established result of Zanchi and Schulz \cite{Zanchi00} that $d$-wave superconductivity will occur in a slightly doped system, where the density wave fluctuations are suppressed.

Our analysis has nevertheless shown that a discretization of 
the Fermi surface in terms of a finite number of patches can 
enhance artificially the coupling between the different
 scattering channels in the low energy regime. In fact, in the numerical studies \cite{Zanchi00,Halboth00,HSFR01} different susceptibilities are found to diverge at a single energy
scale. By contrast, the present results show that only the susceptibility of the dominant instability diverges. This
 behavior has been referred to as the ``moving pole solution'' by
the Russian school \cite{Dzyaloshinskii72,Gorkov74}. 

The decoupling of the RG equations admittedly has only been
established to leading logarithmic order in the energy cutoff
$\Lambda$. Subleading terms of the one-loop RG equation which would couple the competing channels, are neglected. It is argued however that a consistent treatment of subleading terms requires to go beyond the one-loop approximation. 

In order to estimate how small the bare interaction must be, we recall that our approach requires the patch around the saddle point to be small compared to the size of the Brillouin zone ($\rho\ll\pi/2$). On the other hand, the density of states of this patch\footnote{See Eqs. (\ref{p1}) and (\ref{p2}).} has to be big as compared to that of the remaining part of the Brillouin zone (i.e. $\log4/\rho^2\ll\log4\rho^2/\Lambda$). The bare interaction $g\sim U$ must be small enough such that 
$$
2\,U\,B^P_{pp}(\Lambda,\v 0)= \frac U{4\pi^2}\log^2{\frac{4\rho^2}\Lambda}\sim 1.
$$
(The factor $2$ appears because there are two saddle point patches in the Brillouin zone.)  If for each of the two ``$\ll$'' signs above, a factor of 10 is introduced, this amounts to $U\sim0.02\,t$. A less stringent factor of 3 for each of the two inequalities would correspond to $U\sim0.6\, t$.  

To conclude, we have seen that the weak coupling analysis of the 2D Hubbard model reveals two features which are common to the high-$T_c$ cuprates. Namely the SDW phase at half filling, which is continuously transformed into the antiferromagnetic Mott insulator if the coupling is increased, and the existence of a $d$-wave superconducting phase in the doped regime. Unfortunately we have gained no better understanding neither of the under-doped regime nor of the unusual ``normal-state'' properties at optimal doping. In fact, strong- or intermediate coupling seems to be a key ingredient in a theory for these materials.  

It is nevertheless possible that a two-loop calculation would change the picture. The self-energy of the half-filled Hubbard-model does not satisfy the requirements of Fermi liquid theory above the SDW-critical temperature \cite{Rivasseau01}, which indicates that self-energy corrections are important. But a  consistent two-loop calculation for a 2D Fermi system is a very difficult task.

%auto-ignore

\appendix

\chapter{Time reversal invariance}\label{time}
%---------------------------------

In quantum mechanics, the anti-unitary time reversal operator $\cal T$ maps wave functions to their complex conjugate. In a single-band model, it is reasonable to assume that the Wannier wave functions are real, because otherwise there would be two degenerate wave functions per lattice site. The Wannier states are therefore symmetric under time reversal (i. e. ${\cal T}|\v r\rangle=|\v r\rangle$). Including the spin degree of freedom, we find ${\cal T}|\v r\sigma\rangle=\sigma\,|\v r-\sigma\rangle$, where the number $1$ $(-1)$ is assigned to $\up$ ($\down$).

This implies 
\be
{\cal T}c^\dagger_{\v r\sigma}{\cal T}^{-1}=\sigma\,c^\dagger_{\v r-\sigma},\qquad {\cal T}c^{}_{\v r\sigma}{\cal T}^{-1}=\sigma\,c^{}_{\v r-\sigma}
\ee
 and finally after Fourier transformation 
\be
{\cal T}c^\dagger_{\v k\sigma}{\cal T}^{-1}=\sigma\,c^\dagger_{-\v k-\sigma},\qquad {\cal T}c^{}_{\v k\sigma}{\cal T}^{-1}=\sigma\,c^{}_{-\v k-\sigma}.
\ee

Suppose that the system is symmetric under time reversal, i.e. that $H$ commutes with $\cal T$. In order to investigate the consequence of this symmetry for thermal averages, recall that $\langle{\cal T}\psi|{\cal T}\phi\rangle=\langle{\cal T}\psi|{\cal T}|\phi\rangle=\overline{\langle\psi|\phi\rangle}$. Furthermore, if $|n\rangle$ is an orthonormal basis of the Fock space, then is ${\cal T}|n\rangle$. One then shows 
\ba
\langle O\rangle&=&\frac1Z\sum_{n}\langle n|e^{-\beta(H-\mu N)} O|n\rangle\nonumber\\
&=&\frac1Z\sum_{n}\langle{\cal T}^{-1} n|e^{-\beta(H-\mu N)} O|{\cal T}^{-1}n\rangle\nonumber\\
&=&\frac1Z\sum_{n}\overline{\langle n|{\cal T}e^{-\beta(H-\mu N)} O{\cal T}^{-1}|n\rangle}\nonumber\\
&=&{\langle{\cal T}O^\dagger{\cal T}^{-1}\rangle}.
\ea
This equation, applied to the two particle Green's function, leads to Eq. \ref{tgamma}. Note that the hermitian conjugate of $c(\tau)$ is $c^\dagger(-\tau)$.

\chapter{Parquet diagrams: An example}\label{parquetex}
%++++++++++++++++++++++++++++++++++++++++++++++++++++++++++

\begin{figure}  
\centerline{\includegraphics[width=4cm]{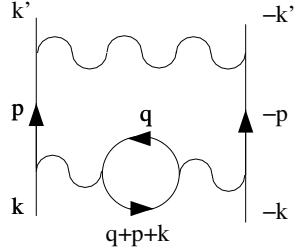}}
\caption{Example of a two loop parquet diagram.}\label{parquetfig}
\end{figure}

As an example of a typical parquet diagram, I consider the diagram shown in Fig. \ref{parquetfig}. For simplicity, I have chosen the total momentum and the external frequencies $k_0$ and $k'_0$ to be zero and I assume a constant coupling function $g$. The value of the diagram (at $T\to0$) is then
\be
g^3\frac1{\beta L^2}\sum_p{ }'\frac{B^{ph}_\Lambda(p+k)}{p_0^2+\xi_{\v p}^2}=g^3\frac1{L^2}\sum_{\v p}{ }'\frac{B^{ph}_\Lambda(p+k)|_{ip_0\to-|\xi_{\v p}|}}{2|\xi_{\v p}|}.
\ee 
The sum $\sum_{\v p}'$ is over all momenta $\v p$ allowed by the infrared cutoff, i.e. satisfying $|\xi_{\v p}|>\Lambda$. The value of the internal p-h bubble is
\ba
-B^{ph}_\Lambda(p+k)|_{ip_0\to-|\xi_{\v p}|}&=&\frac1{\beta L^2}\sum_p{ }'\frac1{(iq_0-\xi_{\v q})(iq_0-|\xi_{\v p}|-\xi_{\v q+\v p+\v k})}\nonumber\\
&=&\frac1{L^2}\sum_{\v q}{ }'\frac{\Theta(\xi_{\v q})-\Theta(|\xi_{\v p}|+\xi_{\v q+\v p+\v k})}{\xi_{\v q}-|\xi_{\v p}|-\xi_{\v q+\v p+\v k}},
\ea
where the sum over $\v q$ is restricted to momenta such that $|\xi_{\v q}|,|\xi_{\v q+\v p+\v k}|>\Lambda$ due to the infrared cutoff. The p-h bubble gives a leading order contribution, if $\v p+\v k$ is a nesting vector. For simplicity, and in order to consider a case where the p-h contribution is most relevant, I assume that  $\xi_{\v q+\v p+\v k}=-\xi_{\v q}$ for a non negligible set of $\v p$-vectors. Within this assumption,
\be
-B^{ph}_\Lambda(p+k)|_{ip_0\to-|\xi_{\v p}|}=\int_\Lambda^W\!\ud\xi'\,\frac{\nu(-\xi')}{2\xi'+|\xi_{\v p}|}+\int_{|\xi_{\v p}|}^W\!\ud\xi'\,\frac{\nu(\xi')}{2\xi'-|\xi_{\v p}|}
\ee
The parquet diagram shown in Fig. \ref{parquetfig} then equals
\be
g^3\int_\Lambda^W\!\ud\xi\,\frac{\tilde\nu(\xi)}{\xi}\left[\int_\Lambda^W\!\ud\xi'\,\frac{\nu(-\xi')}{2\xi'+\xi}+\int_{\xi}^W\!\ud\xi'\,\frac{\nu(\xi')}{2\xi'-\xi}\right],
\ee
where $\tilde\nu(\xi)$ is the density of states of the momenta $\v p$, such that $\v p+\v k$ is a nesting vector.

To the leading logarithmic order and in the absence of van Hove singularities, one can replace the density of states $\tilde\nu(\xi)$ and $\nu(\pm\xi')$  by their value at the Fermi level, so  the diagram of Fig. \ref{parquetfig} is proportional to
\ba
P(\Lambda)&=&\int_\Lambda^W\frac{\ud\xi}{\xi}\left[\int_\Lambda^W\frac{\ud\xi'}{2\xi'+\xi}+\int_{\xi}^W\frac{\ud\xi'}{2\xi'-\xi}\right]\label{2loop}\\
&=&\int_\Lambda^W\frac{\ud\xi}{\xi}\,\frac12\log\frac{4W^2-\xi^2}{\xi(2\Lambda+\xi)}\ \MapRight{\Lambda\to0}\ \frac12\log^2\frac W\Lambda
\ea
The diagram is indeed of the leading order $g^3\log^2\Lambda$. 

To get the contribution to the RG equation, we take the derivative with respect to $\Lambda$
\be
-\partial_\Lambda P(\Lambda)=\frac1{2\Lambda}\log\frac{4W^2-\Lambda^2}{3\Lambda^2}+\int_\Lambda^W\frac{\ud\xi}{\xi(2\Lambda+\xi)}
\ee
The first term is obtained by putting the large loop variable $\xi$ to the energy scale $\Lambda$. The second term comes from the derivative of the internal p-h loop. It is equal to $\frac1{2\Lambda}\log\frac{3W}{W+2\Lambda}\sim1/\Lambda$ and is thus of subleading order, as $\Lambda\to0$. 

This result can be generalized to the following statement. If a diagram is reducible in one channel, the dominant contributions to its $\Lambda$-derivative will be obtained by deriving only the propagators connecting the irreducible blocks and {\it not} the irreducible blocks themselves. This is what is used in the derivation of the one-loop flow equations in section \ref{floweq}.

The same is true in the case of a van Hove singularity. To consider the case, where both, $\nu$ and $\tilde\nu$ are logarithmically diverging at the Fermi level, we multiply the integrand of Eq. \ref{2loop} by $\log\!\xi\log\!\xi'$ and repeat the same arguments. In this case, the leading term is given by $g^2\log^4\Lambda$.

\chapter{Proof of the Polchinski equation}\label{proof}
%========================================

Starting from the definition of the effective interaction, one obtains
\begin{eqnarray*}
e^{-{\cal W}_\Lambda[\chi]}&=&\int\!\ud\mu_{C_\Lambda}[\psi]\,e^{-W[\psi+\chi]}\\
&=&\left.\int\!\ud\mu_{C_\Lambda}[\psi]\,e^{-W[\frac\delta{\delta\bar\eta}]}e^{(\bar\eta,\chi+\psi)+(\bar\chi+\bar\psi,\eta)}\right|_{\eta=0}\\
&=&\left.e^{-W[\frac\delta{\delta\bar\eta}]}e^{(\bar\eta,\chi)+(\bar\chi,\eta)}\underbrace{\int\!\ud\mu_{C_\Lambda}[\psi]\,e^{(\bar\eta,\psi)+(\bar\psi,\eta)}}_{e^{(\bar\eta,C_{\Lambda}\eta)}}\right|_{\eta=0}\\
&=&e^{(\frac\delta{\delta\chi},C_\Lambda\frac\delta{\delta\bar\chi})}e^{-W[\chi]}.
\end{eqnarray*}
For more details about the calculus with Grassmann variables, see \cite{Salmhofer}.  The derivative of the last equation gives
\begin{eqnarray*}
\partial_\Lambda e^{-{\cal W}_\Lambda[\chi]}&=&(\frac\delta{\delta\chi},\dot C_\Lambda\frac\delta{\delta\bar\chi})\,e^{-{\cal W}_\Lambda[\chi]}\\
-\dot{\cal W}_\Lambda[\chi]\,e^{-{\cal W}_\Lambda[\chi]}&=&-\sum_{\sigma,k}\frac\delta{\delta\chi_{\sigma k}}\dot C_\Lambda(k)\frac{\delta{\cal W}_\Lambda[\chi]}{\delta\bar\chi_{\sigma k}}\,e^{-{\cal W}_\Lambda[\chi]}\\
\dot{\cal W}_\Lambda[\chi]&=&\sum_{\sigma,k}\dot C_\Lambda(k)\left(\frac{\delta^2{\cal W}_\Lambda[\chi]}{\delta\!\chi_{\sigma k}\delta\!\bar\chi_{\sigma k}}-\frac{\delta{\cal W}_\Lambda[\chi]}{\delta\!\chi_{\sigma k}}\frac{\delta{\cal W}_\Lambda[\chi]}{\delta\!\bar\chi_{\sigma k}}\right).
\end{eqnarray*}

This is the Polchinski equation \ref{exact}.

\chapter{Estimate of the self-energy in second-order perturbation theory}\markboth{ESTIMATE OF THE SELF-ENERGY...}{}\label{selfenergy}
%========================================================================

In order to decide whether self-energy effects are relevant within the leading-order approximation, I estimate the self-energy corrections to the quasi-particle weight $z=(1-\partial_{ik_0}\Sigma_\Lambda(k))^{-1}$ and to the Fermi velocity $v_F=\nabla_{\v k}(\xi_{\v k}+\Sigma(0,\v k))$, within perturbation theory. For simplicity, I assume a local interaction, i.e. a momentum-independent coupling $g$, but the result remains true for a general short ranged interaction. 

\begin{figure}  
\centerline{\includegraphics[width=4cm]{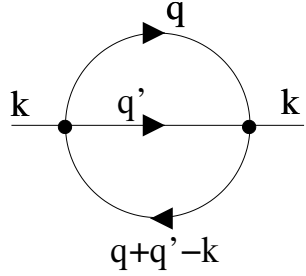}}
\caption{Second-order self-energy diagram.}\label{selffig}
\end{figure}

The first $k$-dependent self-energy contribution comes from the second-order diagram Fig. \ref{selffig}. The interaction lines have been reduced to points, since I assume a local interaction. The analytic expression is\footnote{In fact the diagram shown in Fig. \ref{selffig} stands for two distinct diagrams with wavy interaction lines, one with a Fermion loop and the other without a Fermion loop. A factor of $-2$ is associated to the Fermion loop, hence the minus sign in Eq. \ref{Sigma2}.}
\ba
\Sigma^{(2)}_\Lambda(k)&=&-g^2\frac1{\beta^2L^4}\sum_{q,q'}{ }'C_\Lambda(q)C_\Lambda(q')C_\Lambda(q+q'-k)\label{Sigma2}\\
&=&g^2\frac1{L^4}\sum_{\v q,\v q'}{}'\,\frac{n_{\v q}n_{\v q'}(1-n_{\v q+\v q'-\v k})+(1-n_{\v q})(1-n_{\v q'})n_{\v q+\v q'-\v k}}{ik_0+\xi_{\v q+\v q'-\v k}-\xi_{\v q}-\xi_{\v q'}},\label{Sigma22}
\ea
where $n_{\v p}=(1+e^{\beta\xi_{\v p}})^{-1}$ is the Fermi occupation distribution. The sum in Eq. \ref{Sigma22} is restricted to the momenta allowed by the infra-red cutoff $|\xi_{\v q}|,|\xi_{\v q'}|,|\xi_{\v q+\v q'-\v k}|>\Lambda$.

\section{Quasi-particle weight}
%------------------------------

Differentiating Eq. \ref{Sigma22} with respect to $k_0$ leads to 
\be
\left.\partial_{ik_0}\Sigma^{(2)}_\Lambda(k)\right|_{ik_0=0}=-g^2\frac1{L^4}\sum_{\v q,\v q'}{}'\left(\frac{n_{\v q}n_{\v q'}(1-n_{\v q+\v q'-\v k})}{(\xi_{\v q+\v q'-\v k}-\xi_{\v q}-\xi_{\v q'})^2}+\frac{(1-n_{\v q})(1-n_{\v q'})n_{\v q+\v q'-\v k}}{(\xi_{\v q+\v q'-\v k}-\xi_{\v q}-\xi_{\v q'})^2}\right).\label{z}
\ee
We are going to make a rather crude estimate of this expression. The Fermi function $n_{\v p}$ reduces to a step function $\Theta(-\xi_{\v p})$ at zero temperature. It is then easy to check, that the expression $\xi_{\v q+\v q'-\v k}-\xi_{\v q}-\xi_{\v q'}$ in the denominator is actually a sum of three terms of the same sign. Hence we can use $(\xi_{\v q+\v q'-\v k}-\xi_{\v q}-\xi_{\v q'})^2\geq(\xi_{\v q}+\xi_{\v q'})^2$ to estimate Eq. \ref{z}. Furthermore $n_{\v q+\v q'-\v k}$ and $1-n_{\v q+\v q'-\v k}$ are both bounded by $1$, thus 
\be
\left|\left.\partial_{ik_0}\Sigma^{(2)}_\Lambda(k)\right|_{ik_0=0}\right|\leq g^2\frac1{L^4}\sum_{\v q,\v q'}{}'\left(\frac{n_{\v q}n_{\v q'}}{(\xi_{\v q}+\xi_{\v q'})^2}+\frac{(1-n_{\v q})(1-n_{\v q'})}{(\xi_{\v q}+\xi_{\v q'})^2}\right).
\ee

The estimates used so far are independent of the dispersion $\xi_{\v p}$. In the case of a van Hove singularity, where the density of states is logarithmically diverging, one has
\ba
\left|\left.\partial_{ik_0}\Sigma^{(2)}_\Lambda(k)\right|_{ik_0=0}\right|&<&\mbox{const}\cdot g^2\int_\Lambda^{\Lambda_0}\!\ud\xi\int_\Lambda^{\Lambda_0}\!\ud\xi'\,\frac{\log\frac{\Lambda_0}\xi\log\frac{\Lambda_0}{\xi'}}{(\xi+\xi')^2}\\ 
&<&\mbox{const}\cdot g^2\log^3\frac{\Lambda_0}\Lambda.
\ea

\section{Fermi velocity}
%-----------------------

For an estimate of the self-energy corrections to the Fermi velocity, we evaluate Eq. \ref{Sigma22} at zero frequency and zero Temperature, where it can be written as
\be
\Sigma^{(2)}_\Lambda(0,\v k)=
g^2\frac1{L^4}\sum_{\v q,\v q'}{}'\,\frac{n_{\v q}n_{\v q'}(1-n_{\v q+\v q'-\v k})-(1-n_{\v q})(1-n_{\v q'})n_{\v q+\v q'-\v k}}{|\xi_{\v q+\v q'-\v k}|+|\xi_{\v q}|+|\xi_{\v q'}|}.
\ee

We consider the nearest-neighbor tight-binding band and assume that $\v k$ is close to one of the saddle points ($\v P_1=(\pi,0)$ and $\v P_2=(0,\pi)$), say $\v k=\v P_1+\tilde \v k$, where $\tilde \v k$ is small. Furthermore from the sum $\Sigma_{\v q,\v q'}$, we take into account only the contribution of momenta which are close to one of the two saddle points, because this is where the leading terms are expected to come from.  I.e. $\v q=\v P_1+\tilde \v q$ or   $\v q=\v P_2+\tilde \v q$, where $|\tilde \v q|$ is smaller than some patch size $\rho$ and similarly for $\v q'$. Since each of the two momenta are summed over two patches, $\Sigma^{(2)}_\Lambda(0,\v k)$ is a sum of four terms.
\be
\Sigma^{(2)}_\Lambda(0,\v k)\approx
g^2\sum_{i,j=1}^2\frac1{L^4}\sum_{|\tilde \v q|,|\tilde\v q'|<\rho}\!\!\!\!\!\!{}'\,\,\,\,\,\,\left.\frac{n_{\v q}n_{\v q'}(1-n_{\v q+\v q'-\v k})-(1-n_{\v q})(1-n_{\v q'})n_{\v q+\v q'-\v k}}{|\xi_{\v q+\v q'-\v k}|+|\xi_{\v q}|+|\xi_{\v q'}|}\right|_{\v q=\v P_i+\tilde \v q,\,\v q'=\v P_j+\tilde \v q',}
\ee 
Because of the exact nesting $\xi_{\v P_1+\tilde\v p}=-\xi_{\v P_2+\tilde\v p}$, the denominator is the same for all four terms, but the numerator changes according to $n_{\v P_1+\tilde\v p}=1-n_{\v P_2+\tilde\v p}$. Summing the four contributions, one gets
\be
\Sigma^{(2)}_\Lambda(0,\v P_1+\v k)\approx g^2 \frac1{L^4}\!\sum_{|\v q|,|\v q'|<\rho}\!\!\!\!\!\!{}'\,\,\,\,\,\frac{\sign(\tilde \xi_{\v q+\v q'-\v k})\,\sign(\tilde \xi_{\v q}\,\tilde \xi_{\v q'})}{|\tilde \xi_{\v q+\v q'-\v k}|+|\tilde \xi_{\v q}|+|\tilde \xi_{\v q'}|},\label{vF}
\ee
where I have omitted the tildes on the momenta and introduced the abbreviation $\tilde\xi_{\v p}=\xi_{\v P_1+\v p}$ instead. 

When taking the gradient of Eq. \ref{vF}, we have to remember the implicit infra-red cutoff restriction $|\tilde\xi_{\v q+\v q'-\v k}|>\Lambda$. The sign function $\sign(\tilde \xi_{\v q+\v q'-\v k})$ in Eq. \ref{vF} has to be replaced by $\sign(\tilde \xi_{\v q+\v q'-\v k})\Theta(|\xi_{\v q+\v q'-\v k}|-\Lambda)$, i.e. a function with two steps at $\xi_{\v q+\v q'-\v k}=\pm\Lambda$. Finally one finds
\be
\nabla_{\v k}\Sigma^{(2)}_\Lambda(\v P_1+\v k)\approx g^2 \frac1{L^4}\!\!\!\sum_{|\v q|,|\v q'|<\rho}\!\!\!\!\!\!{}'\,\,\sign(\tilde\xi_{\v q}\tilde\xi_{\v q'})\left(\frac{\delta(|\tilde\xi_{\v q+\v q'-\v k}|-\Lambda)}{\Lambda+|\tilde\xi_{\v q}|+|\tilde\xi_{\v q'}|}-\frac{\Theta(|\tilde\xi_{\v q+\v q'-\v k}|-\Lambda)}{(|\tilde\xi_{\v q+\v q'-\v k}|+|\tilde\xi_{\v q}|+|\tilde\xi_{\v q'}|)^2}\right)\,\cdot\,\nabla_{\v k}\tilde\xi_{\v k-\v q-\v q'}.\label{vF2}
\ee
Because of the hyperbolic form of the dispersion near the saddle points, we have
\be
\nabla_{\v k}\tilde\xi_{\v k-\v q-\v q'}=\nabla_{\v k}\tilde\xi_{\v k}-\nabla_{\v q}\tilde\xi_{\v q}-\nabla_{\v q'}\tilde\xi_{\v q'}.\label{gradient}
\ee
It is clear that we can neglect $\nabla_{\v q}\tilde\xi_{\v q}$ and $\nabla_{\v q'}\tilde\xi_{\v q'}$ for the leading logarithmic contribution to Eq. \ref{vF2}, since  $\nabla_{\v q}\tilde\xi_{\v q}$ vanishes at the saddle point. 

The first term in the parenthesis of Eq. \ref{vF2} is bounded as follows,
\ba
\frac1{L^4}\!\!\!\sum_{|\v q|,|\v q'|<\rho}\!\!\!\!\!\!{}'\,\,\frac{\delta(|\tilde\xi_{\v q+\v q'-\v k}|-\Lambda)}{\Lambda+|\tilde\xi_{\v q}|+|\tilde\xi_{\v q'}|}
&\approx&\frac1{L^4}\!\!\!\sum_{|\v q|,|\v p|<\rho}\!\!\!\!\!\!{}'\,\,\frac{\delta(|\tilde\xi_{\v p}|-\Lambda)}{\Lambda+|\tilde\xi_{\v q}|+|\tilde\xi_{\v p+\v k-\v q}|}\nonumber\\
&<&\frac1{L^4}\!\!\!\sum_{|\v q|,|\v p|<\rho}\!\!\!\!\!\!{}'\,\,\frac{\delta(|\tilde\xi_{\v p}|-\Lambda)}{\Lambda+|\tilde\xi_{\v q}|}\nonumber\\
&<&\mbox{const.}\cdot\log\frac{\Lambda_0}\Lambda\int_\Lambda^{\Lambda_0}\!\ud\xi\,\frac{\log\frac{\Lambda_0}\xi}{\Lambda+\xi}\label{bound3}\\
&<&\mbox{const.}\cdot\log^3\frac{\Lambda_0}\Lambda,\label{vF3}
\ea
where I have introduced the summation variable $\v p=\v q+\v q'-\v k$ (the approximation consists in replacing the constraint $|\v q'|<\rho$ by $|\v p|<\rho$). In \ref{bound3}, I use the fact, that the density of states at the saddle point is proportional to $\log\frac{\Lambda_0}\Lambda$.

The second term in the parenthesis of Eq. \ref{vF2} can be estimated in a similar way as we did in Eq. \ref{z}, 
\be
\frac1{L^4}\!\sum_{|\v q|,|\v q'|<\rho}\!\!\!\!\!\!{}'\,\,\,\,\,\frac{\Theta(|\tilde\xi_{\v q+\v q'-\v k}|-\Lambda)}{(|\tilde\xi_{\v q+\v q'-\v k}|+|\tilde\xi_{\v q}|+|\tilde\xi_{\v q'}|)^2}<\mbox{const.}\cdot\log^3\frac{\Lambda_0}\Lambda\label{vF4}
\ee

The estimate \ref{kSigma} follows immediately from Eqs. \ref{vF2}, \ref{gradient}, \ref{vF3} and \ref{vF4}.

\chapter{Analytic expression of the bubbles for momenta parallel to the Fermi surface}\markboth{ANALYTIC EXPRESSION...}{}\label{Bppapp}
%=========================================================================================

Here I give an approximate analytic expression for the bubbles $\dot B^{pp}(\Lambda,\v k)$ and  $\dot B^{ph}(\Lambda,\v k)$ (defined in Eq. \ref{explicit}) at a momentum $\v k=(\kappa,\kappa)$ (where $\kappa\in[0,\pi]$) parallel to the square Fermi surface of the nearest-neighbor tight-binding band. Since $B^{ph}(\Lambda,\v k)=-B^{pp}(\Lambda,\v Q-\v k)$, we only need to consider the p-p bubble, the quantity which is plotted in Fig. \ref{Bpp}. 

The calculations are rather tedious, using the variables $x_\pm=\tan\frac{p_x\pm p_y}4$ for the integration over the Brillouin zone and distinguishing many cases. For the convenience of notation we introduce the variables 
$$u=\tan\kappa/2$$ 
and 
$$v=\tan\kappa/4$$ 
and we multiply the value of the cutoff by four [i.e. the formulas are given for $\dot B^{pp}(4\Lambda,\kappa,\kappa)$ instead of $\dot B^{pp}(\Lambda,\kappa,\kappa)$]. Finally for an approximate solution of the integrals, a parameter $\epsilon$ was introduced, which is supposed to be
$$
\Lambda\ll\epsilon\ll1.
$$

\subsubsection{Case 1: $0<u<\Lambda$,} i. e. $\kappa$ very small.
\ba
\dot B^{pp}(4\Lambda,\kappa,\kappa)&=&\frac{1+u^2}{(2\pi)^24\Lambda}\left[\log\left|\frac{\sqrt{\epsilon^2-\Lambda^2}+\epsilon}\Lambda\right|-\frac u{\sqrt{\Lambda^2-u^2}}\arccot\left(\frac{u\epsilon+u^2}{\sqrt{(\Lambda^2-u^2)(\epsilon^2-\Lambda^2)}}\right)\right.\nonumber\\
& &\qquad\qquad\ \left.+\frac1{\sqrt{1+u^2}}\log\left|\frac{(\sqrt{1+u^2}+1-\epsilon-u)(\sqrt{1+u^2}-v+u)}{(\sqrt{1+u^2}-1+\epsilon+u)(\sqrt{1+u^2}+v-u)}\right|\right]\nonumber
\ea

\subsubsection{Case 2: $\Lambda<u$ and $v<1-\epsilon$,} i. e. $\kappa$ not too close to $0$ and $\pi$.
\ba
\dot B^{pp}(4\Lambda,\kappa,\kappa)&=&\frac{1+u^2}{(2\pi)^24\Lambda}\left[\log\left|\frac{\sqrt{\epsilon^2-\Lambda^2}+\epsilon}\Lambda\right|+\frac u{\sqrt{u^2-\Lambda^2}}\log\left|\frac{\sqrt{\epsilon^2-\Lambda^2}\sqrt{u^2-\Lambda^2}-u\epsilon-\Lambda^2}{(\epsilon+u)\Lambda}\right|\right.\nonumber\\
& &\qquad\qquad\ \left.+\frac1{\sqrt{1+u^2}}\log\left|\frac{(\sqrt{1+u^2}+1-\epsilon-u)(\sqrt{1+u^2}-v+u)}{(\sqrt{1+u^2}-1+\epsilon+u)(\sqrt{1+u^2}+v-u)}\right|\right]\nonumber
\ea

\subsubsection{Case 3: $1-\epsilon<v<1-\Lambda$ } i. e. $\kappa$ close to $\pi$.
\ba
\dot B^{pp}(4\Lambda,\kappa,\kappa)&=&\frac{1+u^2}{(2\pi)^24\Lambda}\left[\log\left|\frac{\sqrt{(1-v)^2-\Lambda^2}+1-v}\Lambda\right|\right.\nonumber\\
& &\qquad\left.+\frac u{\sqrt{u^2-\Lambda^2}}\log\left|\frac{\sqrt{(1-v)^2-\Lambda^2}\sqrt{u^2-\Lambda^2}-u(1-v)-\Lambda^2}{(1-v+u)\Lambda}\right|\right]\nonumber
\ea

\subsubsection{Case 4: $1-\Lambda<v<1$.}
$$
\dot B^{pp}(4\Lambda,\kappa,\kappa)=0.
$$
(The approximate expression for $B^{pp}(4\Lambda,\kappa,\kappa)$ given here vanishes at $v_c=1-\Lambda$. However it is relatively easy to establish that the exact quantity $B^{pp}(4\Lambda,\kappa,\kappa)$ vanishes at $v_c=\sqrt{\frac{1-\Lambda}{1+\Lambda}}$. The difference is due to the approximation used to solve the integrals.)

\subsubsection{}
The asymptotic behavior for $\kappa\ll1$ can be extracted from the expressions given in the cases 1 and 2
$$
\dot B^{pp}(\Lambda,\kappa,\kappa)\ \MapRight{\kappa\ll1}\ \frac1{(2\pi)^{2}\Lambda}\,\log\frac{16}{\Lambda+4\kappa}.
$$

\chapter{Thermodynamic relations for the charge compressibility}\markboth{THERMODYNAMIC RELATIONS...}{}\label{comprapp}
%===============================================================

Consider a thermodynamic system of $N$ identical particles confined in a volume $V$ in thermal equilibrium with a heat-bath of temperature $T$. The free energy is some function  $F(N,V,T)$. Because $N$, $V$, and $F$ are extensive quantities, the free energy can be written as 
\be
F=\mu N-p V,\label{Euler}
\ee 
where $\mu(N,V,T)=\partial F/\partial N$ is the chemical potential and $p(N,V,T)=-\partial F/\partial V$ is pressure. 

The isothermal compressibility is defined by
\be
\kappa_T=-\frac1V\left(\frac{\partial V}{\partial p}\right)_{N,T}.
\ee
To relate this quantity to the uniform charge susceptibility $\chi_N=\frac1V\left(\frac{\partial N}{\partial \mu}\right)_{V,T}$, we take the partial derivative of Eq. \ref{Euler} with respect to $N$ and obtain
\be
0=N\left(\frac{\partial\mu}{\partial N}\right)_{V,T}-V\left(\frac{\partial p}{\partial N}\right)_{V,T}.\label{derN}
\ee
Similarly we take the partial derivative  of Eq. \ref{Euler} with respect to $V$ to obtain
\be
0=N\left(\frac{\partial\mu}{\partial V}\right)_{N,T}-V\left(\frac{\partial p}{\partial V}\right)_{N,T}.\label{derV}
\ee
The second term of Eq. \ref{derN} and the first term of Eq. \ref{derV} are related by the Maxwell relation
\be
-\left(\frac{\partial p}{\partial N}\right)_{V,T}=\left(\frac{\partial\mu}{\partial V}\right)_{N,T}.\label{Maxwell}
\ee

The relation
\be
\kappa_T=\frac V{N^2}\left(\frac{\partial N}{\partial \mu}\right)_{V,T}=\left(\frac VN\right)^2\chi_N
\ee
is easily obtained from Eqs. \ref{derN}, \ref{derV} and \ref{Maxwell}.

The inverse compressibility is given, up to a multiplying factor by the second derivative of the free energy with respect to the particle number
\be
\kappa_T^{-1}=\frac{N^2}V\frac{\partial^2 F}{\partial N^2}.
\ee
A diverging compressibility is thus a signature for phase separation (the system separates into two phases, if the free energy seizes to be a convex function of $N$), whereas a vanishing compressibility signals a cusp of $F$ as a function of the particle number.

As an example, I calculate the charge compressibility of a system of non-interacting electrons. There,  
\be
N(\mu,V,T)=\sum_{\v k}2 n(\xi_{\v k}),
\ee 
where $n(\xi)$ is the Fermi distribution function. One immediately finds
\be
\chi_N=-2 \int\ud\xi\,\nu(\xi)\frac{\ud n}{\ud\xi}\MapRight{T\to0}\ 2\nu(0).\label{freek}
\ee
In the case, where the density of states diverges logarithmically at the Fermi level, Eq. \ref{freek} is replaced by 
\be
\chi_N\ \MapRight{T\to0}\ 2 \nu(T).
\ee

\chapter{Calculation of the ``uniform susceptibilities''}\label{unifapp}
%=======================================================

The term ``uniform susceptibilities'' is precisely defined in Section \ref{uniform}. In this appendix, the qualitative behavior of each of them is predicted as a superconducting, density-wave, or flux-phase instability is approached.

\section{Relation between uniform susceptibilities and the vertex function}\label{exactsection}
%---------------------------------------------------------------------------

The uniform spin and charge susceptibilities at small momentum depend on the vertex function for forward- and exchange scattering. It is necessary to distinguish the so-called $\v q$- and $\omega$- limits. They are defined as 
\be
\Gamma_{\v q}^f(p,p')=\lim_{\v q\to\v 0}\left.\Gamma(p,p'+q,p')\right|_{q_0=0}
\ee
and
\be
\Gamma_{\omega}^f(p,p')=\lim_{q_0\to0}\left.\Gamma(p,p'+q,p')\right|_{\v q=\v 0}.
\ee
The exchange vertices are defined as $\Gamma_{\v q}^e=X \Gamma_{\v q}^f$ and $\Gamma_{\omega}^e=X \Gamma_{\omega}^f$. Similarly, we define two limits for the vertex function $\Gamma^\eta$, which describes scattering of particle pairs with total momentum close to $\v Q$:
\be
\Gamma_{\v q}^\eta(p,p')=\lim_{\v q\to\v Q}\left.\Gamma(p,q-p,q-p')\right|_{q_0=0},
\ee
\be
\Gamma_{\omega}^\eta(p,p')=\lim_{q_0\to0}\left.\Gamma(p,q-p,q-p')\right|_{\v q=\v Q}.
\ee
The RG procedure with a cutoff $\Lambda$ in the band energy yields $\Gamma_{\omega}$ rather than $\Gamma_{\v q}$. The reason is that at zero temperature, p-h pairs can only be created  with a momentum $\v q$ big enough, such that $|\v v_F\cdot\v q|>\Lambda$, where $\v v_F$ is the Fermi velocity (i.e. the gradient of $\xi$ somewhere on the Fermi surface)\footnote{See also \cite{chitov95} or \cite{Dupuis98}}. Therefore, if not specified differently, I always refer to the $\omega$- limit in this thesis.

The two limits are related by the following Bethe-Salpeter equation\footnote{Eq. \eref{BSlimits} is nothing else than Eq. \eref{scBS} (in the p-h channels) or \eref{BSpp} (in the p-p channel). They have an identical form because of particle-hole symmetry. $\Gamma^\diamond_\omega$ plays the role of the irreducible vertex, since the $\omega$- limit of the p-h bubble vanishes.}
\be
\Gamma_{\v q}^\diamond(p,p')=\Gamma_{\omega}^\diamond(p,p')-\frac1{\beta L^2}\sum_k\Gamma_{\omega}^\diamond(p,k)D^{ph}_{\v q}(k)\Gamma_{\v q}^\diamond(k,p'),\label{BSlimits}
\ee
where $\Gamma^\diamond=-\Gamma^e, 2\Gamma^f-\Gamma^e$ or $\Gamma^\eta$, and $D^{ph}_{\v q}(k)=\left.C(k)C(k+q)\right|_{q_0=0}$. The limit $\v q\to\v 0$ is to be taken after the summation over $k_0$.

From Eqs. \eref{chiZsc} - \eref{Zsc}, we obtain the following identities for the abovementioned spin, charge and pairing susceptibilities
\be
\chi=-\frac1{\beta L^2}\sum_pf^2(\v p)D^{ph}_{\v q}(p)+\frac1{(\beta L^2)^2}\sum_{p,p'}f(\v p)D^{ph}_{\v q}(p)\Gamma^\diamond_{\v q}(p,p')D^{ph}_{\v q}(p')f(\v p'),\label{uniformsusc}
\ee
where $f(\v p)$ is the form factor, i.e. $f=1$ (for $\chi_S$, $\chi_N$ or $\chi_{\eta_s}$) or a $d$-wave function (for $\chi_{S_d}$, $\chi_{N_d}$ or $\chi_{\Pi_d}$). The vertex function is  $\Gamma^\diamond=-\Gamma^e$ (for $\chi_S$ or $\chi_{S_d}$), $2\Gamma^f-\Gamma^e$ (for $\chi_N$ or $\chi_{N_d}$) or $\Gamma^\eta$ (for $\chi_{\eta_s}$ or $\chi_{\Pi_d}$).

There is no infrared cutoff in Eqs. \eref{BSlimits} and \eref{uniformsusc}, but a finite temperature $T$. Nevertheless, we use the  results of the zero temperature RG analysis and replace the cutoff by the temperature, since both are equivalent in the leading logarithmic approximation. 

We will neglect the dependence of $\Gamma_\omega^\diamond(k,k')$ on the frequencies $k_0$ and $k_0'$ from now on. A simple expression for $\chi$ is readily obtained, if we manage to write the vertex functions $\Gamma_\omega^\diamond$ in the form
\be
\Gamma^\diamond_\omega(\v k,\v k')=\sum_n \gamma^\diamond_n\,e^\diamond_n(\v k)e^\diamond_n(\v k'),\label{diagonalvertex}
\ee
where the functions $e^\diamond_n(\v k)$ satisfy the orthogonality relation
\be
-\frac1{\beta L^2}\sum_pD^{ph}_{\v q}(p)\,e^\diamond_n(\v p)e^\diamond_m(\v p)=\delta_{nm}.\label{ortho}
\ee
The $e^\diamond_n(\v k)$ and $\gamma^\diamond_n$ are eigenfunctions and eigenvalues respectively of the operator defined by
\be
h(\v k)\longrightarrow -\frac1{\beta L^2}\sum_p\Gamma^\diamond_\omega(\v k,\v p)D^{ph}_{\v q}(p)h(\v p),\label{GasKernel0}
\ee
where $h$ is any function of $\v k$. Summing the Matsubara frequencies $p_0$ and taking the limit $\v q\to\v 0$, Eq. \eref{GasKernel0} becomes
\be
h(\v k)\longrightarrow -\frac1{L^2}\sum_{\v p}\Gamma^\diamond_\omega(\v k,\v p)n'(\xi_{\v p})\,h(\v p),\label{GasKernel}
\ee
where $n'(\xi)=\ud n/\ud\xi$ is the derivative of the Fermi distribution function.

Using the relations \eref{diagonalvertex} and \eref{ortho}, the Bethe-Salpeter equation \eref{BSlimits} yields
\be
\Gamma_{\v q}^\diamond(\v p,\v p')=\sum_n\frac{\gamma_n}{1-\gamma_n}e_n(\v p)e_n(\v p').
\ee

We expand the form factor $f$ in terms of the eigenfunctions as
\be
f(\v k)=\sum_n a^\diamond_n e^\diamond_n(\v k)\label{formfactordev},
\ee
where $a^\diamond_n$ are coefficients, and obtain the following result for the susceptibilities
\be
\chi=\sum_n\frac{(a^\diamond_n)^2}{1-\gamma^\diamond_n}
\ee
If the form factor $f(\v k)$ is chosen proportional to one of the eigenfunctions, the susceptibility will be given by 
\Rahmen{
\be
\chi=\frac{\chi^{(0)}}{1-\gamma},\label{rparesult}
\ee
}
\noindent where $\chi^{(0)}$ is the value of the susceptibility if the four point vertex is put to zero\footnote{$\chi^{(0)}$ may in general contain self-energy corrections to the single-electron Green's function. In our approach, where self-energy corrections are neglected, $\chi^{(0)}$ is the non-interacting susceptibility.}. 

The eigenvalue $\gamma$ depends on all the parameters of the system including temperature. At high temperature and weak coupling, the eigenvalues $\gamma$ are small numbers. In that case, Eq. \eref{rparesult} gives a small correction to the bare susceptibility. A positive value of $\gamma$ increases the susceptibility and a negative value suppresses it. The susceptibility diverges if $\gamma$ approaches unity. 

In the opposite case, if $\gamma\to-\infty$, the susceptibility is completely suppressed. One should be careful with this conclusion. It can not be justified within perturbation theory, since the infinite series $(1-\gamma)^{-1}=1+\gamma+\gamma^2+\cdots$ does not converge for $\gamma<-1$. The question is whether the Bethe-Salpeter equation \eref{BSlimits} is non-perturbatively true (i.e. if $\Gamma_\omega$ is not small) or not. Although I am not aware of a non perturbative derivation of the Bethe-Salpeter equation, I will assume that $\chi$ is (at least qualitatively) suppressed if $\gamma\to-\infty$.  

In the case $f(\v k)=1$ and approximating the vertex by a constant $\Gamma^\diamond_\omega(\v k,\v k')\to -U$, one obtains the random phase approximation (RPA), where $\gamma=-\chi^{(0)}U$. However the relation \eref{rparesult} goes beyond the RPA. It is exact apart from neglected self-energy effects and the neglected dependence of the vertex $\Gamma^\diamond_\omega(k,k')$ on the two frequencies $k_0$ and $k_0'$. The difficult part is of course to calculate the vertex functions and to write them in the form of Eq. \eref{diagonalvertex}. In the following Sections, we show how our previous result on the behavior of the vertex near an instability can be used to predict the behavior of the susceptibilities.

\section{Diverging peaks in the vertex function}\label{peaksection}
%-----------------------------------------------

In the spirit of Section \ref{2patch}, I propose the following approximations. (The $\omega$- index of the vertices are omitted from now on.)
\begin{enumerate}
\item Due to the diverging density of states near the saddle points, only the momenta within the saddle point patches $P_1$ and $P_2$ are considered.
\item The vertex functions $\Gamma^\diamond(\v k,\v k')$ vary slowly with $\v k$ and $\v k'$ {\bf except} close to the special configurations $\v k+\v k'=\v 0$ and $\v k-\v k'=\v Q$, where they may develop strong peaks.
\end{enumerate}

For the special configurations $\v k+\v k'=\v 0$ and $\v k-\v k'=\v Q$, each of the vertex functions $\Gamma^f$, $\Gamma^e$ and $\Gamma^\eta$ coincides with either $\Gamma^d$, $\Gamma^x$ or $\Gamma^{BCS}$.
In fact\footnote{See Appendix \ref{notation} for the notation.},
\be
\begin{array}{ll}
\Gamma^\eta(\v p,-\v p)=\Gamma^x(\v p,-\v p),&\hspace{1.cm}\Gamma^\eta(\v p,\v p+\v Q)=\Gamma^d(\v p,-\v p),\\
\Gamma^e(\v p,-\v p)=\Gamma^{BCS}(\v p,-\v p),&\hspace{1.cm}\Gamma^e(\v p,\v p+\v Q)=\Gamma^{d}(\v p,\v p),\\
\Gamma^f(\v p,-\v p)=\Gamma^{BCS}(\v p,\v p),&\hspace{1.cm}\Gamma^f(\v p,\v p+\v Q)=\Gamma^{x}(\v p,\v p).
\end{array}
\ee

For example the function $\Gamma^f(\v k,\v k')$ shows the following behavior
\be
\Gamma^f(\v k,\v k') \approx\left\{\begin{array}{ll}
\Gamma_4^f+\Gamma_{4,\,\v k+\v k'}^{BCS}&\mbox{for }\ \v k,\v k'\in P_1\\
\Gamma_2^f+\Gamma_{2,\,\v k-\v k'}^{x}&\mbox{for }\ \v k\in P_1,\  \v k'\in P_2
\end{array}\right..\label{Gfbehavior}
\ee 
where $\Gamma_{\v k+\v k',\,4}^{BCS}$ develops a diverging peak at $\v k+\v k'=\v 0$ in the case of a superconducting instability ($s$- or $d$-wave).
The behavior of the peak close to the instability is given by
\be
\Gamma_{4,\,\v k+\v k'}^{BCS}\approx \frac a{T-T_c+b\left(\frac{\v k+\v k'}T\right)^2 },\label{peakBCS}
\ee
where $a$ and $b$ are positive constants\footnote{This is the result of the ladder approximation at finite temperature (see Section \ref{consistency}).}. This function has a strongly divergent maximum but is completely negligible for $|\v k+\v k'|>\sqrt T$.  
Similarly, the function $\Gamma_{\v q-\v Q,\,2}^{x}$ diverges in the case of a spin instability (SDW or SF) as 
\be 
\Gamma_{2,\,\v q}^{x}\approx \frac {-a}{T-T_c+b\left(\frac{\v q-\v Q}T\right)^2 }.\label{peakx}
\ee
We have thus separated two parts of the forward-scattering vertex function, the regular part (approximated by the constants $\Gamma^f_2$ and $\Gamma^f_4$) and the possible peak close to the instability.

Correspondingly, the vertex functions $\Gamma^e$ and $\Gamma^\eta$ are approximated by
\be
\Gamma^e(\v k,\v k') \approx\left\{\begin{array}{ll}
\Gamma_4^e+\Gamma_{4,\,\v k+\v k'}^{BCS}&\mbox{for }\ \v k,\v k'\in P_1\\
\Gamma_1^e+\Gamma_{1,\,\v k-\v k'}^{d}&\mbox{for }\ \v k\in P_1,\  \v k'\in P_2
\end{array}\right. \label{Gebehavior}
\ee
and
\be
\Gamma^\eta(\v k,\v k') \approx\left\{\begin{array}{ll}
\Gamma_2^\eta+\Gamma_{2,\,\v k+\v k'+\v Q}^{x}&\mbox{for }\ \v k,\v k'\in P_1\\
\Gamma_1^\eta+\Gamma_{1,\,\v k-\v k'}^{d}&\mbox{for }\ \v k\in P_1,\  \v k'\in P_2
\end{array}\right., \label{Getabehavior}
\ee
where $\Gamma_{\v k-\v k',\,1}^{d}$ behaves, close to a spin or charge instability, as
\be
\Gamma_{1,\,\v k-\v k'}^{d}\approx\frac {a}{T-T_c+b\left(\frac{\v k-\v k'-\v Q}T\right)^2 }\cdot\left\{\begin{array}{rl}
-\frac12&\mbox{; SDW or SF}\\
1&\mbox{; CDW or CF}
\end{array}\right..\label{peakd}
\ee

On the basis of these approximations, two approximate eigenfunctions of the operator \eref{GasKernel} are given by
\be 
\mbox{$s$-wave: }f(\v k)=1\qquad\mbox{ and }\qquad\mbox{$d$-wave: } f(\v k)=\left\{\begin{array}{rl}
1&\mbox{ for } \v k\in P_1\\
-1&\mbox{ for } \v k\in P_2
\end{array}\right..
\ee
The approximate eigenvalues are given by
\be
\gamma=\frac{ -\sum_{i=1}^2\frac1{L^2}\sum_{\v p\in P_i}\Gamma^\diamond_\omega(\v k,\v p)n'(\xi_{\v p})\,f(\v p)}{f(\v k)}.\label{explicitg}
\ee
 These eigenvalues are only approximate because the right-hand side of Eq. \eref{explicitg} is only approximately independent of $\v k$.

For example, to calculate the spin susceptibilities $\chi_S$ and $\chi_{S_d}$, we consider the eigenvalue equation for the operator \eref{GasKernel}, where $\Gamma^\diamond=-\Gamma^e$. Using Eqs. \eref{explicitg} and \eref{Gebehavior}, we obtain
\Rahmen{
\be
\gamma_{S,S_d}=-\frac12 \nu(T)\left(\Gamma^e_4\pm\Gamma^e_1\right)+\frac1{L^2}\sum_{\v p\in P_1}n'(\xi_{\v p})\, \Gamma^{BCS}_{4,\,\v k+\v p}\pm\frac1{L^2}\sum_{\v p\in P_2} n'(\xi_{\v p})\, \Gamma^{d}_{1,\,\v k-\v p}.\label{gS}
\ee
}
\noindent Analogous expressions are obtained for $\Gamma^\diamond=2\Gamma^f-\Gamma^x$, namely
\Rahmen{
\ba
\gamma_{N,N_d}&=&\frac12 \nu(T)\left[2\Gamma^f_4-\Gamma^e_4\pm(2\Gamma^f_2-\Gamma^e_1)\right]\nonumber\\
& &\qquad-\frac1{L^2}\sum_{\v p\in P_1}n'(\xi_{\v p})\, \Gamma^{BCS}_{4,\,\v k+\v p}\mp\frac1{L^2}\sum_{\v p\in P_2} n'(\xi_{\v p})\, \left(2\Gamma^{x}_{2,\,\v k-\v p}-\Gamma^{d}_{1,\,\v k-\v p}\right)\label{gN}
\ea
}
\noindent and for $\Gamma^\diamond=\Gamma^\eta$,
\Rahmen{
\be
\gamma_{\eta_s,\Pi_d}=\frac12 \nu(T)\left(\Gamma^\eta_2\pm\Gamma^\eta_1\right)-\frac1{L^2}\sum_{\v p\in P_1}n'(\xi_{\v p})\, \Gamma^{x}_{2,\,\v k+\v p+\v Q}\mp\frac1{L^2}\sum_{\v p\in P_2} n'(\xi_{\v p})\, \Gamma^{d}_{1,\,\v k-\v p}.\label{geta}
\ee
}

The second and third terms on the right-hand side of these equations feel the different instabilities very directly. Using Eqs. \eref{peakBCS}, \eref{peakx} and \eref{peakd} one obtains
\Rahmen{
\be
 \mbox{sSC or dSC instability:}\qquad\frac1{L^2}\sum_{\v p\in P_1}n'(\xi_{\v p})\, \Gamma^{BCS}_{4,\,\v k+\v p}\ \longrightarrow -\infty,\label{e1}
\ee
\be
 \mbox{CDW or CF instability:}\qquad\frac1{L^2}\sum_{\v p\in P_2}n'(\xi_{\v p})\, \Gamma^{d}_{1,\,\v k-\v p}\ \longrightarrow -\infty,
\ee
\be
 \mbox{SDW or SF instability:}\qquad
\frac1{L^2}\sum_{\v p\in P_2}n'(\xi_{\v p})\, \cdot\left\{\begin{array}{c}
\Gamma^{x}_{2,\,\v k-\v p}\\
\Gamma^{d}_{1,\,\v k-\v p}
\end{array}\right.\  \longrightarrow +\infty.\label{divergingpeak}
\ee
}
\noindent The divergence is always logarithmic $\sim \log(T-T_c)$. In the case of a SDW or SF instability, $\Gamma^x$ diverges twice as fast as $\Gamma^d$, such that $\Gamma^c=2\Gamma^d-\Gamma^x$ remains finite. 

The equations \eref{gS}-\eref{divergingpeak} explain already most of the results mentioned in Section \ref{uniform2}.
To obtain the full result, we still have to investigate the first terms of the right-hand side of  Eqs. \eref{gS}-\eref{geta}. The question is thus how the vertices $\Gamma^e_1,\Gamma^\eta_1,\Gamma^f_2,\Gamma^\eta_2,\Gamma^f_4$ and $\Gamma^e_4$ behave as one of the instabilities is approached. We argued in Section \ref{2patch} that general vertices are not dramatically influenced by the instability, i.e. they do not diverge. It turns out that the forward-, exchange- and $\eta$-vertices are not general in that sense but that they respond to $\Gamma^d$, $\Gamma^x$ and $\Gamma^{BCS}$ via a relatively subtle effect. 

\section{Renormalization of the  forward-, exchange- and $\eta$-vertices}\label{renormsection} 
%-----------------------------------------------------------------------

The values of the vertices $\Gamma_1^e,\Gamma_1^\eta,\Gamma_2^f,\ldots$ depend on the energy scale and are determined by the leading-order RG equations. We will use the same formalism as in Section \ref{2patch}, where the energy scale is represented by a sharp infra-red cutoff $\Lambda$. In the end of the calculation we identify, within logarithmic accuracy, the vertex at zero temperature in the presence of a cutoff  $\Lambda$ with the vertex at temperature $T=\Lambda$ (without cutoff). 

As an example we choose two momenta $\v k\in P_1$ and $\v k'\in P_2$ with $|\v k-\v k'-\v Q|>\sqrt \Lambda$ and consider the RG equation of the vertex $\Gamma(\v k,\v k',\v k')=\Gamma^f_2$. It is renormalized mainly from the p-h 1 diagram, since the exchanged momentum transfer $\v k'-\v k$ is in the neighborhood of $\v Q$. The other contributions p-p and p-h 2 are negligible. 

\begin{figure} 
        \centerline{\includegraphics[width=5cm]{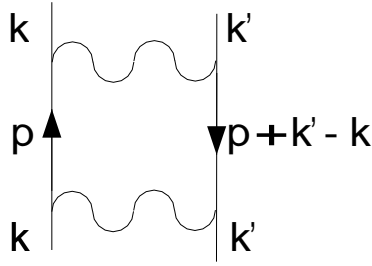}}
\caption{The p-h 1 contribution to  $\partial_\Lambda\Gamma^f(\v k,\v k')$.}\label{examplefwd}
\end{figure}

The p-h 1 diagram is shown in Fig. \ref{examplefwd}. If the internal momentum $\v p$ approaches $\v k+\v Q$, {\it both} vertices of the diagram approach simultaneously the special vertex $\Gamma^d_3$. This is in contrast to the case of a vertex with general external momenta (or the case of Fig. \ref{exdiags}). Similarly, if $\v p$ approaches $-\v k'$, both vertices of the diagram approach simultaneously $\Gamma^{BCS}_3$. 

More quantitatively, the RG equation for $\Gamma^f_2$ reads
\be
\partial_\Lambda\Gamma^f_2=\frac1{L^2}\sum_{\v p}\dot D^{ph}_{\v k'-\v k}(\v p)\, \left(\Gamma(\v k',\v p,\v k)\right)^2,\label{exampleGf}
\ee
where 
\ba
D^{ph}_{\v q}(\v p)&=&\int\!\frac{\ud p_0}{2\pi}\,C_\Lambda(p)C_\Lambda(p+(0,\v q))\\
&=&-\,\frac{\Theta(-\xi_{\v p}\xi_{\v p+\v q})}{|\xi_{\v p}|+|\xi_{\v p+\v q}|}\,\Theta(|\xi_{\v p}|-\Lambda)\Theta(|\xi_{\v p+\v q}|-\Lambda)
\ea
is the p-h pair propagator after the frequency-integration. Its derivative $\dot D=\partial_\Lambda D$ is zero except on the one-dimensional curves $|\xi_{\v p}|,|\xi_{\v p+\v q}|=\Lambda$, where it is positive.

In the case of a charge- or spin- instability, the vertex $\Gamma(\v k',\v p,\v k)$ develops a peak at $\v p=\v k+\v Q$, where it is equal to $\Gamma^d_3$ (see Section \ref{our}). In order to estimate the contribution of this peak to the integral in \eref{exampleGf}, we proceed as in Section \ref{consistency}. The singular contribution to the diagram is of the order $\int_0^{\sqrt\Lambda}\!\ud\kappa\,\left(\Gamma(\Lambda,\kappa,\kappa)\right)^2$, where $\Gamma(\Lambda,\v q)$ is given by Eq. \ref{gq1}. It can be checked, that this integral diverges like $(\Lambda-\Lambda_c)^{-1}$.  Similarly, in the case of superconductivity, the vertex $\Gamma(\v k',\v p,\v k)$ develops a peak close to $\v p=-\v k'$, where it approaches $\Gamma^{BCS}_3$. In both cases, we get
\ba
\partial_\Lambda\Gamma^f_2&\sim& \frac1{\Lambda-\Lambda_c}\\
\Rightarrow\qquad\Gamma^f_2&\sim&\log(\Lambda-\Lambda_c)\ \to\ -\infty.
\ea

We now use this kind of analysis to estimate the behavior of $\Gamma_4^e\pm\Gamma^e_1$, the quantity which enters Eq. \eref{gS}. First, we choose $\v k,\v k'\in P_1$ with $|\v k+\v k'|>\sqrt\Lambda$ and  write the RG equations for $\Gamma_4^e=\Gamma(\v k,\v k',\v k)$ and $\Gamma^e_1=\Gamma(\v k,\v Q-\v k',\v k)$. $\Gamma_4^e$ is renormalized by the p-p diagram and $\Gamma^e_1$ by the p-h 2 diagrams. The RG equation is
\ba
\partial_\Lambda(\Gamma_4^e\pm\Gamma^e_1)&=&\frac1{L^2}\sum_{\v p}\dot D^{pp}_{\v k+\v k'}(\v p)\, \Gamma(\v k',\v k,\v p)\Gamma(\v k,\v k',\v p)\nonumber\\
& &\mp\frac1{L^2}\sum_{\v p}\dot D^{ph}_{\v Q-\v k-\v k'}(\v p)\, \Gamma(\v p,\v Q-\v k',\v k)(2-2X)\Gamma(\v p,\v Q-\v k',\v k)\\
&=&\frac1{L^2}\sum_{\v p}\dot D^{pp}_{\v k+\v k'}(\v p)\left[\Gamma(\v k',\v k,\v p)\Gamma(\v k,\v k',\v p)\pm\Gamma(\v p,\v Q-\v k',\v k)(2-2X)\Gamma(\v p,\v Q-\v k',\v k)  \right],\nonumber
\ea
where $D^{pp}_{\v q}(\v p)$ is the frequency-integrated p-p pair propagator, $X\Gamma(\v p_1,\v p_2,\v p_3)=\Gamma(\v p_2,\v p_1,\v p_3)$ and we have used the nesting property $D^{pp}_{\v q}(\v p)=-D^{ph}_{\v Q-\v q}(\v p)$. The vertex terms in the parenthesis have two possible peaks. One at $\v p= \v k+\v Q$, where the term in the parenthesis equals $\Gamma^d_3\Gamma^x_3\pm2\Gamma_3^x(\Gamma_3^x-\Gamma^d_3)$ and one at $\v p= \v k'+\v Q$, where it equals $\Gamma^d_3\Gamma^x_3\pm2\Gamma_3^{BCS}(\Gamma_{3}^{BCS}-\Gamma^{BCS}_3)$. As one can check, the function $\dot D^{pp}_{\v k+\v k'}(\v p)$ has the same value at both peaks. We can thus simply add the two contributions and write formally
\ba
\partial_\Lambda(\Gamma_4^e+\Gamma^e_1)&\sim& -\int\ 2\left(\Gamma^x_3\right)^2\label{exeq1}\\
\partial_\Lambda(\Gamma_4^e-\Gamma^e_1)&\sim& -\int\ 2\Gamma^x_3\left(2\Gamma^d_3-\Gamma^x_3\right),\label{exeq2}
\ea 
where $\int\ldots$ indicates integration over a peak. If the peak diverges as $(\Lambda-\Lambda_c)^{-2}$, the integral diverges as $(\Lambda-\Lambda_c)^{-1}$ (and if the peak diverges as $(\Lambda-\Lambda_c)^{-1}$, the integral diverges only logarithmically, as pointed out in Section \ref{consistency}).  We therefore obtain
\Rahmen{
\be
 \mbox{SDW or SF instability:}\qquad \Gamma_4^e+\Gamma_1^e\sim -\log(\Lambda-\Lambda_c)\ \longrightarrow +\infty,\label{Eq}
\ee
}
\noindent whereas $\Gamma_4^e\pm\Gamma_1^e$ remains finite in every other case\footnote{Note that in the case of the spin instabilities SDW or SF, $\Gamma^x$ diverges twice as fast as $\Gamma^d$, such that $\Gamma^c=2\Gamma^d-\Gamma^x$ is finite.}. 

There is a competition between the first and the last term of Eq. \eref{gS}. Both terms diverge with opposite signs and our calculation is not precise enough to decide even qualitatively on the behavior of $\gamma_S$, as the SDW or SF is approached. Physically, one suspects that the uniform spin susceptibility is neither suppressed nor diverging in the SDW or SF phase, so that $\gamma_S$ should be a small number. I therefore assume that the two competing terms of  Eq. \eref{gS} essentially compensate each other. But this remains of course to be shown.

The analogous investigation of the quantities $\Gamma_4^f\pm\Gamma_2^f$, which enter Eq. \eref{gN}, leads to
\be
\partial_\Lambda\left(\Gamma_4^f\pm\Gamma_2^f\right)=\frac1{L^2}\sum_{\v p}\dot D^{pp}_{\v k+\v k'}(\v p)\,\left[\Gamma(\v k',\v k,\v p)^2\ \mp\Gamma(\v Q-\v k',\v p,\v k)^2\right]\label{forweq}
\ee
where $\v k,\v k'\in P_1$ with $\v k+\v k'>\sqrt\Lambda$. The vertex $\Gamma_4^f$ is renormalized by the p-p diagram, and $\Gamma_2^f$ by p-h 1. We have used 
$\Gamma_4^f=\Gamma(\v k,\v k',\v k')$, $\Gamma_2^f=\Gamma(\v Q-\v k',\v k,\v k)$ and $\dot D^{ph}_{\v Q-\v k-\v k'}(\v p)=-\dot D^{pp}_{\v k+\v k'}(\v p)$. 
The expression in the parenthesis on the right-hand side of Eq. \eref{forweq} has two possible peaks. One at $\v p=\v k+\v Q$, where the term in the parenthesis equals $(\Gamma^d_3)^2\mp(\Gamma^d_3)^2$ and  $\v p=\v k'+\v Q$ where it equals  $(\Gamma^x_3)^2\mp(\Gamma^{BCS}_3)^2$. As before, the propagator $\dot D^{pp}_{\v k+\v k'}(\v p)$ has the same value at both peaks. Adding the two contributions and using Eqs. \eref{exeq1} and \eref{exeq2}, one obtains 
\ba
\partial_\Lambda\left[2\Gamma_4^f-\Gamma^e_4+2\Gamma_2^f-\Gamma^e_1\right]&\sim&\int \left(\Gamma^{BCS}_3\right)^2\\
\partial_\Lambda\left[2\Gamma_4^f-\Gamma^e_4-2\Gamma_2^f+\Gamma^e_1\right]&\sim&-\int \left[\left(2\Gamma^d_3-\Gamma^x_3\right)^2+3 \left(\Gamma^x_3\right)^2+2\left(\Gamma^{BCS}_3\right)^2\right].
\ea
We conclude that
\Rahmen{
\be
 \mbox{dSC or sSC instability:}\qquad 2\Gamma_4^f-\Gamma^e_4+2\Gamma_2^f-\Gamma^e_1\ \longrightarrow -\infty,\label{e2}
\ee
\be
 \mbox{for any instability:}\qquad 2\Gamma_4^f-\Gamma^e_4-2\Gamma_2^f+\Gamma^e_1\ \longrightarrow +\infty.\label{e3}
\ee
}

Because of Eqs. \eref{e2}, \eref{e1} and \eref{gN}, there is a competition between two terms contributing to $\gamma_N$ in the case of superconductivity. The situation is similar to the spin instability discussed above. On physical grounds, the charge compressibility is neither suppressed nor diverging in the superconducting state, so there is a chance that the competing terms cancel each other and $\gamma_N$ remains small. In contrast, the terms contributing to $\gamma_{N_d}$  (Eq. \eref{gN}) are never in competition but they reinforce each other and lead to a divergence of $\chi_{N_d}$ as any instability is approached. 

An analogous analysis shows that the vertices $\Gamma^\eta_1$ and  $\Gamma^\eta_2$, which enter Eq. \eref{geta} are never divergent.

\chapter{Overview of notations for the vertices}\label{notation}
%===================================================

This appendix gives an overview of the different notations for various vertices, which have been used in chapter \ref{chapter3}.

The main object in this thesis is the 1PI vertex function $\Gamma(k_1,k_2,k_3)$, which was introduced in Section \ref{Green}. Starting from this object, six different vertex functions of two momenta are defined as follows 
\ba
\Gamma^{BCS}(k,k')&=&\Gamma(k,-k,-k'),\\
\Gamma^{d}(k,k')&=&\Gamma(k,k'+Q,k'),\\
\Gamma^{x}(k,k')&=&\Gamma(k,k'+Q,k+Q),\\
\Gamma^{\eta}(k,k')&=&\lim_{q_0\to0}\left.\Gamma(k,q-k,q-k')\right|_{\v q=\v Q},\\
\Gamma^{f}(k,k')&=&\lim_{q_0\to0}\left.\Gamma(k,k'+q,k')\right|_{\v q=\v 0},\\
\Gamma^{e}(k,k')&=&\lim_{q_0\to0}\left.\Gamma(k,k'+q,k+q)\right|_{\v q=\v 0},
\ea
where $Q=(0,\v Q)=(0,\pi,\pi)$. The last three vertex functions sometimes carry the index $\omega$ to precise that the frequency is sent to zero {\it after} the momentum $\v q$ has been set. The opposite succession of the limits is denoted by an index $\v q$. If no index is given, I always refer to the $\omega$-limit. 

Very often, the two vertex functions $\Gamma^x$ and $\Gamma^d$ are transformed into the charge and spin vertices
\ba
\Gamma^c(k,k')&=&2\Gamma^d(k,k')-\Gamma^x(k,k'),\\
\Gamma^s(k,k')&=&-\Gamma^x(k,k').
\ea
The similar combinations $2\Gamma^f-\Gamma^e$ and $-\Gamma^e$ are also important, but no symbol has been introduced for them. 

For every vertex function of two momenta $\Gamma^\diamond(k,k')$, I have introduced the following notation
\be
  \Gamma^\diamond(k,k')=\Gamma^\diamond_{ij}\mbox{ if }\v k\in P_i,\,\v k'\in P_j,
\ee
where $P_1$ and $P_2$ are the two saddle point patches. Every time this notation is used, it is understood that $\v k$ and $\v k'$ are {\it general} momenta in the saddle point patches (i.e. for example $\Gamma^x_{12}\neq\Gamma^d_{12}$, although $\Gamma^x(k,k+Q)=\Gamma^d(k,k+Q)$). Obviously, $\Gamma^\diamond_{11}=\Gamma^\diamond_{22}$ and $\Gamma^\diamond_{12}=\Gamma^\diamond_{21}$. The odd and even combinations are denoted
\be
\Gamma^\diamond_\pm=\Gamma^\diamond_{11}\pm\Gamma^\diamond_{12}.
\ee

An alternative classification of the same special vertices $\Gamma^\diamond_{ij}$ is obtained as follows. If the momenta are restricted to the saddle point patches, the vertex function $\Gamma(k_1,k_2,k_3)$ ``decays'' into four disconnected functions (see Fig. \ref{g14fig})
\be
\Gamma(k_1,k_2,k_3)\equiv\left\{\begin{array}{ll} 
\Gamma_1(k_1,k_2,k_3)\;;\v k_1,\v k_3\in P_1\mbox{ and }\v k_2,\v k_4\in P_2\\ 
\Gamma_2(k_1,k_2,k_3)\;;\v k_1,\v k_4\in P_1\mbox{ and }\v k_2,\v k_3\in P_2\\
\Gamma_3(k_1,k_2,k_3)\;;\v k_1,\v k_2\in P_1\mbox{ and }\v k_3,\v k_4\in P_2\\ 
\Gamma_4(k_1,k_2,k_3)\;;\v k_1,\ldots,\v k_4\in P_1
\end{array}\right..
\ee
In each of these functions I specify one general vertex and three special vertices which correspond to specific configurations of the momenta  
\be 
\Gamma_1(k_1,k_2,k_3)\approx\left\{\begin{array}{ll}
\Gamma_1&\mbox{if }\ |\v k_3-\v k_2- \v Q|,\ |\v k_3-\v k_1 |,\ |\v k_1+\v k_2-\v Q|>O(\sqrt{\Lambda})\\
{ \Gamma_1^{\eta}=\Gamma_{12}^{\eta}}&\mbox{if }\ \v k_1+\v k_2= \v Q\\
{ \Gamma_1^e}=\Gamma_{12}^{e}&\mbox{if }\ \v k_3-\v k_1= \v 0\\
{ \Gamma_1^d}=\Gamma_{11}^{d}&\mbox{if }\ \v k_3-\v k_2= \v Q
\end{array}\right.,
\ee
\be 
\Gamma_2(k_1,k_2,k_3)\approx\left\{\begin{array}{ll}
\Gamma_2&\mbox{if }\ |\v k_3-\v k_2|,\ |\v k_3-\v k_1-\v Q|,\ |\v k_1+\v k_2-\v Q|>O(\sqrt{\Lambda})\\
{ \Gamma_2^{\eta}}=\Gamma_{11}^{\eta}&\mbox{if }\ \v k_1+\v k_2= \v Q\\
{ \Gamma_2^x}=\Gamma_{11}^{x}&\mbox{if }\ \v k_3-\v k_1= \v Q\\
{ \Gamma_2^f}=\Gamma_{12}^{f}&\mbox{if }\ \v k_3-\v k_2= \v 0
\end{array}\right.,
\ee
\be 
\Gamma_3(k_1,k_2,k_3)\approx\left\{\begin{array}{ll}
\Gamma_3&\mbox{if }\ |\v k_3-\v k_2-\v Q|,\ |\v k_3-\v k_1-\v Q|,\ |\v k_1+\v k_2|>O(\sqrt{\Lambda})\\
{ \Gamma_3^{BCS}}=\Gamma_{12}^{BCS}&\mbox{if }\ \v k_1+\v k_2= \v 0\\
{ \Gamma_3^x}=\Gamma_{12}^{x}&\mbox{if }\ \v k_3-\v k_1= \v Q\\
{ \Gamma_3^d}=\Gamma_{12}^{d}&\mbox{if }\ \v k_3-\v k_2= \v Q
\end{array}\right.,
\ee
\be 
\Gamma_4(k_1,k_2,k_3)\approx\left\{\begin{array}{ll}
\Gamma_4&\mbox{if }\ |\v k_3-\v k_2|,\ |\v k_3-\v k_1|,\ |\v k_1+\v k_2|>O(\sqrt{\Lambda})\\
{ \Gamma_4^{BCS}}=\Gamma_{11}^{BCS}&\mbox{if }\ \v k_1+\v k_2= \v 0\\
{ \Gamma_4^e}=\Gamma_{11}^{e}&\mbox{if }\ \v k_3-\v k_1= \v 0\\
{ \Gamma_4^f}=\Gamma_{11}^{f}&\mbox{if }\ \v k_3-\v k_2= \v 0
\end{array}\right..
\ee

The combinations, which are important for the different susceptibilities are the following
\be
\begin{array}{rll}
SDW/SF:&\ \Gamma_\pm^s&=-\Gamma^x_2\mp \Gamma^x_3,\\
CDW/CF:&\ \Gamma_\pm^c&=2\Gamma^d_1-\Gamma^x_2\pm (2\Gamma^d_3-\Gamma^x_3),\\
sSC/dSC:&\ \Gamma_\pm^{BCS}&=\Gamma^{BCS}_4\pm \Gamma^{BCS}_3,\\
S/S_d:&\ -\Gamma^e_\pm&=-\Gamma^e_4\mp\Gamma^e_1,\\
N/N_d:&\ 2\Gamma^f_\pm-\Gamma^e_\pm&=2\Gamma^f_4-\Gamma^e_4\pm(2\Gamma^f_2-\Gamma^e_1),\\
\eta_s/\Pi_d:&\ \Gamma^\eta_\pm&=\Gamma^\eta_2\pm\Gamma^\eta_1.
\end{array}
\ee

%\nocite{*}
\bibliographystyle{alpha}
\bibliography{references}

\end{document}